# Generalized Principle of Limiting 4-Dimensional Symmetry. Solution of the "Two-Spaceship Paradox"

Jaykov Foukzon
Israel Institute of Technology

jaykovfoukzon@list.ru

**Abstract:** A "Two-Spaceship Paradox" in special relativity is resolved and discussed. We demonstrate a nonstandard resolution to the "two-spaceship paradox" by explicit calculation using Generalized Principle of limiting 4-dimensional symmetry proposed in previous paper [1]. The physical and geometrical meaning of the nonholonomic transformations used in special relativity is determined.

# Table of contents

# I.Introduction.

Imagine that a spaceship with a proper length $L$ stays still at first in the inertial frame $F_I$, and then starts accelerating to reach a steady speed $u$. After the steady state is reached, the length of the spaceship observed from $F_I$ contracts from $L$ to $L'$, where

$$L' = L\left(\sqrt{1 - \frac{u^2}{c^2}}\right). \tag{1.1.1}$$

This is the Lorentz contraction. Now, let us imagine two spaceships of the same type, **A** and **B**, which stay still at first in the inertial frame $S$, the distance between the two spaceships being $L$. At $t = 0$ these spaceships start accelerating in the same direction along the line joining **A** and **B**, undergo the same acceleration for the same duration, stop accelerating at the same time and reach a steady speed $u$, all viewed from $S$. Now we ask the question: "What is the distance between these two spaceships after they reach the steady speed $u$, observed from $S$? Is it $L$ or $L'$?" The answer is obvious: it is $L$. The distance between two spaceships does not undergo Lorentz contraction contrarily to the length of one spaceship.

Since the problem has been made famous by the famous physicist John Bell (1976), it is now widely known as **Bell's problem** or **Bell's spaceship paradox.**

According to the testimony of John Bell (1976), a polemic over this old problem that was started once between him and a distinguished experimental physicist in the CERN canteen was eventually passed on to a significantly broader forum for arbitration: the **CERN** theory division. A clear consensus emerged, testifies Bell, that the thread would not break.

It is now accepted, however, that the answer is wrong. The elementary explication, in Bell's formulation, runs as follows: 'if the thread is just long enough to span the required distance initially, then as the rockets speed up, it will become too short, because of its need to Fitzgerald contract, and must finally break. It must break when, at a sufficiently high velocity, the artificial prevention of the natural contraction imposes intolerable stress' (Bell 1976, emphasis added).

In the paper [2], by Jong-Ping Hsu and Nobuhiro Suzuki was shown that one can obtain a spacetime transformation between an inertial frame $F_I(t_I, x_I, y_I, z_I)$ and a frame undergoing constant linear acceleration (CLA) $F(t, x, y, z)$ characterized by a metric with the form $ds^2 = W^2 dw^2 - dx^2 - dy^2 - dz^2$, and that such a transformation can resolve the apparent paradox.

*However the physical meaning of constant-linear-acceleration of this reference frame is as follows:* the relativistic momentum $p_{Ix}$ (or energy $p_{I0}$) of such a particle satisfies $dp_{Ix}/dt_I =$ constant (or $dp_{I0}/dx_I =$ constant), as measured in an inertial frame [2].

**Remark** *Thus this property agrees only with the motion of particles in a high energy linear accelerator which has a constant potential dropper unit length along the particle beam, but dont agrees with original* **Bell's problem**

## II. The Rockets-and-String Paradox.
## II.1. Historical background.

One of the most important new physical insights given in Einstein's seminal paper on Special Relativity (SR) [10] was the realisation that the Lorentz-Fitzgerald Contraction (LFC), which had previously been interpreted by Lorentz and Poincaré as an electrodynamical effect, was most easily understood as a simple consequence of the space-time Lorentz Transformation (LT), i.e. as a geometrical effect. The 'Rockets-and-String' and

'Pole-and-Barn' as well as the similar 'Man-and-Grid' and 'Rod-and-Hole' paradoxes have all been extensively used in text books on SR, for example in Taylor and Wheeler (1966) and, more recently, by Tipler and Lewellyn (2000).

There is nothing in Einstein's 1905 SR paper to suggest that the LFC should be considered as a 'real' rather than an 'apparent' effect. Specifically, Einstein wrote [11]:

> *Thus, whereas the $X$ and $Z$ dimensions of the sphere (and therefore every rigid body of no matter what form) do not appear modified by the motion, the $X$ dimension appears shortened in the ratio $1 : \sqrt{1 - \mathbf{v}^2/c^2}$, i.e. the greater the value of $\mathbf{v}$ the greater the shortening.*

In the original German the crucial phrase is: 'erscheint die X-Dimension im

Verhältnis
$1 : \sqrt{1 - \mathbf{v}^2/c^2}$ ... '. The verb 'erscheinen', translated into English [12] means 'to appear'. Einstein never stated, in Reference [11], that the LFC is a 'real' effect.

Let us consider a three small spaceships **A**, **B** and **C** drift freely in a region of space remote from other matter, without rotation and relative motion, with **B** and **C** equidistant from A . The spaceships are at rest relative to an inertial frame $F_I = F_I(t_I, x_I, y_I, z_I)$.

At one moment two identical signals from **A** are emitted towards **B** and **C**. On simultaneous (with respect to $F_I$) reception of these signals the motors of **B** and **C** ignited and they accelerate gently along the straight line connecting them. Let ships **B** and **C** be identical, and have identical acceleration programmes. Then each point of **B** will have at every moment the same velocity as the corresponding point of **C**, and thus any two corresponding points of the ships will always be at the same distance from one another, all measured in $F_I$. Let us suppose that a fragile thread connects two identical projections placed exactly at the midpoints of **B** and **C** before the motors were started. If the thread with no stress is just long enough to span the initial distance in question, then as the ships accelerate the thread travels with them. Assume that the thread does not affect the motion of the ships. Will the thread break when **B** and **C** reach a sufficiently high speed?

This fascinating riddle was devised by Dewan and Beran (1959) as an illustration of the reality of the Fitzgerald–Lorentz contraction and especially of the reality of stress effects due to artificial prevention of the relativistic length contraction. Dewan and Beran's original formulation of the riddle was corrected by Evett and Wangsness (1960), and was recently criticized by Cornwell (2005).

In the 'tough' variant of the problem, the acceleration of ships **B** and **C** may never cease and their speed may increase indefinitely approaching c, as measured in $F_I$ (Dewan and Beran 1959, Bell 1976, Gershteïn and Logunov 1998, Flores 2005, 2008).   In its 'mild' variant, at an instant of the $F_I$ time the ships' acceleration ceases and they coast with the same constant velocity as measured in $F_I$ (Dewan 1963, Evett 1972, Tartaglia and Ruggiero 2003, Matsuda and Kinoshita 2004, Styer 2007).

It was correctly concluded that the real distance between the objects in $F_I$ would remain unchanged throughout the aceleration procedure. However, it was not stated that, after acceleration, the proper separation between the rockets is the same as their original separation in $F_I$. Dewan and Beran then introduced a continous string attached between the rockets during the acceleration and drew a distinction between two distances:

*(a) the distance between two ends of a connected rod and*
*(b) the distance between two objects which are not connected but each of which independently and simultaneously moves with the same velocity with respect to an inertial frame*

It was then stated (without any supporting argument) that the distance (a) is subject to the LFC and (b) not. Replacing the 'connected rod' of (a) by a continous string attached between the rockets, it was concluded that the string would be stressed and ultimately break, since the distance between the rockets does not change, whereas the string shrinks due to the LFC. Several arguments were given why the real distance between the rockets does not change. The conclusion that the string would be stressed and break was due to the failure to discriminate between the real separation of the rockets (correctly calculated) and the apparent contraction due to the LFC, which as
correctly pointed out by Nawrocki (1962) applies equally to the distance between the ends of the string and that between the points on the rockets to which it is attached. Dewan and Beran's distinction between the distances (a) and (b) is then wrong. Both the distance between the points of attachment of the string and the length of the string undergo the same apparent LFC. There is no stress in the string. It does not break.

Dewan did not respond to Nawrocki's objection that stated the equivalence of the length of an extended object and the distance between two independant objects separated by a distance equal to this length, but instead introduced a new argument, claiming that Nawrocki had not correctly taken into account the relativity of simultaneity of SR. Dewan considered the sequence of events corresponding to the firing of the rockets as observed in $F'_I$, the co-moving frame of the rockets after acceleration. It was concluded that the spring breaks because, in this case, the final separation of the rockets is $\gamma L$ :

$$\gamma L = \frac{L}{\sqrt{1 - \mathbf{v}^2/c^2}}. \qquad (2.1.1)$$

## II.2. The generalized inertial non-orthogonal frames of reference. Physical distance affine and coordinate distances.

Einstein's theory of special relativity is based on two postulates:
▲ The speed of light is constant for all observers,
▲ The laws of physics are the same in any inertial frame of reference.
▲ Additionally, one of the fundamental assertions of Einstein's relativity is that lengths and time can be measured by a set of imaginary arbitrary elements called **"measuring rigid rods"** and **"clocks"**. Unfortunately, these elements do not have proper metrics based on physical reality, because Einsteinian relativity do not discuss what basic concepts of distance, time, and mass are, and how the relativistic mechanism is ensured in the smallest scale.
▲ Additionally, our hypothesis suggests that every observer is able to determine a concrete definition of the metric, which emerges as a property of the whole space geometry by observing its own state in smallest scale. In fact, this concrete definition is the physical formulation of Einstein's **"measuring rigid rods"** and **"clocks"**. As a result, the proper definition of the spatial metric explains the relativistic correlation between fundamental concepts like distance and time.

**Claim**  2.2.1 Hence (a) **physical proper distance (lengths)** $l^{ph}$ this is the lengths which was experimentally measured by Einstein's **rigid rod** only [31 p.188-190].

(b) **physical proper time** $\tau^{ph}$ this is the time which was experimentally measured by Einstein's **clocks** only [31].

As known [30], constructing the covariant SRT one should exactly distinguish a *coordinate velocity* $dx/dt$ of a particle. The latter is defined as the ratio of the "physical" (in some sense formal-mathematical) distance $dl$ and "physical" (in some sense formal-mathematical) time $d\tau$ :

$$ds^2 = g_{ij}dx^i dx^j = c^2 d\tau^2 - dl^2; i,j = 0,1,2,3,$$

$$d\tau_{\underline{ph}} = \sqrt{g_{00}}\, dt + \frac{g_{0\alpha}dx^\alpha}{c \cdot \sqrt{g_{00}}},$$

$$dl = dl_{\underline{ph}} = \left(-g_{\alpha\beta} + \frac{g_{0\alpha}g_{0\beta}}{g_{00}}\right)dx^\alpha dx^\beta = \kappa_{\alpha\beta}dx^\alpha dx^\beta,$$

$$\alpha, \beta = 1,2,3.$$

(2.2.1)

The "physical" distance $dl_{\underline{ph}}$ these is only a version of the affine distances defined in [45].

**Remark**  2.2.1 In general case (a) the "physical" distance $dl_{\underline{ph}}$ doesn't coincide with a **physical proper distance (lengths) $dl^{ph}$** : $dl^{ph} \neq dl_{\underline{ph}}$

(b) the "physical" time $d\tau_{\underline{ph}}$ doesn't coincide with a **physical proper time $d\tau^{ph}$** : $d\tau^{ph} \neq d\tau_{\underline{ph}}$. (see Theorem 2.2.1).

To measure distances in GRT formalism we have to measure "lengths" of light rays, as in the Introduction [45]. But light rays are lightlike curves and they have no length.  To measure distances and angles, an observer has to project these light rays onto its physical space (i.e. the orthogonal space of its 4-velocity). This idea drives us to the  next definition of distance:

**Definition**  (definition 7.[45]) Let be a light ray from $q$ to $p$ and let $u$ be an observer at $p$. The affine distance from $q$ to $p$ observed by $u$, $d_u(q,p)$, is the module of the projection of $\exp_p^{-1}(q)$ onto $u^\perp$.

**Definition**  (definition 10.[45]) Let $\beta, \beta'$ be two observers. The affine distance from $\beta'$ to $\beta$ observed by $\beta$ is a real positive function $d_\beta$ defined on $\beta$ such that, given $p \in \beta, d_\beta(p)$ is the affine distance from $q$ to $p$ observed by $u$, where $u$ is the 4-velocity of at p, and $q$ is the

*unique event of ' such that there exists a light ray from $q$ to $p$.*

**Example** *2.2.1 In the Minkowski metric with rectangular coordinates $ds^2 = -dt^2 + dx^2 + dy^2 + dz^2$, $c \triangleq 1$, let us consider an event $q = (t_1, x_1, y_1, z_1)$ observed at $p = (t_2, x_2, y_2, z_2)$ by an observer*

$$u = \gamma\left(\left[\frac{\partial}{\partial t}\right]_p + v^x\left[\frac{\partial}{\partial x}\right]_p + v^y\left[\frac{\partial}{\partial y}\right]_p + v^z\left[\frac{\partial}{\partial z}\right]_p\right),\text{ where } \gamma$$

*is the gamma factor given by $1/\sqrt{1-(v^x)^2-(v^y)^2-(v^z)^2}$.*
*Then, using Eq. $d_u(q,p) = g(exp_p^{-1}(q), u)$, we have the general expression for the affine distance from $q$ to $p$ observed by $u$:*

$$d_u(q,p) = g(q-p, u) = \gamma((t_2-t_1) + v^x(x_1-x_2) + v^y(y_1-y_2) + v^z(z_1-z_2)).$$

Note that $(t_2 - t_1) = \sqrt{(x_1-x_2)^2 + (y_1-y_2)^2 + (z_1-z_2)^2}$ because there is a light ray from $q$ to $p$. There exists a known method to measure distances between an observer $\beta$ (that we can suppose parameterized by its proper time $\tau$) and an observed event $q$, called "radar method", consisting on emitting a light ray from $\beta(\tau_1)$ to $q$, that bounces and arrives at $p = \beta(\tau_2)$. The radar distance between $\beta$ and $q$ observed by is given by $1/2(\tau_2 - \tau_1)$, $c \triangleq 1$ [46]. So, considering a geodesic observer $\beta$ passing through $p$ with 4-velocity

$$u = \gamma\left(\left[\frac{\partial}{\partial t}\right]_p + v^x\left[\frac{\partial}{\partial x}\right]_p + v^y\left[\frac{\partial}{\partial y}\right]_p + v^z\left[\frac{\partial}{\partial z}\right]_p\right) \text{ at } p \text{ we have that}$$

$\beta(\tau) = (\gamma(\tau-\tau_2)+t_2, \gamma v^x(\tau-\tau_2)+x_2, v^y(\tau-\tau_2)+y_2, \gamma v^z(\tau-\tau_2)+z_2)$ is the parameterization by its proper time. Setting out that $q - \beta(\tau_1)$ is lightlike and $\tau_2 - \tau_1 \neq 0$, finally we obtain:

$$\frac{1}{2}(\tau_2 - \tau_1) = \gamma((t_2-t_1) + v^x(x_1-x_2) + v^y(y_1-y_2) + v^z(z_1-z_2)).$$

Thus we state that affine distance coincides with radar distance for geodesic observers in Minkowski space-time.

**Definition** *2.2.2 In SRT, we postulate that in the whole space there is a physical frame of reference (FR) called an inertial orthogonal (orthogonal Galilee) one in which the interval between events of this space is written as*

$$ds^2 = c^2 dT^2 - dX^2 - dY^2 - dZ^2. \tag{2.2.2}$$

**Definition** *2.2.3 In general, in the Minkowski space-time any FR in which the interval $ds^2$ has the general form*

$$ds^2 = g_{ik}(x)dx^i dx^k, x \in \mathbb{R}^4,$$
$$x = (x^0, x^1, x^2, x^3) = (x^0, \mathbf{x}),$$
$$\mathbf{x} \in \mathbb{R}^3; i,j = 0,1,2,3. \quad (2.2.3)$$
$$g_{00}(x) > 0, g_{\alpha\beta}(x)dx^\alpha dx^\beta < 0,$$
$$\alpha, \beta = 1,2,3$$

*The curvature tensor becomes zero identically: $R_{iklm} = 0$ in any FR of the Minkowski space, including non-inertial (accelerated) one.*

**Definition** 2.2.4 (a) *The metric (2.2.3) with $g_{0\alpha}(x) \neq 0$ gives a general non-orthogonal frame of reference in the SRT including non-inertial (accelerated) one.*

*(b) The metric (2.2.3) with $g_{0\alpha}(x) = 0, \alpha = 1,2,3$ gives a general orthogonal frame of reference in the SRT including non-inertial (accelerated) one.*

**Theorem** **Theorem** 2.2.1(a) *The "physical" distance $dl_{ph}$ coincide*
**Remark** *with a **physical proper distance (lengths) $dl^{ph}$** iff the metric (2.2.3) gives a general orthogonal frame of reference, i.e.*

$$\mathbf{dl^{ph}} = dl_{\underline{ph}} \Leftrightarrow g_{0\alpha}(x) = 0, \alpha = 1,2,3 \quad (a)$$

**Remark** *(b) the "physical" time $d\tau_{ph}$ coincide with a physical proper time $d\tau^{ph}$ iff the metric (2.2.3) gives a general orthogonal frame of reference, i.e.*

$$\mathbf{d\tau^{ph}} = d\tau_{\underline{ph}} \Leftrightarrow g_{0\alpha}(x) = 0, \alpha = 1,2,3 \quad (b)$$

Let us pass from Galilee's coordinates $X^i = (X,Y,Z,T)$ with the metric (2.2.2) to coordinates $x^i = (x,y,z,t)$ by arbitrary linear transformation. This transformation is equivalent up to a space axis rotation to a transformation in plane $X,T$:

$$X = ax + bt; T = qx + pt; Y = y; Z = z. \quad (2.2.4)$$

Substituting (2.2.4) in (2.2.2), in coordinates $x^i$ the metric gets the form:

$$ds^2_{GI} = c^2 g_{00}dt^2 + 2cg_{01}dtdx + g_{11}dx^2 - dy^2 - dz^2,$$
$$g_{00} = p^2 - b^2/c^2; g_{01} = c(pq - ab/c^2); g_{11} = c^2 q^2 - a^2. \quad (2.2.5)$$

The transformation (2.2.4) describes the rotation of the axes $x,t$ in the plane $X,T$, with after the rotation the axis $x$ can be not orthogonal to the axis $t$, i.e. $x$ and $t$ rotate on angles, which may differ.

**Definition** 2.2.5 *The metric (2.2.5) gives a general inertial non-orthogonal frame of reference in the SRT.*

**Definition** 2.2.6 Let us consider the metric $ds^2(x^0)$ (2.2.3) with $g_{0\alpha}(x^0,\mathbf{x}) \neq 0$, $g_{00}(x^0,\mathbf{x}) = g_{00}(x^0)$ which gives a non-inertial (accelerated) non-orthogonal frame $FR_{x^0}$ of reference and the metric $ds^2_{GI}$ (2.2.5) which gives a general inertial non-orthogonal frame $GIFR$ of reference.

We define general inertial non-orthogonal frame $GIFR$ **instantly coincides** with non-inertial (accelerated) non-orthogonal frame $FR_{x^0}$ at instant $\bar{x}^0$ by equality: $ds^2(\bar{x}^0) \equiv ds^2_{GI}$.

**Definition** 2.2.7 Let us consider the metric $ds^2(x^0)$ (2.2.3) with $g_{0\alpha}(x^0,\mathbf{x}) \neq 0$, $g_{00}(x^0,\mathbf{x}) = g_{00}(x^0)$ which gives a non-inertial (accelerated) non-orthogonal frame $FR_{x^0}$ of reference. We define **instant physical proper distance (lengths)** $l^{ph}(x^0,\mathbf{x}_1,\mathbf{x}_2)$ between two points $\alpha_1,\alpha_2 \in FR_{x^0}$ with coordinates $\mathbf{x}_1,\mathbf{x}_2 \in \mathbb{R}^3$ by **physical proper distance (lengths)** $l^{ph}(\mathbf{x}_1,\mathbf{x}_2)$ which measured by an observer in general inertial non-orthogonal frame $GIFR$ **instantly coincides** with non-inertial (accelerated) non-orthogonal frame $FR_{x^0}$ at instant $x^0$.

**Remark** 2.2.2 The Lorentz transformations are a particular case of the general linear transformations (2.2.4), corresponding to the choice $g_{00} = 1, g_{01} = 0, g_{11} = -1$ in (2.2.5). Hence, the metric (2.2.5), in contrast to (2.2.2), is not forminvariant with respect to the Lorentz transformations. But the metric (2.2.5) is forminvariant that with respect to the so-called generalized inertial Lorentz-Poincare group $\hat{L}^i_m$ connected with the classic Lorentz-Poincare group $L^i_m$ one by the relation:

$$\hat{L}^n_k x^k = [\Lambda^n_i L^i_m (\Lambda^{-1})^m_k] x^k,$$

where $\Lambda^i_k$ is the matrix of the linear (non-orthogonal) transformations (2.2.4)

$$x^i = \Lambda^i_k x'^k.$$

forming the generalized inertial FR (2.2.5).

**Definition** 2.2.8 The simplest generalized inertial non-orthogonal FR is the FR connected with the inertial one (2.2.2) by the classic Galilee transformation: $X = x + Vt; t = T$, and corresponding to rotation of the axis $T$ with fixed orientation of the axis $X$.

Let us consider the simplest dependence $t = T$ at the part of the inertial motion by the law

$$X = x + \mathbf{v}t. \tag{2.2.6}$$

Thus, the metric (2.2.2) get the form:

$$ds^2 = \left(1 - \frac{\mathbf{v}^2}{c^2}\right)c^2 dt^2 - 2\mathbf{v}\,dt\,dx - dx^2 - dy^2 - dz^2. \tag{2.2.7}$$

From Eqs. (2.2.1) and (2.2.7) we then have (for motion along the axis $x$):

$$dl_{\underline{ph}} = \frac{dx}{\sqrt{1 - \frac{\mathbf{v}^2}{c^2}}},$$

$$l_{\underline{ph}} = \frac{x_2 - x_1}{\sqrt{1 - \frac{\mathbf{v}^2}{c^2}}}. \tag{2.2.8}$$

**Remark** *2.2.3 It is obvious, that the metric (2.2.7) unfortunately gives inertial **non-orthogonal** FR. Thus from Theorem 2.2.1(a) we conclude that:*

$$l^{ph} \neq l_{\underline{ph}}. \;!!!$$

Let a relativistic uniformly accelerated frame $F_a$ with the coordinates $(x, t)$ move without initial velocity along the axis $X$ of an inertial Galilee rame (2.2.2) with the coordinates $(X, T)$ and at $t = T = 0$ their origins coincide. Then the coordinate transformation formulas $x$ have the form [30]:

$$x = X - \frac{c^2}{a}\left[\sqrt{1 + \frac{a^2 T^2}{c^2}} - 1\right]. \tag{2.2.9}$$

In this subsection we consider the simplest possible dependence $t = T$. In this case, substituting (2.2.9) and $t = T$ into the metrics (2.2.2) gives the expression of the metrics of a uniformly accelerated FR

$$ds^2 = \frac{c^2}{1 + \frac{a^2 t^2}{c^2}} dt - \frac{2at}{\sqrt{1 + \frac{a^2 t^2}{c^2}}} dt\,dx - dx^2 - dy^2 - dz^2. \tag{2.2.10}$$

From (2.2.1) by simple calculation we obtain:

$$dl_{ph} = dx\sqrt{1 + \frac{a^2 t^2}{c^2}},$$

$$l_{ph} = (x_2 - x_1)\sqrt{1 + \frac{a^2 t^2}{c^2}}. \tag{2.2.11}$$

**Remark** *2.2.4 It is obvious, that the metric (2.2.7) unfortunately gives a non-inertial non-orthogonal frame. Thus from Theorem 2.2.1(a)*

*we can conclude that:*

$$I^{ph} \neq l_{\underline{ph}}. \;!!!$$

**Theorem** **Theorem** *2.2.2 Let us consider an non-inertial (accelerated) observer in*

*the FR*
*with the metric $ds^2(x^0)$ (2.2.3) $g_{ij} = g_{ij}(x^0, \boldsymbol{x})$, which gives an non-inertial (accelerated) frame $FR_{x^0}$ of reference*

**Definition** *and two points $\alpha_1, \alpha_2 \in FR_{x^0}$ with coordinates $\boldsymbol{x}_1, \boldsymbol{x}_2 \in \mathbb{R}^3$.*

**Theorem** *Let us consider instantaneously comoving fundamental inertial observer in the **orthogonal** frame $\boldsymbol{S}'_{t^0} \triangleq \boldsymbol{S}'_{t^0}(t', x')$.*
*Then the distance $d(t^0, \alpha_1, \alpha_2)$ which measured by an observer in $\boldsymbol{S}'_{t^0}$ coincide with **instant physical proper distance (lengths)** $I^{ph}(x^0, \boldsymbol{x}_1, \boldsymbol{x}_2)$ between two points $\alpha_1, \alpha_2 \in FR_{x^0}$ with coordinates $\boldsymbol{x}_1, \boldsymbol{x}_2 \in \mathbb{R}^3$ iff the frame $FR_{x^0}$ is **orthogonal** in instant $x^0$, i.e. $g_{0\alpha}(x^0, \boldsymbol{x}) = 0, \alpha = 1, 2, 3$.*

**Corollary** **Theorem** *2.2.3 Let us consider the metric $ds^2_{GI}$ (2.2.5) with $g_{ij} = const$,*

**Definition** *which gives an inertial frame $GIFR$ of reference. Let us consider instantaneously comoving fundamental inertial*

**Theorem** *observer in the **orthogonal** frame $\boldsymbol{S}'_{t^0} \triangleq \boldsymbol{S}'_{t^0}(t', x')$.*
*Then the distance $d(t^0, \alpha_1, \alpha_2)$ which measured by an observer in $\boldsymbol{S}'_{t^0}$ coincide with **physical proper distance (lengths)** $I^{ph}(x^0, \boldsymbol{x}_1, \boldsymbol{x}_2)$ between two points $\alpha_1, \alpha_2 \in GIFR$ with coordinates $\boldsymbol{x}_1, \boldsymbol{x}_2 \in \mathbb{R}^3$ iff the frame $GIFR$ is **orthogonal**, i.e. $g_{0\alpha} = 0, \alpha = 1, 2, 3$.*

**Example** *2.2.2 Let us now consider a situation where two equal rockets are initially placed on the $x_I$ axis at a distance $L$ from one another. Every rocket carries on board a scientist to make measurements, and an engineer to control the thrust of the rocket. The engineers carry identical (initially) synchronized clocks and have the same instructions for the regime of the engines. Let us call **B** the front rocket (and moving observer), and **A** the rear rocket, with its observer. **A** and **B** are not physically connected, so that they move exactly with the same velocity **v** at any time.*

Let $F_I = F_I(t_I, x_I)$ be a stationary inertial frame and $F_{\mathbf{v}}$ the generalized inertial frame of an observers on the rockets. We assume that we know the functions

$$x_A(t_A) = \mathbf{v} t_A \qquad (2.2.12)$$

and

$$x_B(t_A) = L + \mathbf{v} t_A, \qquad (2.2.13)$$

so we have

$$\mathbf{v}(t_A) = \frac{dx_A(t_A)}{dt_A} = \mathbf{v}. \tag{2.2.14}$$

The Eq. (2.2.14) defines the infinite sequence of instantaneously comoving inertial *orthogonal frames* $\mathbf{S}'_{t_A} = \mathbf{S}'_{t_A}(T,X) \triangleq \mathbf{S}'_{t_A}(t',x')$. The rocket **A** is instantaneously at rest for an observer in this frames $\mathbf{S}'_{t_A}$. Thus for coordinate distance $L'(t')$ between rockets **A** and **B** which measured by observer on the frame $\mathbf{S}'_{t_A}$ at the instant $t'$ we obtain:

$$L'(t') = x'_B(t') - x'_A(t'), \tag{2.2.15}$$

From Lorentz transformation by usual way we obtain:

$$x_A = \gamma(\mathbf{v})[x'_A + \mathbf{v}t'],$$

$$t_A = \gamma(\mathbf{v})\left[t' + \frac{\mathbf{v}}{c^2}x'_A\right],$$

$$x_B(t_B) = \gamma(\mathbf{v})[x'_B + \mathbf{v}t'], \tag{2.2.16}$$

$$t_B = \gamma(\mathbf{v})\left[t' + \frac{\mathbf{v}}{c^2}x'_B\right],$$

where $\gamma(\mathbf{v}) = 1/\sqrt{1-\mathbf{v}^2/c^2}$, $x'_A \triangleq x'_A(t'), x'_B \triangleq x'_B(t'), x_A \triangleq x_A(t_A), x_B \triangleq x_B(t_B)$. By setting $x'_A(t') = 0$, from (2.2.16) we obtain:

$$x_A(t_A) = x_A = \gamma(\mathbf{v})\mathbf{v}t', \quad (2.2.17.a)$$

$$t_A = \gamma(\mathbf{v})t', \quad (2.2.17.b)$$

$$x_B(t_B) = x_B = \gamma(\mathbf{v})[x'_B + \mathbf{v}t'], \quad (2.2.17.c)$$

$$t_B = \gamma(\mathbf{v})\left[t' + \frac{\mathbf{v}}{c^2}x'_B\right]. \quad (2.2.17.d)$$

(2.2.17)

$$x'_B = L'(t'), L'(0) = L.$$

Thus

$$x_B(t_B) - x_A(t_A) = \gamma(\mathbf{v})x'_B = \gamma(\mathbf{v})L'(t'). \quad (2.2.18)$$

From Eqs.(2.2.17)-(2.2.16) we obtain:

$$L + \mathbf{v}t_B - \mathbf{v}t_A = \gamma(\mathbf{v})L'(t'),$$

$$L + \mathbf{v}(t_B - t_A) = \gamma(\mathbf{v})L'(t'). \quad (2.2.19)$$

From Eqs.(2.2.19),(2.2.17) we obtain:

$$L + \mathbf{v}\left(\gamma(\mathbf{v})\left[t' + \frac{\mathbf{v}}{c^2}x'_B\right] - \gamma(\mathbf{v})t'\right) =$$

$$= L + \gamma(\mathbf{v})\frac{\mathbf{v}^2}{c^2}x'_B = L + \gamma(\mathbf{v})\frac{\mathbf{v}^2}{c^2}L'(t').$$

(2.2.20)

$$L + \gamma(\mathbf{v})\frac{\mathbf{v}^2}{c^2}L'(t') = \gamma(\mathbf{v})L'(t').$$

$$L = \gamma(\mathbf{v})L'(t')\left(1 - \frac{\mathbf{v}^2}{c^2}\right) = \gamma^{-1}(\mathbf{v})L'(t').$$

Thus

$$L'(t') = \frac{L}{\sqrt{1 - \frac{\mathbf{v}^2}{c^2}}}. \quad (2.2.21)$$

From Eqs.(2.2.12),(2.2.16) we obtain:

$$x_A(t_A) = \mathbf{v}t_A,$$

$$\gamma(\mathbf{v})[x'_A + \mathbf{v}t'], t_A = \gamma(\mathbf{v})\left[t' + \frac{\mathbf{v}}{c^2}x'_A\right],$$

$$\gamma(\mathbf{v})[x'_A + \mathbf{v}t'] = \mathbf{v}\gamma(\mathbf{v})\left[t' + \frac{\mathbf{v}}{c^2}x'_A\right],$$

$$x'_A + \underline{\mathbf{v}t'} = \underline{\mathbf{v}t'} + \frac{\mathbf{v}^2}{c^2}x'_A,$$

(2.2.22)

$$x'_A = \frac{\mathbf{v}^2}{c^2}x'_A$$

$$\frac{dx'_A}{dt'} = \frac{\mathbf{v}^2}{c^2}\frac{dx'_A}{dt'},$$

$$\frac{dx'_A}{dt'}\left(1 - \frac{\mathbf{v}^2}{c^2}\right) = 0.$$

Hence

$$\frac{dx'_A}{dt'} \equiv 0. \qquad (2.2.23)$$

From Eqs.(2.2.13),(2.2.17) we obtain:

$$x_B(t_B) = L + \mathbf{v}t_B,$$

$$\gamma(\mathbf{v})[x'_B + \mathbf{v}t'], t_B = L + \gamma(\mathbf{v})\left[t' + \frac{\mathbf{v}}{c^2}x'_B\right],$$

$$\gamma(\mathbf{v})[x'_B + \mathbf{v}t'] = L + \mathbf{v}\gamma(\mathbf{v})\left[t' + \frac{\mathbf{v}}{c^2}x'_B\right],$$

$$\gamma(\mathbf{v})x'_B + \underline{\gamma(\mathbf{v})\mathbf{v}t'} = L + \underline{\gamma(\mathbf{v})\mathbf{v}t'} + \gamma(\mathbf{v})\frac{\mathbf{v}^2}{c^2}x'_B,$$

$$\gamma(\mathbf{v})x'_B = L + \gamma(\mathbf{v})\frac{\mathbf{v}^2}{c^2}x'_B, \tag{2.2.24}$$

$$\gamma(\mathbf{v})\frac{dx'_B}{dt'} = \gamma(\mathbf{v})\frac{\mathbf{v}^2}{c^2}\frac{dx'_B}{dt'},$$

$$\frac{dx'_B}{dt'}\gamma(\mathbf{v})\left(1 - \frac{\mathbf{v}^2}{c^2}\right) = 0,$$

$$\frac{dx'_B}{dt'}\sqrt{1 - \frac{\mathbf{v}^2}{c^2}} = 0.$$

Hence

$$\frac{dx'_B}{dt'} \equiv 0. \tag{2.2.25}$$

From Eqs.(2.2.23),(2.2.25) we is concluded the next simple result**:**
**Theorem 2.2.1.** The bouth rockets **A** and **B** is instantaneously at rest for an observer in instantaneously comoving inertial *orthogonal* frame $\mathbf{S}'_{t'_A}(t', x')$.

    **Remark**     *2.2.5 It is obvious, that the metric (2.2.7) unfortunately gives inertial non-orthogonal frame. Thus from Corollary 2.2.1 we conclude that:*

$$\mathbf{l}^{ph} \neq L'(t') = \frac{L}{\sqrt{1 - \frac{\mathbf{v}^2}{c^2}}} \tag{2.2.26}$$

# II.3.Coordinate distance between two equal accelerated rockets measured by observer on the

instantly comoving inertial frame.

Let us now consider a situation where two equal rockets are initially placed on the $x_I$ axis at a distance $L$ from one another. Every rocket carries on board a scientist to make measurements, and an engineer to control the thrust of the rocket. The engineers carry identical (initially) synchronized clocks and have the same instructions for the regime of the engines. Let us call **B** the front rocket (and moving observer), and **A** the rear rocket, with its observer. **A** and **B** are not physically connected, so that they move exactly with the same proper acceleration at any time.

Let $F_I = F_I(t_I, x_I)$ be a stationary inertial frame and $F_a$ the accelerated frame of an observers on the rockets. We assume that we know the functions $x_A(t_A)$ and $x_B(t_A)$, so we also know the velocity

$$\mathbf{v}(t_A) = \frac{dx_A(t_A)}{dt_A}. \qquad (2.3.1)$$

The function (2.3.1) defines the infinite sequence of comoving inertial frames $\mathbf{S}'_{t_A} = \mathbf{S}'_{t_A}(t', x')$. The rocket **A** is instantaneously at rest for an observer in this frames $\mathbf{S}'_{t_A}$.

Thus for coordinate distance between rockets **A** and **B** which measured by observer on the frame $\mathbf{S}'_{t_A}$ at the instant $t'$ we obtain:

$$L'(t') = x'_B(t') - x'_A(t'), \qquad (2.3.2)$$

where $x'_A(t')$ and $x'_B(t')$ is the coordinates of the rockets **A** and **B**, which measured by observer on the frame $\mathbf{S}'_{t_A}$ at the same instant $t'$.

From Lorentz transformation by usual way we obtain:

$$x_A = \gamma(t_A)[x'_A + \mathbf{v}_A t'],$$

$$t_A = \gamma(t_A)\left[t' + \frac{\mathbf{v}_A}{c^2} x'_A\right],$$

$$x_B(t_B) = \gamma(t_A)[x'_B + \mathbf{v}_A t'], \qquad (2.3.3)$$

$$t_B = \gamma(t_A)\left[t' + \frac{\mathbf{v}_A}{c^2} x'_B\right],$$

where
$\gamma(t_A) \triangleq \gamma[\mathbf{v}(t_A)] = 1/\sqrt{1 - \mathbf{v}^2(t_A)/c^2}, \mathbf{v}_A \triangleq \mathbf{v}(t_A), x'_A \triangleq x'_A(t'), x'_B \triangleq x'_B(t'), x_A \triangleq x_A(t_A),$
$x_B \triangleq x_B(t_B).$ By setting $x'_A(t') = 0,$ from (2.3.3) we obtain:

$$x_A(t_A) = x_A = \gamma(t_A)\mathbf{v}_A t', \quad (2.3.4.a)$$

$$t_A = \gamma(t_A)t', \quad (2.3.4.b)$$

$$x_B(t_B) = x_B = \gamma(t_A)[x'_B + \mathbf{v}_A t'], \quad (2.3.4.c)$$

$$t_B = \gamma(t_A)\left[t' + \frac{\mathbf{v}_A}{c^2}x'_B\right]. \quad (2.3.4.d)$$

$$x'_B = L'(t'), L'(0) = L.$$

(2.3.4)

Thus

$$x_B(t_B) - x_A(t_A) = \gamma(t_A)x'_B = \gamma(t_A)L'(t'). \quad (2.3.5)$$

By setting

$$x_A(t) = \frac{c^2}{a}\left[\sqrt{1 + \frac{a^2 t^2}{c^2}} - 1\right],$$

$$x_B(t) = L + \frac{c^2}{a}\left[\sqrt{1 + \frac{a^2 t^2}{c^2}} - 1\right],$$

(2.3.6)

we obtain:

$$x_A(t_A) = \frac{c^2}{a}\left[\sqrt{1 + \frac{a^2 t_A^2}{c^2}} - 1\right],$$

$$x_B(t_B) = L + \frac{c^2}{a}\left[\sqrt{1 + \frac{a^2 t_B^2}{c^2}} - 1\right],$$

(2.3.7)

and

$$\mathbf{v}_A = \frac{at_A}{\sqrt{1 + \frac{a^2 t_A^2}{c^2}}}, \quad (2.3.8.a)$$

$$\gamma(t_A) = \sqrt{1 + \frac{a^2 t_A^2}{c^2}}. \quad (2.3.8.b)$$

(2.3.8)

From Eq.(2.3.4.d) and Eq.(2.3.8) we obtain:

$$t_B = t'\sqrt{1 + \frac{a^2 t_A^2}{c^2}} + \frac{at_A}{c^2}L'(t'). \qquad (2.3.9)$$

From Eqs.(2.3.5.),(2.3.7)-(2.3.8.b) we obtain:

$$\gamma(t_A)L'(t') = x_B(t_B) - x_A(t_A) = \gamma(t_A)x'_B.$$

$$L'(t')\sqrt{1 + \frac{a^2 t_A^2}{c^2}} = L + \frac{c^2}{a}\left[\sqrt{1 + \frac{a^2 t_B^2}{c^2}} - 1\right] - \frac{c^2}{a}\left[\sqrt{1 + \frac{a^2 t_A^2}{c^2}} - 1\right] =$$

$$= L + \frac{c^2}{a}\sqrt{1 + \frac{a^2 t_B^2}{c^2}} - \frac{c^2}{a}\sqrt{1 + \frac{a^2 t_A^2}{c^2}}. \qquad (2.3.10)$$

Thus we obtain:

$$\left[L'(t') + \frac{c^2}{a}\right]\sqrt{1 + \frac{a^2 t_A^2}{c^2}} = L + \frac{c^2}{a}\sqrt{1 + \frac{a^2 t_B^2}{c^2}}. \qquad (2.3.11)$$

From Eq.(2.3.4.b) and Eq.(2.3.8.b) we obtain:

$$t_A = \gamma(t_A)t' = t'\sqrt{1 + \frac{a^2 t_A^2}{c^2}},$$

$$t_A^2 = t'^2\left(1 + \frac{a^2 t_A^2}{c^2}\right),$$

$$t_A = \frac{t'}{\sqrt{1 - \frac{a^2 t'^2}{c^2}}}, 0 \leq t' < \frac{c}{a}. \qquad (2.3.12)$$

From Eq.(2.3.9.) and Eq.(2.3.12.) we obtain:

$$t_B = t'\sqrt{1 + \frac{a^2 t_A^2}{c^2}} + \frac{at_A}{c^2}L'(t') =$$

$$= t'\sqrt{1 + \frac{a^2}{c^2}\frac{t'^2}{1 - \frac{a^2 t'^2}{c^2}}} + L'(t')\frac{a}{c^2}\frac{t'}{\sqrt{1 - \frac{a^2 t'^2}{c^2}}}.$$

$$1 + \frac{a^2}{c^2}\frac{t'^2}{1 - \frac{a^2 t'^2}{c^2}} = 1 + \frac{a^2}{c^2}\frac{c^2 t'^2}{c^2 - a^2 t'^2} = 1 + \frac{a^2 t'^2}{c^2 - a^2 t'^2} =$$

$$= \frac{c^2 - a^2 t'^2}{c^2 - a^2 t'^2} + \frac{a^2 t'^2}{c^2 - a^2 t'^2} = \frac{c^2}{c^2 - a^2 t'^2} = \frac{1}{1 - \frac{a^2 t'^2}{c^2}}. \quad (2.3.13)$$

$$t_B = \frac{t'}{\sqrt{1 - \frac{a^2 t'^2}{c^2}}} + L'(t')\frac{a}{c^2}\frac{t'}{\sqrt{1 - \frac{a^2 t'^2}{c^2}}}.$$

$$t_B = \frac{t'\left[1 + L'(t')\frac{a}{c^2}\right]}{\sqrt{1 - \frac{a^2 t'^2}{c^2}}}, \quad t_B^2 = \frac{t'^2\left[1 + L'(t')\frac{a}{c^2}\right]^2}{1 - \frac{a^2 t'^2}{c^2}}$$

By substitution Eq.(2.3.12) and Eq.(2.3.13) into Eq.(2.3.11) we obtain:

$$\left[L'(t') + \frac{c^2}{a}\right]\sqrt{1 + \frac{a^2 t_A^2}{c^2}} = L + \frac{c^2}{a}\sqrt{1 + \frac{a^2 t_B^2}{c^2}}.$$

$$\frac{\left[L'(t') + \frac{c^2}{a}\right]}{\sqrt{1 - \frac{a^2 t'^2}{c^2}}} = L + \frac{c^2}{a}\sqrt{1 + \frac{a^2}{c^2}\frac{t'^2\left[1 + L'(t')\frac{a}{c^2}\right]^2}{1 - \frac{a^2 t'^2}{c^2}}}.$$

$$\frac{\left[L'(t') + \frac{c^2}{a}\right]^2}{1 - \frac{a^2 t'^2}{c^2}} = \left\{L + \frac{c^2}{a}\sqrt{1 + \frac{a^2}{c^2}\frac{t'^2\left[1 + L'(t')\frac{a}{c^2}\right]^2}{1 - \frac{a^2 t'^2}{c^2}}}\right\}^2 =$$

$$= L^2 + 2L\frac{c^2}{a}\sqrt{1 + \frac{a^2}{c^2}\frac{t'^2\left[1 + L'(t')\frac{a}{c^2}\right]^2}{1 - \frac{a^2 t'^2}{c^2}}} +$$

$$+ \frac{c^4}{a^2}\left[1 + \frac{a^2}{c^2}\frac{t'^2\left[1 + L'(t')\frac{a}{c^2}\right]^2}{1 - \frac{a^2 t'^2}{c^2}}\right] =$$

$$= L^2 + 2L\frac{c^2}{a}\sqrt{1 + \frac{a^2}{c^2}\frac{t'^2\left[1 + L'(t')\frac{a}{c^2}\right]^2}{1 - \frac{a^2 t'^2}{c^2}}} +$$

(2.3.14)

$$+ \frac{c^4}{a^2} + \frac{c^2 t'^2\left[1 + L'(t')\frac{a}{c^2}\right]^2}{1 - \frac{a^2 t'^2}{c^2}} =$$

$$L^2 + 2L\frac{c^2}{a}\sqrt{1 + \frac{a^2}{c^2}\frac{t'^2\left[1 + L'(t')\frac{a}{c^2}\right]^2}{1 - \frac{a^2 t'^2}{c^2}}} + \frac{c^4}{a^2} + \frac{\frac{a^2}{c^2}t'^2\left[\frac{c^2}{a} + L'(t')\right]^2}{1 - \frac{a^2 t'^2}{c^2}}.$$

$$\frac{\left[L'(t') + \frac{c^2}{a}\right]^2}{1 - \frac{a^2 t'^2}{c^2}} - \frac{\frac{a^2}{c^2}t'^2\left[\frac{c^2}{a} + L'(t')\right]^2}{1 - \frac{a^2 t'^2}{c^2}} =$$

Thus we obtain:

$$\left[L'(t') + \frac{c^2}{a}\right]^2 = L^2 + 2L\sqrt{\frac{c^4}{a^2} + \frac{\frac{a^2}{c^2}t'^2\left[\frac{c^2}{a} + L'(t')\right]^2}{1 - \frac{a^2 t'^2}{c^2}}} + \frac{c^4}{a^2},$$

$$z = z(t') \triangleq \left[L'(t') + \frac{c^2}{a}\right]^2.$$

$$z(0) = \left(L + \frac{c^2}{a}\right).$$

$$z = L^2 + 2L\sqrt{\frac{c^4}{a^2} + \frac{\frac{a^2}{c^2}t'^2 z}{1 - \frac{a^2 t'^2}{c^2}}} + \frac{c^4}{a^2},$$

(2.3.15)

$$\left(z - L^2 - \frac{c^4}{a^2}\right)^2 = 4L^2\left(\frac{c^4}{a^2} + \frac{\frac{a^2}{c^2}t'^2 z}{1 - \frac{a^2 t'^2}{c^2}}\right).$$

$$z^2 - 2z\left(L^2 + \frac{c^4}{a^2}\right) + \left(L^2 + \frac{c^4}{a^2}\right)^2 = 4L^2\frac{c^4}{a^2} + \frac{4L^2\frac{a^2}{c^2}t'^2 z}{1 - \frac{a^2 t'^2}{c^2}}.$$

$$z^2 - z\left[2\left(L^2 + \frac{c^4}{a^2}\right) + \frac{4L^2\frac{a^2}{c^2}t'^2}{1 - \frac{a^2 t'^2}{c^2}}\right] + \left(L^2 + \frac{c^4}{a^2}\right)^2 - 4L^2\frac{c^4}{a^2} = 0.$$

From Eq.(2.3.15) we obtain:

$$z_{1,2}(t') = \left(L^2 + \frac{c^4}{a^2}\right) + \frac{2L^2\frac{a^2}{c^2}t'^2}{1 - \frac{a^2 t'^2}{c^2}} \pm$$

(2.3.16)

$$\pm \sqrt{\left[\left(L^2 + \frac{c^4}{a^2}\right) + \frac{2L^2\frac{a^2}{c^2}t'^2}{1 - \frac{a^2 t'^2}{c^2}}\right]^2 - \left(L^2 + \frac{c^4}{a^2}\right)^2 + 4L^2\frac{c^4}{a^2}}.$$

For instant $t' = 0$ we obtain:

$$z_{1,2}(0) = \left(L^2 + \frac{c^4}{a^2}\right) \pm$$

$$\pm \sqrt{\left[\left(L^2 + \frac{c^4}{a^2}\right)\right]^2 - \left(L^2 + \frac{c^4}{a^2}\right)^2 + 4L^2\frac{c^4}{a^2}}. \quad (2.3.17)$$

$$= L^2 + \frac{c^4}{a^2} \pm 2L\frac{c^2}{a} = \left(L \pm \frac{c^2}{a}\right)^2.$$

Finally we obtain:

$$L'(t') = \left[\left(L^2 + \frac{c^4}{a^2}\right) + \frac{2L^2\frac{a^2}{c^2}t'^2}{1 - \frac{a^2 t'^2}{c^2}} + \right.$$

$$\left. + \sqrt{\left[\left(L^2 + \frac{c^4}{a^2}\right) + \frac{2L^2\frac{a^2}{c^2}t'^2}{1 - \frac{a^2 t'^2}{c^2}}\right]^2 - \left(L^2 + \frac{c^4}{a^2}\right)^2 + 4L^2\frac{c^4}{a^2}}\right]^{1/2} - \frac{c^2}{a}. \quad (2.3.18)$$

From Eqs.(2.3.6) we obtain:

$$x_A + \frac{c^2}{a} = \frac{c^2}{a}\sqrt{1 + \frac{a^2 t^2}{c^2}} = \sqrt{\frac{c^4}{a^2} + c^2 t^2},$$

$$x_B - L + \frac{c^2}{a} = \frac{c^2}{a}\sqrt{1 + \frac{a^2 t^2}{c^2}} = \sqrt{\frac{c^4}{a^2} + c^2 t^2}. \quad (2.3.19)$$

And

$$\left(x_A + \frac{c^2}{a}\right)^2 = \frac{c^4}{a^2} + c^2 t_A^2, \quad (a)$$

$$\left(x_B - L + \frac{c^2}{a}\right)^2 = \frac{c^4}{a^2} + c^2 t_B^2. \quad (b) \quad (2.3.20)$$

From Lorentz transformation by usual way we obtain:

$$x_A = \gamma(\mathbf{v})[x'_A + \mathbf{v}t'],$$

$$t_A = \gamma(\mathbf{v})\left[t' + \frac{\mathbf{v}}{c^2}x'_A\right],$$

$$x_B(t_B) = \gamma(\mathbf{v})[x'_B + \mathbf{v}t'], \tag{2.3.21}$$

$$t_B = \gamma(\mathbf{v})\left[t' + \frac{\mathbf{v}}{c^2}x'_B\right],$$

where
$\mathbf{v} = \mathbf{v}_A = \mathbf{v}(t_A), \gamma(\mathbf{v}) = 1/\sqrt{1 - \mathbf{v}^2/c^2}, x'_A \triangleq x'_A(t'), x'_B \triangleq x'_B(t'), x_A \triangleq x_A(t_A), x_B \triangleq x_B(t_B)$.
By substitution (2.3.21) into (2.3.20.a) we obtain:

$$\left(\gamma(\mathbf{v})[x'_A + \mathbf{v}t'] + \frac{c^2}{a}\right)^2 = \frac{c^4}{a^2} + c^2\gamma^2(\mathbf{v})\left[t' + \frac{\mathbf{v}}{c^2}x'_A\right]^2, \tag{2.3.22}$$

By differentiation Eq.(2.3.22) we obtain:

$$2\gamma(\mathbf{v})\left(\gamma(\mathbf{v})[x'_A + \mathbf{v}t'] + \frac{c^2}{a}\right)\left(\frac{dx'_A}{dt'} + \mathbf{v}\right) =$$

$$= 2c^2\gamma^2(\mathbf{v})\left(t' + \frac{\mathbf{v}}{c^2}x'_A\right)\left(1 + \frac{\mathbf{v}}{c^2}\frac{dx'_A}{dt'}\right). \tag{2.3.23}$$

By setting $x'_A(t') = 0$, for instant $t'$ we obtain:

$$\left(\gamma(\mathbf{v})\mathbf{v}t' + \frac{c^2}{a}\right)\left(\frac{dx'_A}{dt'} + \mathbf{v}\right) = c^2\gamma(\mathbf{v})t'\left(1 + \frac{\mathbf{v}}{c^2}\frac{dx'_A}{dt'}\right),$$

$$\frac{dx'_A}{dt'}\gamma(\mathbf{v})\mathbf{v}t' + \frac{dx'_A}{dt'}\frac{c^2}{a} + \mathbf{v}\left(\gamma(\mathbf{v})\mathbf{v}t' + \frac{c^2}{a}\right) = c^2\gamma(\mathbf{v})t' + \frac{dx'_A}{dt'}\gamma(\mathbf{v})\mathbf{v}t',$$

$$\frac{dx'_A}{dt'}\frac{c^2}{a} = c^2\gamma(\mathbf{v})t' - \gamma(\mathbf{v})\mathbf{v}^2 t' - \mathbf{v}\frac{c^2}{a} = c^2\gamma(\mathbf{v})t'\left(1 - \frac{\mathbf{v}^2}{c^2}\right) - \mathbf{v}\frac{c^2}{a} =$$

$$= c^2 t'\sqrt{1 - \frac{\mathbf{v}^2}{c^2}} - \mathbf{v}\frac{c^2}{a}.$$

$$\frac{dx'_A}{dt'} = at'\sqrt{1 - \frac{\mathbf{v}^2}{c^2}} - \mathbf{v}. \tag{2.3.24}$$

$$at'\sqrt{1 - \frac{\mathbf{v}^2}{c^2}} - \mathbf{v} = 0,$$

$$a^2 t'^2\left(1 - \frac{\mathbf{v}^2}{c^2}\right) = \mathbf{v}^2,$$

$$a^2 t'^2 = \mathbf{v}^2\left(1 + \frac{a^2 t'^2}{c^2}\right),$$

$$\mathbf{v} = \frac{at'}{\sqrt{1 + \frac{a^2 t'^2}{c^2}}}. \quad (a)$$

Hence

$$\frac{dx'_A(t')}{dt'} = 0 \Leftrightarrow \mathbf{v} = \frac{at'}{\sqrt{1 + \frac{a^2 t'^2}{c^2}}}. \tag{2.3.25}$$

By substitution (2.3.21) into (2.3.20.b) we obtain:

$$\left(\gamma(\mathbf{v})[x'_B + \mathbf{v}t'] - L + \frac{c^2}{a}\right)^2 = \frac{c^4}{a^2} + c^2\gamma^2(\mathbf{v})\left[t' + \frac{\mathbf{v}}{c^2}x'_B\right]^2.$$

$$\gamma^2(\mathbf{v})[x'_B + \mathbf{v}t']^2 - 2\gamma(\mathbf{v})[x'_B + \mathbf{v}t']\left(L - \frac{c^2}{a}\right) + \left(L - \frac{c^2}{a}\right)^2 =$$

$$= \frac{c^4}{a^2} + c^2\gamma^2(\mathbf{v})\left(t'^2 + 2\frac{\mathbf{v}}{c^2}t'x'_B + \frac{\mathbf{v}^2}{c^4}x'^2_B\right),$$

$$\gamma^2(\mathbf{v})(x'^2_B + 2\mathbf{v}t'x'_B + \mathbf{v}^2t'^2) - 2\gamma(\mathbf{v})\left(L - \frac{c^2}{a}\right)x'_B -$$

$$-2\gamma(\mathbf{v})\mathbf{v}t'\left(L - \frac{c^2}{a}\right) + \left(L - \frac{c^2}{a}\right)^2 = \frac{c^4}{a^2} + c^2\gamma^2(\mathbf{v})\left(t'^2 + 2\frac{\mathbf{v}}{c^2}t'x'_B + \frac{\mathbf{v}^2}{c^4}x'^2_B\right),$$

$$\underline{\underline{\gamma^2(\mathbf{v})x'^2_B}} + \underline{2\gamma^2(\mathbf{v})\mathbf{v}t'x'_B} + \underbrace{\gamma^2(\mathbf{v})\mathbf{v}^2t'^2} - \underline{2\gamma(\mathbf{v})\left(L - \frac{c^2}{a}\right)x'_B} -$$

$$-2\gamma(\mathbf{v})\mathbf{v}t'\left(L - \frac{c^2}{a}\right) + \left(L - \frac{c^2}{a}\right)^2 - \frac{c^4}{a^2} - c^2\gamma^2(\mathbf{v})t'^2 -$$

$$-\underline{2\gamma^2(\mathbf{v})\mathbf{v}t'x'_B} - \underbrace{\gamma^2(\mathbf{v})\mathbf{v}^2t'^2} - \underline{\underline{\gamma^2(\mathbf{v})\frac{\mathbf{v}^2}{c^2}x'^2_B}} = 0. \qquad (2.3.26)$$

$$x'^2_B\gamma^2(\mathbf{v})\left(1 - \frac{\mathbf{v}^2}{c^2}\right) - 2\gamma(\mathbf{v})\left(L - \frac{c^2}{a}\right)x'_B - c^2\gamma^2(\mathbf{v})t'^2 -$$

$$-2\gamma(\mathbf{v})\mathbf{v}t'\left(L - \frac{c^2}{a}\right) + \left(L - \frac{c^2}{a}\right)^2 - \frac{c^4}{a^2} = 0.$$

$$x'^2_B - 2\gamma(\mathbf{v})\left(L - \frac{c^2}{a}\right)x'_B - c^2\gamma^2(\mathbf{v})t'^2 - 2\gamma(\mathbf{v})\mathbf{v}t'\left(L - \frac{c^2}{a}\right) +$$

$$+\left(L - \frac{c^2}{a}\right)^2 - \frac{c^4}{a^2} = 0.$$

$$x'_B(t') = \gamma(\mathbf{v})\left(L - \frac{c^2}{a}\right) \pm$$

$$\pm\sqrt{\gamma^2(\mathbf{v})\left(L - \frac{c^2}{a}\right)^2 + c^2\gamma^2(\mathbf{v})t'^2 + 2\gamma(\mathbf{v})\mathbf{v}t'\left(L - \frac{c^2}{a}\right) - \left(L - \frac{c^2}{a}\right)^2 + \frac{c^4}{a^2}}. \quad (a)$$

By differentiation Eq.(2.3.26.a) we obtain:

$$\frac{dx'_B(t')}{dt'} =$$

$$= \frac{2c^2\gamma^2(\mathbf{v})t' + 2\gamma(\mathbf{v})\mathbf{v}\left(L - \frac{c^2}{a}\right)}{\sqrt{\gamma^2(\mathbf{v})\left(L - \frac{c^2}{a}\right)^2 + c^2\gamma^2(\mathbf{v})t'^2 + 2\gamma(\mathbf{v})\mathbf{v}t'\left(L - \frac{c^2}{a}\right) - \left(L - \frac{c^2}{a}\right)^2 + \frac{c^4}{a^2}}}. \qquad (2.3.27)$$

By substitution (2.3.24.a) into (2.3.27.) we obtain:

$$\frac{dx'_B(t')}{dt'} =$$

$$= \frac{2c^2\gamma^2(\mathbf{v})t' + 2\gamma(\mathbf{v})\mathbf{v}\left(L - \frac{c^2}{a}\right)}{\sqrt{\gamma^2(\mathbf{v})\left(L - \frac{c^2}{a}\right)^2 + c^2\gamma^2(\mathbf{v})t'^2 + 2\gamma(\mathbf{v})\mathbf{v}t'\left(L - \frac{c^2}{a}\right) - \left(L - \frac{c^2}{a}\right)^2 + \frac{c^4}{a^2}}}. \qquad (2.3.28)$$

## II.4. Radar distance between two equal accelerated rockets.

Let us now consider a situation where two equal rockets are initially placed on the $x_I$ axis at a distance $L$ from one another. Every rocket carries on board a scientist to make measurements, and an engineer to control the thrust of the rocket. The engineers carry identical (initially) synchronized clocks and have the same instructions for the regime of the engines. Let us call **B** the front rocket (and moving observer), and **A** the rear rocket, with its observer. **A** and **B** are not physically connected, so that they move exactly with the same proper acceleration at any time.

**Claim** *We have also assumed in this subsection that the sizes of physical systems we considered, were small enough not to incur into troubles with horizons and other difficulties typical of extended accelerated reference frames.*

The way used to monitor the reciprocal positions is the exchange of light rays.

The infinitesimal proper time interval $d\tau$ is given, in terms of the coordinate time interval $dt$, by

$$d\tau = dt\sqrt{1 - \frac{\mathbf{v}^2(t)}{c^2}} = \gamma^{-1}(\mathbf{v})dt \qquad (2.4.1)$$

where we have introduced the Lorentz factor $\gamma(\mathbf{v}) = 1/\sqrt{1 - \frac{\mathbf{v}^2(t)}{c^2}}$. Hence, substituting **v** from (2.3.8.a) in (2.4.1), we obtain:

$$d\tau = \frac{dt}{\sqrt{1 + \frac{a^2 t^2}{c^2}}}, \qquad (2.4.2)$$

in terms of the coordinate time $t$. By integrating ($\tau = 0$ when $t = 0$), the proper time lapse turns out to be

$$\tau = \frac{c}{a}\sinh^{-1}\left(\frac{a \cdot t}{c}\right). \qquad (2.4.3)$$

hence, we obtain

$$t = \frac{c}{a}\sinh\left(\frac{a \cdot \tau}{c}\right). \qquad (2.4.4)$$

Now, if we substitute in (2.4.1) this expression (2.4.4) of coordinate time, as a function of the elapsed proper time $\tau$, we see that the coordinate velocity $\mathbf{v}(t)$ can be written as

$$\mathbf{v}(\tau) = c\tanh\left(\frac{a \cdot \tau}{c}\right) \qquad (2.4.5)$$

At a given predetermined $\tau_0$ the engines are stopped on both rockets. From that moment on, both for **A** and **B**, the flight continues at a constant coordinate speed

$$\mathbf{v}_0(\tau_0) = \frac{dx}{dt} = c\tanh\left(\frac{a \cdot \tau_0}{c}\right) \qquad (2.4.6)$$

and the corresponding Lorentz factor $\gamma(\mathbf{v}_0)$ is

$$\gamma(\mathbf{v}_0) = \cosh\left(\frac{a \cdot \tau_0}{c}\right) \qquad (2.4.7)$$

The round trip of light between the space ships corresponds to a coordinate time interval

$$\Delta t = \frac{2l}{c}\cosh^2\left(\frac{a \cdot \tau_0}{c}\right) \qquad (2.4.8)$$

and in terms of proper time of the rear rocket **A**

$$\Delta \tau = \frac{2l}{c}\cosh\left(\frac{a \cdot \tau_0}{c}\right),$$
$$\Delta t = \gamma(\mathbf{v}_0)\Delta \tau. \qquad (2.4.9)$$

The *radar distance [45]* $R_d$ is usually defined as:

$$R_d = \frac{1}{2}c\Delta \tau. \qquad (2.4.10)$$

From (2.4.10) we obtain:

$$R_d = \frac{1}{2}c\Delta \tau = l\cosh\left(\frac{a \cdot \tau_0}{c}\right) \qquad (2.4.11)$$

which, considering (2.4.7) gives:

$$R_d = \frac{l}{\sqrt{1 - \frac{\mathbf{v}_0^2}{c^2}}}. \tag{2.4.12}$$

Hence

$$R_d = l\sqrt{1 + \frac{a^2 t^2}{c^2}}. \tag{2.4.13}$$

Let a relativistic uniformly accelerated frame $F_a$ with the coordinates $(x, \tau)$ move without initial velocity along the axis $X$ of an inertial Galilee rame (2.2.2) with the coordinates $(X, T)$ and at $\tau = T = 0$ their origins coincide. Then the coordinate transformation formulas $X, T$ have the form [30]:

$$X = x + \frac{c^2}{a}\left[\sqrt{1 + \frac{a^2 T^2}{c^2}} - 1\right],$$
$$T = \frac{c}{a}\sinh\left(\frac{a \cdot \tau}{c}\right). \tag{2.4.14}$$

In this case, substituting (2.4.14) into the metrics (2.2.2) gives the expression of the metrics of a uniformly accelerated FR

$$ds^2 = c^2 d\tau^2 - 2c\sinh\left(\frac{a \cdot \tau}{c}\right)d\tau dx - dx^2 - dy^2 - dz^2. \tag{2.4.15}$$

Coordinate velocities $\check{c} = c_\pm : c_+, c_-$ of the ligt in a directions of an axis $\pm \vec{x}$, we obtain by usual way from the equation:

$$ds^2 = 0. \tag{2.4.16}$$

Thus

$$ds^2 = c^2 d\tau^2 - g_{01} d\tau dx - dx^2 = 0,$$
$$g_{01} = 2c\sinh\left(\frac{a \cdot \tau}{c}\right). \tag{2.4.17}$$

and

$$s^2 = \left(c^2 - g_{01}\frac{dx}{d\tau} - \frac{dx^2}{d\tau^2}\right)dt^2 = 0,$$
$$c^2 - g_{01}\frac{dx}{d\tau} - \frac{dx^2}{d\tau^2} = 0$$
$$\frac{dx}{dt} \triangleq \check{c}, \tag{2.4.18}$$
$$c^2 - g_{01}\check{c} - \check{c}^2 = 0.$$

Hence

$$\check{c}^2 + g_{01}\check{c} - c^2 = 0. \tag{2.4.19}$$

From (2.4.17) and (2.4.19) we obtain:

$$\check{c} = c_\pm = -\frac{1}{2}g_{01} \pm \sqrt{\frac{1}{4}g_{01}^2 + c^2}.$$

$$c_\pm = -c\sinh\left(\frac{a\cdot\tau}{c}\right) \pm c\sqrt{1 + \sinh^2\left(\frac{a\cdot\tau}{c}\right)} =$$
$$= -c\sinh\left(\frac{a\cdot\tau}{c}\right) \pm c\cosh\left(\frac{a\cdot\tau}{c}\right)$$
$$c_+(\tau) = -c\left[\sinh\left(\frac{a\cdot\tau}{c}\right) - \cosh\left(\frac{a\cdot\tau}{c}\right)\right] = \quad (2.4.20)$$
$$= \exp\left(-\frac{a\cdot\tau}{c}\right).$$
$$c_-(\tau) = -c\left[\sinh\left(\frac{a\cdot\tau}{c}\right) + \cosh\left(\frac{a\cdot\tau}{c}\right)\right] =$$
$$-\exp\left(\frac{a\cdot\tau}{c}\right).$$

## II.5. The horizon.

Let a relativistic uniformly accelerated frame $F_a$ with the coordinates $(x,t)$ move without initial velocity along the axis $X$ of an inertial Galilee rame (2.2.2) with the coordinates $(X,T)$ and at $t = T = 0$ their origins coincide. Then the coordinate transformation formulas $X$ have the form [30]:

$$X = x + \frac{c^2}{a}\left[\sqrt{1 + \frac{a^2T^2}{c^2}} - 1\right]. \quad (2.5.1)$$

In this subsection we consider the simplest possible dependence $t = T$. In this case, substituting (2.5.1) and $t = T$ into the metrics (2.2.2) gives the expression of the metrics of a uniformly accelerated FR

$$ds^2 = \frac{c^2}{1 + \frac{a^2t^2}{c^2}}dt - \frac{2at}{\sqrt{1 + \frac{a^2t^2}{c^2}}}dtdx - dx^2 - dy^2 - dz^2. \quad (2.5.2)$$

Thus

$$ds^2 = c^2g_{00}dt^2 - 2at\sqrt{g_{00}}\,dxdt - dx^2,$$
$$g_{00} = \frac{1}{1 + \frac{a^2t^2}{c^2}}, \quad (2.5.3)$$

and

$$ds^2 = \left(c^2g_{00} - 2at\sqrt{g_{00}}\frac{dx}{dt} - \frac{dx^2}{dt^2}\right)dt^2. \quad (2.5.4)$$

Coordinate velocities $\check{c} = c_\pm : c_+, c_-$ of the ligt in a directions of an axis $\pm\vec{x}$, we obtain by usual way from the equation:

$$ds^2 = 0. \quad (2.5.5)$$

From Eqs.(2.5.4)-(2.5.5) we obtain:

$$c^2 g_{00} - 2at\sqrt{g_{00}}\,\frac{dx}{dt} - \frac{dx^2}{dt^2} = 0,$$

$$\frac{dx}{dt} \triangleq \check{c},$$

(2.5.6)

$$c^2 g_{00} - 2at\sqrt{g_{00}}\,\check{c} - \check{c}^2 = 0,$$

$$\check{c}^2 + 2at\sqrt{g_{00}}\,\check{c} - c^2 g_{00} = 0.$$

From Eq.(2.5.6) we obtain:

$$\check{c}(t) = c_{\pm}(t) = -at\sqrt{g_{00}} \pm \sqrt{a^2 t^2 g_{00} + c^2 g_{00}} =$$

$$= -at\sqrt{g_{00}} \pm \sqrt{g_{00}}\sqrt{a^2 t^2 + c^2} =$$

$$= \sqrt{g_{00}}\left(-at \pm \sqrt{a^2 t^2 + c^2}\right) =$$

$$= \frac{-at \pm \sqrt{a^2 t^2 + c^2}}{\sqrt{1 + \frac{a^2 t^2}{c^2}}} = \frac{-at \pm c\sqrt{1 + \frac{a^2 t^2}{c^2}}}{\sqrt{1 + \frac{a^2 t^2}{c^2}}} =$$

(2.5.7)

$$= \frac{-at}{\sqrt{1 + \frac{a^2 t^2}{c^2}}} \pm c.$$

$$c_{\pm}(t) = \frac{-at}{\sqrt{1 + \frac{a^2 t^2}{c^2}}} \pm c.$$

Thus

$$c_+(t) = \frac{-at}{\sqrt{1 + \frac{a^2 t^2}{c^2}}} + c.$$

$$c_-(t) = \frac{-at}{\sqrt{1 + \frac{a^2 t^2}{c^2}}} - c.$$

(2.5.8)

By simple calculation we obtain:

$$\lim_{t \to \infty} \frac{-at}{\sqrt{1 + \frac{a^2 t^2}{c^2}}} = \lim_{t \to \infty} \frac{-1}{\sqrt{\frac{1}{a^2 t^2} + \frac{1}{c^2}}} = -c. \qquad (2.5.9)$$

From (2.5.8)-(2.5.9) we obtain:

$$\begin{aligned} c_-(t) &= -2c. \\ c_+(t) &= 0. \end{aligned} \qquad (2.5.10)$$

Thus for the length $Lr_\pm(t,x)$ of a way passed ligt rays in a directions of an axis $\pm\vec{x}$, we obtain:

$$Lr_+(t,x) = \int_0^t c_+(t) dt = \int_0^t \left( c + \frac{-at}{\sqrt{1 + \frac{a^2 t^2}{c^2}}} \right) dt,$$

$$Lr_-(t,x) = \int_0^t c_-(t) dt = -\int_0^t \left( c + \frac{at}{\sqrt{1 + \frac{a^2 t^2}{c^2}}} \right) dt.$$

(2.5.11)

By simple calculation from (2.5.11) we obtain:

$$Lr_+(t,x) = ct - \frac{c^2}{a}\left[ \sqrt{1 + \frac{a^2 t^2}{c^2}} - 1 \right],$$

$$Lr_-(t,x) = -ct - \frac{c^2}{a}\left[ \sqrt{1 + \frac{a^2 t^2}{c^2}} - 1 \right].$$

(2.5.12)

Thus

$$Lr_+(t,x) = ct - \frac{c^2}{a}\left[\sqrt{1+\frac{a^2t^2}{c^2}} - 1\right] = ct - \sqrt{\frac{c^4}{a^2}+c^2t^2} + \frac{c^2}{a} =$$

$$= ct\left(1 - \sqrt{1+\frac{c^4}{a^2(c^2t^2)}}\right) + \frac{c^2}{a} =$$

$$= ct\left[1 - \left(1 + \frac{1}{2}\frac{c^4}{a^2(c^2t^2)} - \frac{1}{8}\frac{c^8}{a^4(c^2t^2)^2} + \ldots\right)\right] + \frac{c^2}{a} = \quad (2.5.13)$$

$$= ct\left(-\frac{1}{2}\frac{c^4}{a^2(c^2t^2)} + \frac{1}{8}\frac{c^8}{a^4(c^2t^2)^2} - \ldots\right) + \frac{c^2}{a} =$$

$$= \frac{c^2}{a} - \frac{1}{2}\frac{c^3}{a^2 t} + \frac{1}{8}\frac{c^5}{a^4 t^4} - \ldots \propto \frac{c^2}{a}, t \to \infty.$$

## III. Generalized Principle of limiting 4-dimensional symmetry for the case of the uniformly accelerated frames of reference. Generalized Lorentz length contractions.

### III.1. Generalized Principle of limiting 4-dimensional symmetry for the case of the uniformly accelerated frames of reference.

Let us considered (acceleration) transformations between the two relativistic frames one of which inertial Minkowski frame $F_I(t_I,x_I,y_I,z_I) = F_I(t',x',y',z')$ will be considered to be at "rest", while another one uniformly linearly accelerated (ULA) frame $F_a = F_a(t,x,y,z)$ will move with respect to the first one by the law:

$$x' = x + \int_0^t \mathbf{v}(\tau)d\tau, y = y', z = z', t = t', \quad (3.1.1)$$

or in equivalent infinitesimal forms

$$dx' = dx + \mathbf{v}(t)dt, \mathbf{v}(t) < c; dy = dy', dz = dz', dt' = dt. \quad (3.1.2)$$

Thus metric for the frame $F_a(t,x,y,z)$ gets the form:

$$ds^2 = c^2\left(1 - \frac{\mathbf{v}^2(t)}{c^2}\right)dt^2 - 2\mathbf{v}(t)dtdx - dx^2 - dy'^2 - dz'^2. \quad (3.1.3)$$

In the limit of zero acceleration $\lim_{t \to T} \vec{a}(t) \to 0$, we have $\mathbf{v}(t) = \mathbf{v}(T) = \mathbf{v} = const, t \geqslant T$

and transformations (3.1) becomes to the form

$$\Lambda(\mathbf{v}) : x = x' - \mathbf{v}t', t = t' \qquad (3.1.4)$$

and metric (3.3) gets the form:

$$ds^2 = c^2\left(1 - \frac{\mathbf{v}^2}{c^2}\right)dt^2 - 2\mathbf{v}dtdx - dx^2 - dy^2 - dz^2. \qquad (3.1.5)$$

**Remark** *We stress that the distance $x_2 - x_1$ can be the length of a space-ship or the distance between the two spaceships **A** and **B** which are at rest in the ULA frame $F_a$.*

**Theorem.(see [1]) (Generalized Principle of limiting 4-dimensional symmetry  for the case of the uniformly accelerated frames of reference).**

The metrics (3.1.3) of the uniformly accelerated noninertial FR in the limit of zero accelerationis is forminvariant that with respect to the so-called generalized inertial Lorentz-Poincare group $\hat{L}_m^i$ connected with the classic Lorentz-Poincare group $L_m^i$ one by the relation:

$$\hat{L}_k^n x^k = [\Lambda_i^n L_m^i (\Lambda^{-1})_k^m] x^k, \qquad (3.1.6)$$

where $\Lambda_k^i$ is the matrix of the linear (non-orthogonal) transformations

$$x^i = \Lambda_k^i x'^k. \qquad (3.1.7)$$

forming the generalized inertial FR (3.1.5).

## III.2. Generalized Lorentz transformations between an inertial Minkowski frame $F_I(t_I, x_I, y_I, z_I)$ and a generalized inertial frame $F_\mathbf{v}(t, x, y, z)$.

Let us calculate the generalized Lorentz transformations $\hat{L}(\mathbf{v})$ of space and time between an inertial Minkowski frame $F_I(t_I, x_I, y_I, z_I) = F_I(t', x', y', z')$ and a generalized inertial frame $F_\mathbf{v}(t, x, y, z)$ with a metric (3.1.5) forming by the linear (non-orthogonal) transformations $\Lambda(\mathbf{v})$ (3.1.4).

From [3] (see [3] Theorem in subsection I)  we obtain transformations $\hat{L}(\mathbf{v})$:

$$\hat{L}(\mathbf{v}) = \Lambda(\mathbf{v}) \circ L(\mathbf{v}).$$
$$\Lambda(\mathbf{v}) : (t',x') \to (t,x)$$
$$x = x' - \mathbf{v}t', t = t' \quad (3.2.1.a)$$

$$L(\mathbf{v}) : \begin{cases} x' \to \gamma(\mathbf{v})(x' - \mathbf{v}t') & (3.2.1b) \\ t' \to \gamma(\mathbf{v})\left(t' - \frac{\mathbf{v}}{c^2}x'\right) & (3.2.1.c) \end{cases}$$

(3.2.1)

By substitution (3.2.1.b)-(3.2.1.c) into (3.2.1.a) we obtain the generalized Lorentz transformations $\hat{L}(\mathbf{v})$ :

$$x = \gamma(\mathbf{v})\left[x'\left(1 + \frac{\mathbf{v}^2}{c^2}\right) - 2\mathbf{v}t'\right],$$
$$t = \gamma(\mathbf{v})\left(t' - \frac{\mathbf{v}}{c^2}x'\right).$$

(3.2.2)

Or in equivalent infinitesimal form:

$$dx = \gamma(\mathbf{v})\left[dx'\left(1 + \frac{\mathbf{v}^2}{c^2}\right) - 2\mathbf{v}dt'\right],$$
$$dt = \gamma(\mathbf{v})\left(dt' - \frac{\mathbf{v}}{c^2}dx'\right).$$

(3.2.3)

Substitution (3.2.3) into (3.1.5) gives:

$$ds^2 = c^2\left(1 - \frac{\mathbf{v}^2}{c^2}\right)dt^2 - 2\mathbf{v}dtdx - dx^2 =$$

$$c^2\left(dt' - \frac{\mathbf{v}}{c^2}dx'\right)^2 - 2\mathbf{v}\gamma^2(\mathbf{v})\left(dt' - \frac{\mathbf{v}}{c^2}dx'\right)\left[dx'\left(1 + \frac{\mathbf{v}^2}{c^2}\right) - 2\mathbf{v}dt'\right] -$$

$$-\gamma^2(\mathbf{v})\left[dx'\left(1 + \frac{\mathbf{v}^2}{c^2}\right) - 2\mathbf{v}dt'\right]^2 =$$

$$c^2\left(dt' - \frac{\mathbf{v}}{c^2}dx'\right)^2 -$$

$$-\gamma^2(\mathbf{v})\left[dx'\left(1 + \frac{\mathbf{v}^2}{c^2}\right) - 2\mathbf{v}dt'\right] \times$$

$$\times\left[2\mathbf{v}\left(dt' - \frac{\mathbf{v}}{c^2}dx'\right) + \left[dx'\left(1 + \frac{\mathbf{v}^2}{c^2}\right) - 2\mathbf{v}dt'\right]\right] = \quad (3.2.4)$$

$$c^2\left(dt' - \frac{\mathbf{v}}{c^2}dx'\right)^2 -$$

$$-\gamma^2(\mathbf{v})\left[dx'\left(1 + \frac{\mathbf{v}^2}{c^2}\right) - 2\mathbf{v}dt'\right]\left[2\mathbf{v}dt' - \frac{2\mathbf{v}^2}{c^2}dx' + dx' + \frac{\mathbf{v}^2}{c^2}dx' - 2\mathbf{v}dt'\right] =$$

$$c^2\left(dt' - \frac{\mathbf{v}}{c^2}dx'\right)^2 - \gamma^2(\mathbf{v})dx'\left[dx'\left(1 + \frac{\mathbf{v}^2}{c^2}\right) - 2\mathbf{v}dt'\right]\left(1 - \frac{\mathbf{v}^2}{c^2}\right) =$$

$$c^2\left(dt' - \frac{\mathbf{v}}{c^2}dx'\right)^2 - dx'\left[dx'\left(1 + \frac{\mathbf{v}^2}{c^2}\right) - 2\mathbf{v}dt'\right] =$$

$$c^2\left(dt' - \frac{\mathbf{v}}{c^2}dx'\right)^2 - dx'^2 - dx'^2\frac{\mathbf{v}^2}{c^2} + 2\mathbf{v}dt'dx' =$$

$$c^2dt' - 2\mathbf{v}dt'dx' + \frac{\mathbf{v}^2}{c^2}dx'^2 - dx'^2 - dx'^2\frac{\mathbf{v}^2}{c^2} + 2\mathbf{v}dt'dx' = c^2dt'^2 - dx'^2 = ds'^2.$$

Thus, the invariant interval $ds^2$ under the generalized Lorentz transformation (3.2.4) is

$$ds^2 = c^2\left(1 - \frac{\mathbf{v}^2}{c^2}\right)dt^2 - 2\mathbf{v}dtdx - dx^2 = c^2dt'^2 - dx'^2 = ds'^2. \quad (3.2.5)$$

From the generalized Lorentz transformations (3.2.2) by usual way we obtain the generalized Lorentz length contractions:

$$x_2 - x_1 = \gamma(\mathbf{v})\left(1 + \frac{\mathbf{v}^2}{c^2}\right)(x'_2 - x'_1) =$$

$$= \frac{1 + \frac{\mathbf{v}^2}{c^2}}{\sqrt{1 - \frac{\mathbf{v}^2}{c^2}}}(x'_2 - x'_1). \quad (3.2.6)$$

This result is similar to that in standard special relativity: i.e., roughly speaking, a "moving" meter stick (at rest in $F_\mathbf{v}$) appears to be shorter, as measured by observers in the frame $F_I$, as shown in (3.2.6).

From [3] (see [3] Theorem in subsection I) we obtain transformations $\hat{L}(\mathbf{v})$:

$$\hat{L}(\mathbf{v}) = \Lambda(\mathbf{v}) \circ L(\mathbf{v}).$$
$$\Lambda(\mathbf{v}) : (t',x') \to (t,x)$$
$$x = x' - \mathbf{v}t',\, t = t' \quad (3.2.1.a)$$

$$L(\mathbf{v}) : \begin{cases} x' \to \gamma(\mathbf{v})(x' + \mathbf{v}t') & (3.2.1b) \\ t' \to \gamma(\mathbf{v})\left(t' + \dfrac{\mathbf{v}}{c^2}x'\right) & (3.2.1.c) \end{cases}$$

(3.2.7)

By substitution (3.2.1.b)-(3.2.1.c) into (3.2.1.a) we obtain the generalized Lorentz transformations $\hat{L}(\mathbf{v})$ :

$$\begin{aligned} x &= \gamma(\mathbf{v})(x' + \mathbf{v}t') - \mathbf{v}\gamma(\mathbf{v})\left(t' + \frac{\mathbf{v}}{c^2}x'\right) = \\ &\gamma(\mathbf{v})\left(1 - \frac{\mathbf{v}^2}{c^2}\right)x' = \gamma^{-1}(\mathbf{v})x' \end{aligned} \quad (3.2.8)$$

Thus

$$\begin{aligned} x &= \gamma^{-1}(\mathbf{v})x', \\ t &= \gamma(\mathbf{v})\left(t' + \frac{\mathbf{v}}{c^2}x'\right). \end{aligned} \quad (3.2.9)$$

Or in equivalent infinitesimal form:

$$\begin{aligned} dx &= \gamma^{-1}(\mathbf{v})dx' = dx'\sqrt{1 - \frac{\mathbf{v}^2}{c^2}}, \\ dt &= \gamma(\mathbf{v})\left(dt' + \frac{\mathbf{v}}{c^2}dx'\right). \end{aligned} \quad (3.2.10)$$

Substitution (3.2.10) into (3.1.5) gives:

$$\begin{aligned} ds^2 &= c^2\left(1 - \frac{\mathbf{v}^2}{c^2}\right)dt^2 - 2\mathbf{v}dtdx - dx^2 = \\ &c^2\left(1 - \frac{\mathbf{v}^2}{c^2}\right)\gamma^2(\mathbf{v})\left(dt' + \frac{\mathbf{v}}{c^2}dx'\right)^2 - 2\mathbf{v}\gamma(\mathbf{v})\left(dt' + \frac{\mathbf{v}}{c^2}dx'\right)\gamma^{-1}(\mathbf{v})dx' - \\ &-\left(1 - \frac{\mathbf{v}^2}{c^2}\right)^2 dx'^2 = \\ &c^2\left(dt' + \frac{\mathbf{v}}{c^2}dx'\right)^2 - 2\mathbf{v}\left(dt' + \frac{\mathbf{v}}{c^2}dx'\right)dx' - \left(1 - \frac{\mathbf{v}^2}{c^2}\right)dx'^2 = \\ &c^2 dt'^2 + \underline{2\mathbf{v}dt'dx'} + \frac{\mathbf{v}^2}{c^2}dx'^2 - \underline{2\mathbf{v}dt'dx'} - 2\frac{\mathbf{v}^2}{c^2}dx'^2 - \\ &-dx'^2 + \underline{\frac{\mathbf{v}^2}{c^2}dx'^2} = c^2 dt'^2 - dx'^2 = ds'^2. \end{aligned} \quad (3.2.11)$$

Thus, the invariant interval $ds^2$ under transformation (3.2.10) is

$$ds^2 = c^2\left(1 - \frac{\mathbf{v}^2}{c^2}\right)dt^2 - 2\mathbf{v}dtdx - dx^2 = c^2 dt'^2 - dx'^2 = ds'^2. \quad (3.2.12)$$

Let us calculate now from generalized Lorentz transformations $\hat{L}(\mathbf{v})$ (3.2.3) the

transformations $L^*(\mathbf{v})$, which keeping the metric (3.1.5) forminvariant:

$$L^*(\mathbf{v}) = \check{L}(\mathbf{v}) \circ \Lambda^{-1}(\mathbf{v}),$$

$$\check{L}(\mathbf{v}) : (t', x') \to (t_1, x_1) = \hat{L}(\mathbf{v})(t', x').$$

$$\Lambda^{-1}(\mathbf{v}) : (t, x) \to (t', x')$$

$$x' = x + \mathbf{v}t,\ t' = t. \quad (3.2.13.a) \quad\quad (3.2.13)$$

$$\hat{L}(\mathbf{v}) : \begin{cases} x_1 = \gamma(\mathbf{v})\left[ x'\left(1 + \frac{\mathbf{v}^2}{c^2}\right) - 2\mathbf{v}t' \right], & (3.2.14.b) \\ t_1 = \gamma(\mathbf{v})\left( t' - \frac{\mathbf{v}}{c^2} x' \right). & (3.2.15.c) \end{cases}$$

By substitution (3.2.13.a) into (3.2.13.b)-(3.2.13.c) we obtain the transformations which keeping the metric (3.1.5) forminvariant (were first obtained in [30], but in a much more complicated way — by solving a system of partial differential equations):

$$x_1 = \gamma(\mathbf{v})\left[ x'\left(1 + \frac{\mathbf{v}^2}{c^2}\right) - 2\mathbf{v}t' \right] = \gamma(\mathbf{v})\left[ (x + \mathbf{v}t)\left(1 + \frac{\mathbf{v}^2}{c^2}\right) - 2\mathbf{v}t \right] =$$

$$\gamma(\mathbf{v})\left[ x + \frac{\mathbf{v}^2}{c^2}x + \mathbf{v}t + \frac{\mathbf{v}^3}{c^2}t - 2\mathbf{v}t \right] = \gamma(\mathbf{v})\left[ \left(1 + \frac{\mathbf{v}^2}{c^2}\right)x - \left(\mathbf{v} - \frac{\mathbf{v}^3}{c^2}\right)t \right].$$

$$t_1 = \gamma(\mathbf{v})\left( t' - \frac{\mathbf{v}}{c^2}x' \right) = \gamma(\mathbf{v})\left[ t - \frac{\mathbf{v}}{c^2}(x + \mathbf{v}t) \right] =$$

$$\gamma(\mathbf{v})\left[ \left(1 - \frac{\mathbf{v}^2}{c^2}\right)t - \frac{\mathbf{v}}{c^2}x \right].$$

$$(3.2.14)$$

Thus

$$x_1 = \frac{\left(1 + \frac{\mathbf{v}^2}{c^2}\right)x - \left(\mathbf{v} - \frac{\mathbf{v}^3}{c^2}\right)t}{\sqrt{1 - \frac{\mathbf{v}^2}{c^2}}} =$$

$$\frac{\left(1 + \frac{\mathbf{v}^2}{c^2}\right)x - \left(1 - \frac{\mathbf{v}^2}{c^2}\right)\mathbf{v}t}{\sqrt{1 - \frac{\mathbf{v}^2}{c^2}}}, \quad (3.2.15)$$

$$t_1 = \frac{\left(1 - \frac{\mathbf{v}^2}{c^2}\right)t - \frac{\mathbf{v}}{c^2}x}{\sqrt{1 - \frac{\mathbf{v}^2}{c^2}}}.$$

Or in equivalent infinitesimal form:

$$dx_1 = \gamma(\mathbf{v})\left[ \left(1 + \frac{\mathbf{v}^2}{c^2}\right)dx - \left(\mathbf{v} - \frac{\mathbf{v}^3}{c^2}\right)dt \right]$$

$$dt_1 = \gamma(\mathbf{v})\left[ \left(1 - \frac{\mathbf{v}^2}{c^2}\right)dt - \frac{\mathbf{v}}{c^2}dx \right]$$

$$(3.2.16)$$

Substitution (3.2.16) into (3.1.5) gives:

$$ds_1^2 = c^2\left(1 - \frac{\mathbf{v}^2}{c^2}\right)dt_1^2 - 2\mathbf{v}dt_1dx_1 - dx_1^2 =$$

$$\gamma^2(\mathbf{v})c^2\left(1 - \frac{\mathbf{v}^2}{c^2}\right)\left[\left(1 - \frac{\mathbf{v}^2}{c^2}\right)dt - \frac{\mathbf{v}}{c^2}dx\right]^2 -$$

$$-2\mathbf{v}\gamma^2(\mathbf{v})\left[\left(1 - \frac{\mathbf{v}^2}{c^2}\right)dt - \frac{\mathbf{v}}{c^2}dx\right]\left[\left(1 + \frac{\mathbf{v}^2}{c^2}\right)dx - \left(\mathbf{v} - \frac{\mathbf{v}^3}{c^2}\right)dt\right] -$$

$$-\gamma^2(\mathbf{v})\left[\left(1 + \frac{\mathbf{v}^2}{c^2}\right)dx - \left(\mathbf{v} - \frac{\mathbf{v}^3}{c^2}\right)dt\right]^2 =$$

$$c^2\left[\left(1 - \frac{\mathbf{v}^2}{c^2}\right)dt - \frac{\mathbf{v}}{c^2}dx\right]^2 -$$

$$-\gamma^2(\mathbf{v})\left\{2\mathbf{v}\left[\left(1 - \frac{\mathbf{v}^2}{c^2}\right)dt - \frac{\mathbf{v}}{c^2}dx\right] + \left[\left(1 + \frac{\mathbf{v}^2}{c^2}\right)dx - \left(\mathbf{v} - \frac{\mathbf{v}^3}{c^2}\right)dt\right]\right\} \times$$

$$\times \left[\left(1 + \frac{\mathbf{v}^2}{c^2}\right)dx - \left(\mathbf{v} - \frac{\mathbf{v}^3}{c^2}\right)dt\right] =$$

$$c^2\left[\left(1 - \frac{\mathbf{v}^2}{c^2}\right)dt - \frac{\mathbf{v}}{c^2}dx\right]^2 -$$

$$-\gamma^2(\mathbf{v})\left(2\mathbf{v}dt - \frac{2\mathbf{v}^3}{c^2}dt - \frac{2\mathbf{v}^2}{c^2}dx + dx + \frac{\mathbf{v}^2}{c^2}dx - \mathbf{v}dt + \frac{\mathbf{v}^3}{c^2}dt\right) \times$$

$$\times \left[\left(1 + \frac{\mathbf{v}^2}{c^2}\right)dx - \left(\mathbf{v} - \frac{\mathbf{v}^3}{c^2}\right)dt\right] =$$

$$c^2\left[\left(1 - \frac{\mathbf{v}^2}{c^2}\right)dt - \frac{\mathbf{v}}{c^2}dx\right]^2 -$$

$$-\gamma^2(\mathbf{v})\left(\mathbf{v}dt - \frac{\mathbf{v}^3}{c^2}dt - \frac{\mathbf{v}^2}{c^2}dx + dx\right)\left[\left(1 + \frac{\mathbf{v}^2}{c^2}\right)dx - \left(\mathbf{v} - \frac{\mathbf{v}^3}{c^2}\right)dt\right] =$$

$$c^2\left[\left(1 - \frac{\mathbf{v}^2}{c^2}\right)dt - \frac{\mathbf{v}}{c^2}dx\right]^2 -$$

$$-\gamma^2(\mathbf{v})\left[\mathbf{v}dt\left(1 - \frac{\mathbf{v}^2}{c^2}\right) - dx\left(1 - \frac{\mathbf{v}^2}{c^2}\right)\right] \times \qquad (3.2.17)$$

$$\times \left[\left(1 + \frac{\mathbf{v}^2}{c^2}\right)dx - \left(\mathbf{v} - \frac{\mathbf{v}^3}{c^2}\right)dt\right] =$$

$$c^2\left[\left(1 - \frac{\mathbf{v}^2}{c^2}\right)dt - \frac{\mathbf{v}}{c^2}dx\right]^2 -$$

$$-(\mathbf{v}dt - dx)\left[\left(1 + \frac{\mathbf{v}^2}{c^2}\right)dx - \left(\mathbf{v} - \frac{\mathbf{v}^3}{c^2}\right)dt\right] =$$

$$c^2\left(1 - \frac{\mathbf{v}^2}{c^2}\right)^2 dt^2 - 2\mathbf{v}\left(1 - \frac{\mathbf{v}^2}{c^2}\right)dtdx + \frac{\mathbf{v}^2}{c^2}dx^2 -$$

$$-\mathbf{v}\left(1 + \frac{\mathbf{v}^2}{c^2}\right)dtdx + \mathbf{v}^2\left(1 - \frac{\mathbf{v}^2}{c^2}\right)dt^2 - \left(1 + \frac{\mathbf{v}^2}{c^2}\right)dx^2 +$$

$$+\mathbf{v}\left(1 - \frac{\mathbf{v}^2}{c^2}\right)dtdx =$$

$$c^2\left(1 - \frac{\mathbf{v}^2}{c^2}\right)^2 dt^2 - 2\mathbf{v}\left(1 - \frac{\mathbf{v}^2}{c^2}\right)dtdx + \frac{\mathbf{v}^2}{c^2}dx^2 - \underline{\mathbf{v}dtdx} -$$

$$-\frac{\mathbf{v}^3}{c^2}dtdx + \mathbf{v}^2dt^2 -$$

$$-\frac{\mathbf{v}^4}{c^2}dt^2 - dx^2 - \underline{\frac{\mathbf{v}^2}{c^2}dx^2} + \underline{\mathbf{v}dtdx} - \frac{\mathbf{v}^3}{c^2}dtdx =$$

Thus under transformations (3.2.16) the metric (3.1.5) is forminvariant.

# III.3. Generalized Lorentz transformations between a generalized inertial frame $F_\mathbf{v}(t,x,y,z)$ and an inertial Minkowski frame $F_I(t_I,x_I,y_I,z_I)$.

Let us calculate the generalized Lorentz transformations $\hat{L}^{-1}(\mathbf{v})$ of space and time between a generalized inertial frame $F_\mathbf{v}(t,x,y,z)$ with a metric (3.1.5) forming by the linear (non-orthogonal) transformations $\Lambda(\mathbf{v})$ (3.1.4) and an inertial Minkowski frame $F_I(t_I,x_I,y_I,z_I) = F_I(t',x',y',z')$

From [3] (see [3] Theorem in subsection I) we obtain:

$$\hat{L}^{-1}(\mathbf{v}) = \check{L}(\mathbf{v}) \circ \Lambda^{-1}(\mathbf{v}). \qquad (3.3.1.a)$$

$$\Lambda^{-1}(\mathbf{v}) : (t,x) \to (t',x')$$

$$x' = x + \mathbf{v}t, t = t'. \qquad (3.3.1.b)$$

$$\check{L}(\mathbf{v}) : (t',x') \to L(\mathbf{v})(t',x') \qquad (3.3.1)$$

$$L(\mathbf{v}) : \begin{cases} x' \to \gamma(\mathbf{v})(x' + \mathbf{v}t') \\ t' \to \gamma(\mathbf{v})\left(t' + \dfrac{\mathbf{v}}{c^2}x'\right) \end{cases} \qquad (3.3.1.c)$$

By substitution (3.3.1.b) into (3.3.1.c) we obtain the generalized Lorentz transformations $\hat{L}^{-1}(\mathbf{v})$ :

$$x' = \gamma(\mathbf{v})(x + 2\mathbf{v}t),$$

$$t' = \gamma(\mathbf{v})\left(t + \dfrac{\mathbf{v}}{c^2}(x + \mathbf{v}t)\right) = \qquad (3.3.2)$$

$$\gamma(\mathbf{v})\left[t\left(1 + \dfrac{\mathbf{v}^2}{c^2}\right) + \dfrac{\mathbf{v}}{c^2}x\right]$$

Or in equivalent infinitesimal form:

$$dx' = \gamma(\mathbf{v})(dx + 2\mathbf{v}dt),$$

$$dt' = \gamma(\mathbf{v})\left[dt\left(1 + \dfrac{\mathbf{v}^2}{c^2}\right) + \dfrac{\mathbf{v}}{c^2}dx\right]. \qquad (3.3.3)$$

Substitution (3.3.3) into Minkowski frame $ds'^2 = c^2 dt'^2 - dx'^2$ gives:

$$c^2dt'^2 - dx'^2 = \gamma(\mathbf{v})\left\{c^2\left[dt\left(1+\frac{\mathbf{v}^2}{c^2}\right)+\frac{\mathbf{v}}{c^2}dx\right]^2 - (dx+2\mathbf{v}dt)^2\right\} =$$

$$\gamma(\mathbf{v})\left\{c^2\left[dt^2\left(1+\frac{\mathbf{v}^2}{c^2}\right)^2 + 2dt\frac{\mathbf{v}}{c^2}dx\left(1+\frac{\mathbf{v}^2}{c^2}\right) + \frac{\mathbf{v}^2}{c^4}dx^2\right]\right.$$
$$-(dx^2 + 4\mathbf{v}dtdx + 4\mathbf{v}^2dt^2)\} =$$

$$\gamma(\mathbf{v})\left\{c^2dt^2 + 2dt^2\mathbf{v}^2 + dt^2\frac{\mathbf{v}^4}{c^2} + 2\mathbf{v}dtx + 2dtdx\frac{\mathbf{v}^3}{c^2} + \frac{\mathbf{v}^2}{c^2}dx\mathbf{v}^2\right.$$
$$-dx^2 - 4dtdx - 4\mathbf{v}^2dt^2\} = \qquad (3.3.4)$$

$$\gamma(\mathbf{v})\left\{c^2dt^2 - 2dt^2\mathbf{v}^2 + dt^2\frac{\mathbf{v}^4}{c^2} - 2\mathbf{v}dtdx + 2dtdx\frac{\mathbf{v}^3}{c^2} + \frac{\mathbf{v}^2}{c^2}dx^2 - dx^2\right\} =$$

$$= \gamma(\mathbf{v})\left\{c^2dt^2\left(1-\frac{\mathbf{v}^2}{c^2}\right)^2 - 2\mathbf{v}dtdx\left(1-\frac{\mathbf{v}^2}{c^2}\right) - dx^2\left(1-\frac{\mathbf{v}^2}{c^2}\right)\right\} =$$

$$= c^2dt^2\left(1-\frac{\mathbf{v}^2}{c^2}\right) - 2\mathbf{v}dtdx - dx^2 = ds^2.$$

Thus, the invariant interval $ds'^2$ under the generalized Lorentz transformation (3.3.3) is

$$ds'^2 = c^2dt'^2 - dx'^2 = c^2\left(1-\frac{\mathbf{v}^2}{c^2}\right)dt^2 - 2\mathbf{v}dtdx - dx^2 = ds^2. \qquad (3.3.5)$$

Let us calculate now the generalized Lorentz transformations $\hat{L}_{\mathbf{M}}^{-1}(\mathbf{v})$:

$$\hat{L}_{\mathbf{M}}^{-1}(\mathbf{v}) = \check{L}(\mathbf{v}) \circ \Lambda^{-1}(\mathbf{v}). \quad (3.3.6.a)$$
$$\Lambda^{-1}(\mathbf{v}) : (t,x) \rightarrow (t',x')$$
$$x' = x + \mathbf{v}t, t' = t. \quad (3.3.6.b)$$
$$\check{L}(\mathbf{v}) : (t',x') \rightarrow L(\mathbf{v})(t',x'), \qquad (3.3.6)$$
$$\text{where} \quad L(\mathbf{v}) : \begin{cases} x' \rightarrow \gamma(\mathbf{v})(x' - \mathbf{v}t') \\ t' \rightarrow \gamma(\mathbf{v})\left(t' - \frac{\mathbf{v}}{c^2}x'\right) \end{cases} \quad (3.3.6.c)$$

By substitution (3.3.6.b) into (3.3.6.c) we obtain the generalized Lorentz transformations $\hat{L}_{\mathbf{M}}^{-1}(\mathbf{v})$:

$$x' = \gamma(\mathbf{v})x,$$
$$t' = \gamma(\mathbf{v})\left[t\left(1-\frac{\mathbf{v}^2}{c^2}\right) - \frac{\mathbf{v}}{c^2}x\right]. \qquad (3.3.7)$$

were which first obtained by C. Møller in [31].Or in equivalent infinitesimal form:

$$dx' = \gamma(\mathbf{v})dx,$$
$$dt' = \gamma(\mathbf{v})\left[dt\left(1 - \frac{\mathbf{v}^2}{c^2}\right) - \frac{\mathbf{v}}{c^2}dx\right].$$
(3.3.8)

Substitution (3.17) into Minkowski frame $ds'^2 = c^2 dt'^2 - dx'^2$ gives:

$$ds'^2 = c^2 dt'^2 - dx'^2 =$$
$$c^2 \gamma^2(\mathbf{v})\left[dt\left(1 - \frac{\mathbf{v}^2}{c^2}\right) - \frac{\mathbf{v}}{c^2}dx\right]^2 - \gamma^2(\mathbf{v})dx^2 =$$
$$\gamma^2(\mathbf{v})\left\{c^2\left[dt\left(1 - \frac{\mathbf{v}^2}{c^2}\right) - \frac{\mathbf{v}}{c^2}dx\right]^2 - dx^2\right\} =$$
$$\gamma^2(\mathbf{v})\left[c^2 dt^2\left(1 - \frac{\mathbf{v}^2}{c^2}\right)^2 - 2\mathbf{v}\left(1 - \frac{\mathbf{v}^2}{c^2}\right)dtdx + \frac{\mathbf{v}^2}{c^2}dx^2 - dx^2\right] =$$
$$\gamma^2(\mathbf{v})\left[c^2 dt^2\left(1 - \frac{\mathbf{v}^2}{c^2}\right)^2 - 2\mathbf{v}\left(1 - \frac{\mathbf{v}^2}{c^2}\right)dtdx - \left(1 - \frac{\mathbf{v}^2}{c^2}\right)dx^2\right] =$$
$$c^2 dt^2\left(1 - \frac{\mathbf{v}^2}{c^2}\right) - 2\mathbf{v}dtdx - dx^2 = ds^2.$$
(3.3.9)

Let us calculate now from generalized Lorentz transformations $\hat{L}^{-1}(\mathbf{v})$ (3.3.3) the transformations $L^{*-1}(\mathbf{v})$, which keeping the metric (3.1.5) forminvariant

$$L^{*-1}(\mathbf{v}) = \Lambda(\mathbf{v}) \circ \hat{L}^{-1}(\mathbf{v}),$$
$$\hat{L}^{-1}(\mathbf{v}) : (t_1, x_1) \to (t', x') = \hat{L}^{-1}(\mathbf{v})(t_1, x_1).$$
$$\Lambda(\mathbf{v}) : (t', x') \to (t, x)$$
$$x' = x + \mathbf{v}t, t' = t \text{ i.e. } x = x' - \mathbf{v}t', t = t'. \quad (3.3.10.a) \quad (3.3.10)$$

$$\hat{L}^{-1}(\mathbf{v}) : \begin{cases} x' = \gamma(\mathbf{v})(x_1 + 2\mathbf{v}t_1), & (3.3.10.b) \\ t' = \gamma(\mathbf{v})\left[t_1\left(1 + \frac{\mathbf{v}^2}{c^2}\right) + \frac{\mathbf{v}}{c^2}x_1\right]. & (3.3.10.c) \end{cases}$$

By substitution (3.3.10.b)-(3.3.10.c) into (3.3.10.a) we obtain the transformations which keeping the metric (3.1.5) forminvariant, were first obtained in [30], but in a much more complicated way — by solving a system of a partial differential equations:

$$x = x' - \mathbf{v}t' =$$
$$\gamma(\mathbf{v})(x_1 + 2\mathbf{v}t_1) - \mathbf{v}\gamma(\mathbf{v})\left[t_1\left(1 + \frac{\mathbf{v}^2}{c^2}\right) + \frac{\mathbf{v}}{c^2}x_1\right] =$$
$$\gamma(\mathbf{v})\left[(x_1 + 2\mathbf{v}t_1) - \mathbf{v}t_1\left(1 + \frac{\mathbf{v}^2}{c^2}\right) - \frac{\mathbf{v}^2}{c^2}x_1\right] =$$
$$\gamma(\mathbf{v})\left[\left(1 - \frac{\mathbf{v}^2}{c^2}\right)x_1 + \left(1 - \frac{\mathbf{v}^2}{c^2}\right)\mathbf{v}t_1\right].$$
(3.3.11)

Thus

$$x = \frac{\left(1 - \frac{\mathbf{v}^2}{c^2}\right)x_1 + \left(1 - \frac{\mathbf{v}^2}{c^2}\right)\mathbf{v}t_1}{\sqrt{1 - \frac{\mathbf{v}^2}{c^2}}},$$

$$t = \frac{\left(1 + \frac{\mathbf{v}^2}{c^2}\right)t_1 + \frac{\mathbf{v}}{c^2}x_1}{\sqrt{1 - \frac{\mathbf{v}^2}{c^2}}}.$$

(3.3.12)

## III.4. Adding velocities.

Let us consider the generalized Lorentzin transformations (3.2.3) in equivalent infinitesimal form:

$$dx = \gamma(\mathbf{v})\left[dx'\left(1 + \frac{\mathbf{v}^2}{c^2}\right) - 2\mathbf{v}dt'\right], dy = dy', dz = dz',$$

$$dt = \gamma(\mathbf{v})\left(dt' - \frac{\mathbf{v}}{c^2}dx'\right).$$

(3.4.1)

From (3.4.1) we obtain

$$\frac{dx}{dt} = \frac{dx'\left(1 + \frac{\mathbf{v}^2}{c^2}\right) - 2\mathbf{v}dt'}{dt' - \frac{\mathbf{v}}{c^2}dx'},$$

$$\frac{dy}{dt} = \frac{dy'\sqrt{1 - \frac{\mathbf{v}^2}{c^2}}}{dt' - \frac{\mathbf{v}}{c^2}dx'},$$

(3.4.2)

$$\frac{dz}{dt} = \frac{dz'\sqrt{1 - \frac{\mathbf{v}^2}{c^2}}}{dt' - \frac{\mathbf{v}}{c^2}dx'},$$

From (3.4.2) by usual way we obtain

$$\frac{dx}{dt} = \frac{\frac{dx'}{dt'}\left(1 + \frac{\mathbf{v}^2}{c^2}\right) - 2\mathbf{v}}{1 - \frac{\mathbf{v}}{c^2}\frac{dx'}{dt'}},$$

$$\frac{dy}{dt} = \frac{\frac{dy'}{dt'}\sqrt{1 - \frac{\mathbf{v}^2}{c^2}}}{1 - \frac{\mathbf{v}}{c^2}\frac{dx'}{dt'}}, \qquad (3.4.3)$$

$$\frac{dz}{dt} = \frac{\frac{dz'}{dt'}\sqrt{1 - \frac{\mathbf{v}^2}{c^2}}}{1 - \frac{\mathbf{v}}{c^2}\frac{dx'}{dt'}}.$$

By setting

$$\mathbf{v}_x = \frac{dx}{dt}, \mathbf{v}'_x = \frac{dx'}{dt'},$$

$$\mathbf{v}_y = \frac{dy}{dt}, \mathbf{v}'_y = \frac{dy'}{dt'}, \qquad (3.4.4)$$

$$\mathbf{v}_z = \frac{dz}{dt}, \mathbf{v}'_z = \frac{dz'}{dt'},$$

from (3.4.3) we obtain the formulae relating velocities

$$\mathbf{v}_x = \frac{\mathbf{v}'_x\left(1 + \frac{\mathbf{v}^2}{c^2}\right) - 2\mathbf{v}}{1 - \frac{\mathbf{v}\mathbf{v}'_x}{c^2}},$$

$$\mathbf{v}_y = \frac{\mathbf{v}'_y\sqrt{1 - \frac{\mathbf{v}^2}{c^2}}}{1 - \frac{\mathbf{v}\mathbf{v}'_x}{c^2}}, \qquad (3.4.5)$$

$$\mathbf{v}_z = \frac{\mathbf{v}_z\sqrt{1 - \frac{\mathbf{v}^2}{c^2}}}{1 - \frac{\mathbf{v}\mathbf{v}'_x}{c^2}}.$$

Let us consider now the generalized Lorentzin transformations (3.3.2) in equivalent infinitesimal form:

$$dx' = \gamma(\mathbf{v})(dx + 2\mathbf{v}dt), dy' = dy, dz' = dz,$$

$$dt' = \gamma(\mathbf{v})\left[dt\left(1 + \frac{\mathbf{v}^2}{c^2}\right) + \frac{\mathbf{v}}{c^2}dx\right]. \qquad (3.4.6)$$

From (3.4.6) we obtain

$$\frac{dx'}{dt'} = \frac{\frac{dx}{dt} + 2\mathbf{v}}{1 + \frac{\mathbf{v}^2}{c^2} + \frac{\mathbf{v}}{c^2}\frac{dx}{dt}},$$

$$\frac{dy'}{dt'} = \frac{\frac{dy}{dt}\sqrt{1 - \frac{\mathbf{v}^2}{c^2}}}{1 + \frac{\mathbf{v}^2}{c^2} + \frac{\mathbf{v}}{c^2}\frac{dx}{dt}}, \qquad (3.4.7)$$

$$\frac{dz'}{dt'} = \frac{\frac{dz}{dt}\sqrt{1 - \frac{\mathbf{v}^2}{c^2}}}{1 + \frac{\mathbf{v}^2}{c^2} + \frac{\mathbf{v}}{c^2}\frac{dx}{dt}}.$$

From (3.4.4),(3.4.7) we obtain the formulae relating velocities

$$\mathbf{v}'_x = \frac{\mathbf{v}_x + 2\mathbf{v}}{1 + \frac{\mathbf{v}^2}{c^2} + \frac{\mathbf{v}\mathbf{v}_x}{c^2}},$$

$$\mathbf{v}'_y = \frac{\mathbf{v}_y\sqrt{1 - \frac{\mathbf{v}^2}{c^2}}}{1 + \frac{\mathbf{v}^2}{c^2} + \frac{\mathbf{v}\mathbf{v}_x}{c^2}}, \qquad (3.4.8)$$

$$\mathbf{v}'_z = \frac{\mathbf{v}_z\sqrt{1 - \frac{\mathbf{v}^2}{c^2}}}{1 + \frac{\mathbf{v}^2}{c^2} + \frac{\mathbf{v}\mathbf{v}_x}{c^2}}.$$

By setting $\mathbf{v}_x = 0$ we obtain

$$\mathbf{v}'_x = \frac{2\mathbf{v}}{1 + \frac{\mathbf{v}^2}{c^2}}. \qquad (3.4.9)$$

## III.5. Length contraction.

Again we consider two such inertial reference frame $F_{\mathbf{v}}(t,x,y,z)$ with a metric (3.1.5) forming by the linear (non-orthogonal) transformations $\Lambda(\mathbf{v})$ (3.1.4) and Minkowski frame $F_I(t_I,x_I,y_I,z_I) = F'_I(t',x',y',z')$, one of which $F_{\mathbf{v}}$ will be considered to be at rest, while another $F'_I$ one will move with respect to the first one.

Consider measurement, in a moving reference frame $F'_I$, of the length of a rod that is at rest in another reference frame $F_{\mathbf{v}}$. We first determine the method for measuring the length of a moving rod. Consider an observer in the moving reference frame $F'_I$, who records the ends of the rod, $x'_1$ and $x'_2$, at the same moment of time

$$t'_1 = t'_2, \qquad (3.5.1)$$

this permits to reduce the Minkowski interval $s'^2_{1,2}$ in the moving reference frame $F'_I$ to the spatial part only

$$s'^2_{1,2} = -(x'_2 - x'_1)^2 = -l'^2. \qquad (3.5.2)$$

Thus, in our method of determining the length of a moving rod, by usual way to consider the quantity $l'$ as its *length* in a moving reference frame $F'_I$.

The same interval in the reference frame at rest $F_{\mathbf{v}}$, where the rod is in the state of rest, is given as follows

$$s_{1,2}^2 = c^2\left(1 - \frac{\mathbf{v}^2}{c^2}\right)(t_2 - t_1)^2 - 2\mathbf{v}(t_2 - t_1)(x_2 - x_1) - (x_2 - x_1)^2. \quad (3.5.3)$$

But, in accordance with the generalized Lorentz transformations (3.3.3) we have

$$t_2' - t_1' = \gamma(\mathbf{v})\left[\left(1 + \frac{\mathbf{v}^2}{c^2}\right)(t_2 - t_1) + \frac{\mathbf{v}}{c^2}(x_2 - x_1)\right]. \quad (3.5.4)$$

whence for our case (3.5.1) we find

$$\left(1 + \frac{\mathbf{v}^2}{c^2}\right)(t_2 - t_1) + \frac{\mathbf{v}}{c^2}(x_2 - x_1) = 0,$$

$$t_2 - t_1 = -\frac{\mathbf{v}}{c^2}\frac{x_2 - x_1}{1 + \frac{\mathbf{v}^2}{c^2}} = -\frac{\mathbf{v}}{c^2}\frac{l_0}{1 + \frac{\mathbf{v}^2}{c^2}}, \quad (3.5.5)$$

where $l_0 = x_2 - x_1$ is length of the rod in the reference frame at rest. Substituting this expression into (3.5.3) we obtain:

$$s_{1,2}^2 = c^2\left(1 - \frac{\mathbf{v}^2}{c^2}\right)(t_2 - t_1)^2 - 2\mathbf{v}l_0(t_2 - t_1) - l_0^2 =$$

$$\left(1 - \frac{\mathbf{v}^2}{c^2}\right)\frac{\mathbf{v}^2}{c^2}\frac{l_0^2}{\left(1 + \frac{\mathbf{v}^2}{c^2}\right)^2} + \frac{2\mathbf{v}^2}{c^2}\frac{l_0^2}{1 + \frac{\mathbf{v}^2}{c^2}} - l_0^2 =$$

$$l_0^2\left(1 + \frac{\mathbf{v}^2}{c^2}\right)^{-2}\left[\left(1 - \frac{\mathbf{v}^2}{c^2}\right)\frac{\mathbf{v}^2}{c^2} + \frac{2\mathbf{v}^2}{c^2}\left(1 + \frac{\mathbf{v}^2}{c^2}\right) - \left(1 + \frac{\mathbf{v}^2}{c^2}\right)^2\right] =$$

$$l_0^2\left(1 + \frac{\mathbf{v}^2}{c^2}\right)^{-2}\left[\frac{\mathbf{v}^2}{c^2} - \frac{\mathbf{v}^4}{c^4} + \frac{2\mathbf{v}^2}{c^2} + \frac{2\mathbf{v}^4}{c^4} - 1 - 2\frac{\mathbf{v}^2}{c^2} - \frac{\mathbf{v}^4}{c^4}\right] = \quad (3.5.6)$$

$$-l_0^2\frac{1 - \frac{\mathbf{v}^2}{c^2}}{\left(1 + \frac{\mathbf{v}^2}{c^2}\right)^2},$$

$$s_{1,2}^2 = -l_0^2\frac{1 - \frac{\mathbf{v}^2}{c^2}}{\left(1 + \frac{\mathbf{v}^2}{c^2}\right)^2}.$$

Comparing (3.5.2) and (3.5.6) we find

$$l = l_0\frac{\sqrt{1 - \frac{\mathbf{v}^2}{c^2}}}{1 + \frac{\mathbf{v}^2}{c^2}}. \quad (3.5.7)$$

Eq.(3.5.7) coincides with Eq.(3.2.6).

## IV. Uniformly accelerated frames of reference. Non-holonomic Minkovsky reference frames. Non-holonomic Lorentz transformations. Local Lorentz length and time contractions. Relativistic contraction of an accelerated rod from non-holonomic Lorentz transformations.

### IV.1. Uniformly accelerated frames of reference.

Let a relativistic uniformly accelerated frame $F_a$ with the coordinates $(x,t)$ move without initial velocity along the axis $x'$ of an inertial Galilee rame $F_I = F_I(x',t')$ with the coordinates $(x',t')$ and at $t = t' = 0$ their origins coincide. Then the coordinate transformation formulas $x$ have the form [30]:

$$x = x' - \frac{c^2}{|\vec{a}|}\left[\sqrt{1 + \frac{|\vec{a}|^2 t'^2}{c^2}} - 1\right]. \tag{4.1.1}$$

Or in equivalent infinitesimal form:

$$\begin{aligned} dx &= dx' + \frac{\vec{a}t'}{\sqrt{1 + \frac{|\vec{a}|^2 t'^2}{c^2}}} dt' = \\ &= dx' + \vec{v}(t')dt', \\ \vec{v}(t') &= \frac{|\vec{a}|t'}{\sqrt{1 + \frac{|\vec{a}|^2 t'^2}{c^2}}}. \\ dt &= dt'. \end{aligned} \tag{4.1.2}$$

The motion law, i.e. the transformation to FR stringently connected with a uniformly accelerated moving body, is already defined, it means that the space coordinate transformation is already fixed (up to rotations and shifts). The covariance of the SRT description finds here its manifestation in arbitrariness of the dependence $t = t(x', t')$ of the time coordinate. We considered now the simplest possible dependence $t = t'$. In this case, substituting (4.1.2) and $t = t'$ into the Minkovsky metrics gives the expression of the metrics of a uniformly accelerated frame $F_a$:

$$ds^2 = \frac{c^2}{1 + \frac{a^2 t^2}{c^2}} dt - \frac{2at}{\sqrt{1 + \frac{a^2 t^2}{c^2}}} dtdx - dx^2 - dy^2 - dz^2. \tag{4.1.3}$$

In a holonomic reference system all geometric objects are represented as functions of the coordinates $x_a, a = 0, 1, 2, 3$ in a new holonomic system they become functions of the new coordinates instead of the old. In a non-holonomic reference frame, however, we do not have a complete set of new coordinates; we have only a set of new differentials. Thus, relative to a non-holonomic reference system, geometric objects are expressed as functions not only of the new differentials but also of the old coordinates. In general, one must carry the original holonomic coordinates along with the new differentials. Consequently, there will here be nothing in the appearance of the transformed geometric objects to characterize them as being in a non-holonomic reference frame [32],[33],[34].

From the Eq.(4.1.1) with the initial condition $x'(t' = 0) = 0$ we obtain

$$x'(t') = \frac{c^2}{|\vec{a}|}\sqrt{1 + \frac{|\vec{a}|^2 t'^2}{c^2}} - \frac{c^2}{|\vec{a}|} \qquad (4.1.4)$$

Let us consider a rigid rod whose velocity and acceleration $\vec{a}$ are directed along its length, but are otherwise arbitrary functions of time. Let $F_I = F_I(t_I, x_I) = F_I(t', x')$ be a stationary inertial frame and $F_{\mathbf{V}(x')}(t, x)$ the accelerated frame of an observer on the rod. In particular if the rod is rigid and does not rotate, it is enough to know how one particular point $x_A(t'_A)$ of the rod labeled by $A$ changes its position with time $t'_A$ by the law $x_A(t'_A)$. The rigid rod is instantaneously at rest for an observer in the frame $F_{\mathbf{V}(x')}(t, x)$. This means that he observes no contraction, i.e., $\mathbf{L}_0 = x_B - x_A$, where $\mathbf{L}_0$ is the proper length of the rod, while **A** and **B** label the back and front ends of the rod, respectively. He can observes both ends at the any instants of the time $t_A$ and $t_B$, so $t_B - t_A \neq 0$.

**Claim** *(A) Suppose that*:

(1)

$$x'_A(t'_A) = \frac{c^2}{|\vec{a}|}\sqrt{1 + \frac{|\vec{a}|^2 t'^2_A}{c^2}} - \frac{c^2}{|\vec{a}|}. \qquad (4.1.5)$$

(2) *for the case when the force is applied to the back end of the rigid rod*

*(pushed rod)*

$$x'(t'_A, \bar{L}_0) = \frac{c^2}{|\vec{a}(\bar{L}_0)|}\sqrt{1 + \frac{|\vec{a}(\bar{L}_0)|^2 t'^2_A}{c^2}} - \frac{c^2}{|\vec{a}(\bar{L}_0)|}. \qquad (4.1.6)$$

*where*

$$\vec{a}(\bar{L}_0) = \frac{\vec{a}}{1 + \frac{|\vec{a}|\bar{L}_0}{c^2}}, 0 \leq \bar{L}_0 \leq \mathbf{L}_0, \qquad (4.1.7)$$

(see.[40] Eq.(27)) *where the quantity $\mathbf{L}_0$ is the proper length of the pushed rod.*

Substitution (4.1.7) into (4.1.6) with the initial condition $x'(t' = 0, \bar{L}_0) = \bar{L}_0$ gives:

$$x'(t'_A, \bar{L}_0) = \frac{c^2}{|\vec{a}|}\left(1 + \frac{|\vec{a}|\bar{L}_0}{c^2}\right) \times$$

$$\times \sqrt{1 + \left(\frac{\vec{a}}{1 + \frac{|\vec{a}|\bar{L}_0}{c^2}}\right)^2 \frac{t'^2_A}{c^2}} - \frac{c^2}{|\vec{a}|}\left(1 + \frac{|\vec{a}|\bar{L}_0}{c^2}\right) + \bar{L}_0 =$$

(4.1.8)

$$= \left(\frac{c^2}{|\vec{a}|} + \bar{L}_0\right)\sqrt{1 + \frac{\frac{|\vec{a}|^2 t'^2_A}{c^2}}{\left(1 + \frac{|\vec{a}|\bar{L}_0}{c^2}\right)^2}} - \frac{c^2}{|\vec{a}|}.$$

Thus we find

$$x'_B(t'_A) = x'(t'_A, \mathbf{L}_0) = \left(\frac{c^2}{|\vec{a}|} + \mathbf{L}_0\right)\sqrt{1 + \frac{\frac{|\vec{a}|^2 t'^2_A}{c^2}}{\left(1 + \frac{|\vec{a}|\mathbf{L}_0}{c^2}\right)^2}} - \frac{c^2}{|\vec{a}|}. \quad (4.1.9)$$

Thus the length $L(t')$ as a function of time measured by inertial observer is given by

$$L(t') = x'_B(t'_A) - x'_A(t'_A) = \left(\frac{c^2}{|\vec{a}|} + \mathbf{L}_0\right)\sqrt{1 + \frac{\frac{|\vec{a}|^2 t'^2_A}{c^2}}{\left(1 + \frac{|\vec{a}|\mathbf{L}_0}{c^2}\right)^2}} - \frac{c^2}{|\vec{a}|} -$$

$$-\frac{c^2}{|\vec{a}|}\sqrt{1 + \frac{|\vec{a}|^2 t'^2_A}{c^2}} + \frac{c^2}{|\vec{a}|} =$$

(4.1.10)

$$\left(\frac{c^2}{|\vec{a}|} + \mathbf{L}_0\right)\sqrt{1 + \frac{\frac{|\vec{a}|^2 t'^2_A}{c^2}}{\left(1 + \frac{|\vec{a}|\mathbf{L}_0}{c^2}\right)^2}} - \frac{c^2}{|\vec{a}|}\sqrt{1 + \frac{|\vec{a}|^2 t'^2_A}{c^2}}.$$

$$L(t'_A) = \left(\frac{c^2}{|\vec{a}|} + \mathbf{L}_0\right)\sqrt{1 + \frac{\frac{|\vec{a}|^2 t'^2_A}{c^2}}{\left(1 + \frac{|\vec{a}|\mathbf{L}_0}{c^2}\right)^2}} - \frac{c^2}{|\vec{a}|}\sqrt{1 + \frac{|\vec{a}|^2 t'^2_A}{c^2}}.$$

## IV.2. Non-holonomic Lorentz transformations. Non-holonomic Minkovsky reference frames.

The essence of the SRT, its basic and actually unique postulate can be formulated in the following way: "all physical processes run in the unit Minkowski time-space, the geometry of which is pseudo-Euclidean". In other words in SRT we postulate that in the whole space there is a physical holonomic frame $F_I = F_I(x',t')$ of reference (HFR) called an inertial (Galilee) one in which the infinitesimal interval between two infinitely close events of this space is written as:

$$ds_I^2 = c^2 dt'^2 - dx'^2 - dy'^2 - dz'^2. \qquad (4.2.1)$$

Or in global form:

$$s_I^2 = c^2(t_1' - t_2')^2 - (x_1' - x_2')^2 - (y_1' - y_2')^2 - (z_1' - z_2')^2. \qquad (4.2.2)$$

The metrics $ds_I^2$ and $s_I^2$ of the inertial (Galilee) FR (4.2.1),(4.2.2) is forminvariant with respect to the classical *holonomic* infinitesimal Lorentz transformations $\mathbf{dL}_m^i[\mathbf{V}]$, $\mathbf{V} = (V_x, V_y, V_z)$.

**Definition. 4.2.1.** Generalized infinitesimal Lorentz transformations $\mathbf{dL}_m^i[\mathbf{V}(\mathbf{t}, \mathbf{x})]$, $\mathbf{V}(\mathbf{t}, \mathbf{x}) = (V_x(\mathbf{t}, \mathbf{x}), V_y(\mathbf{t}, \mathbf{x}), V_z(\mathbf{t}, \mathbf{x}))$ called an *non-holonomic Lorentz transformations.*

In a classical holonomic reference system all geometric objects are represented as functions of the coordinates $x_a, a = 0, 1, 2, 3$ in a new holonomic system they become functions of the new coordinates instead of the old. In a non-holonomic reference frame, however, we do not have a complete set of new coordinates; we have only a set of new differentials. Thus, relative to a non-holonomic reference system, geometric objects are expressed as functions not only of the new differentials but also of the old coordinates. In general, one must carry the original holonomic coordinates along with the new differentials. Consequently, there will here be nothing in the appearance of the transformed geometric objects to characterize them as being in a non-holonomic reference frame [32],[33],[34].

## IV.2.1. Non-holonomic Lorentz transformations of the type I.

Let a relativistic accelerated frame $F_{a(t',x')} = F_{a(t',x')}(t,x)$ with the coordinates $(x,t)$ move along the axis $x'$ of an inertial Galilee rame $F_I = F_I(x',t')$ with the coordinates $(x',t')$ and at $t = t' = 0$ their origins coincide.

**(I)** Let us consider the non-holonomic transformations of the type **(I)** of space and time between an inertial inertial Galilee frame $F_I = F_I(x',t')$ and an accelerated frame $F_a^{(I)} = F_a^{(I)}(t,x) = F_{a(t',x')}(t,x)$ :

$$\begin{aligned} dx &= \gamma[\mathbf{V}(t',x')][dx' + \mathbf{V}(t',x')dt'], \\ dt &= \gamma[\mathbf{V}(t',x')]\left[dt' + \frac{\mathbf{V}(t',x')}{c^2}dx'\right]. \end{aligned} \qquad (4.2.3)$$

**Theorem 4.2.1.** The non-holonomic transformations (4.2.3) keeping the Minkowski metric $ds^2 = c^2 dt^2 - dx^2$ forminvariant.
**Proof.** From Eqs. (4.2.3) we obtain

$$ds^2 = c^2dt^2 - dx^2 =$$

$$= c^2\gamma^2[\mathbf{V}(t',x')]\left[dt' + \frac{\mathbf{V}(t',x')}{c^2}dx'\right]^2 - \gamma^2[\mathbf{V}(t',x')][dx' + \mathbf{V}(t',x')dt']^2 =$$

$$= \gamma^2[\mathbf{V}(t',x')]\left[\underline{c^2dt'^2} + \boxed{2\mathbf{V}(t',x')dt'dx'} + \frac{\mathbf{V}^2(t',x')}{c^2}dx'^2 - dx'^2 \right.$$

$$\left. - \boxed{2\mathbf{V}(t',x')dt'dx'} - \underline{\mathbf{V}^2(t',x')dt'^2}\right] =$$

$$= \gamma^2[\mathbf{V}(t',x')]\left[c^2\left(1 - \frac{\mathbf{V}^2(t',x')}{c^2}\right)dt'^2 - \left(1 - \frac{\mathbf{V}^2(t',x')}{c^2}\right)dx'^2\right] =$$

$$= c^2dt'^2 - dx'^2 = ds'^2.$$

(4.2.4)

Thus the non-holonomic transformations (4.2.3) keeping the Minkowski metric $ds^2 = c^2dt^2 - dx^2$ forminvariant.

**Corollary 4.2.1.** The interval between two infinitely close events of the frame $F_a^{(I)}$

is written as $ds^2 = c^2dt^2 - dx^2$.

**Definition 4.2.2.** The pair $\langle F_a^{(I)}, F_I \rangle$ there is a physical non-holonomic frame of reference (NHFR) called an (non-inertial) non-holonomic Minkowski frame of the type (I) (I-NHMF).

## IV.2.2. Non-holonomic Lorentz transformations of the type **II**.

**(II)** Let us consider the non-holonomic transformations of the type (**II**) of space and time between an inertial inertial Galilee frame $F_I = F_I(x',t')$ and an accelerated frame $F_a^{(II)}(t,x) = F_{a(t,x)}(t,x)$ :

$$dx = \gamma[\mathbf{V}(t,x)][dx' + \mathbf{V}(t,x)dt'],$$

$$dt = \gamma[\mathbf{V}(t,x)]\left[dt' + \frac{\mathbf{V}(t,x)}{c^2}dx'\right].$$

(4.2.5)

**Theorem 4.2.2.** The non-holonomic transformations (4.2.5) keeping the Minkowski metric $ds^2 = c^2dt^2 - dx^2$ forminvariant.

**Proof.** From Eqs. (4.2.5) we obtain:

$$ds^2 = c^2dt^2 - dx^2 =$$

$$= c^2\gamma^2[\mathbf{V}(t,x)]\left[dt' + \frac{\mathbf{V}(t,x)}{c^2}dx'\right]^2 - \gamma^2[\mathbf{V}(t,x)][dx' + \mathbf{V}(t,x)dt']^2 =$$

$$= \gamma^2[\mathbf{V}(t,x)]\left[\underline{c^2dt'^2} + \underbrace{2\mathbf{V}(t,x)dt'dx'} + \frac{\mathbf{V}^2(t,x)}{c^2}dx'^2 - dx'^2\right.$$

$$\left.\underbrace{-2\mathbf{V}(t,x)dt'dx'} - \underline{\mathbf{V}^2(t,x)dt'^2}\right] = \quad (4.2.6)$$

$$= \gamma^2[\mathbf{V}(t,x)]\left[c^2\left(1 - \frac{\mathbf{V}^2(t,x)}{c^2}\right)dt'^2 - \left(1 - \frac{\mathbf{V}^2(t,x)}{c^2}\right)dx'^2\right] =$$

$$= c^2dt'^2 - dx'^2 = ds'^2.$$

Thus the non-holonomic transformations (4.2.5) keeping the metric $ds^2 = c^2dt^2 - dx^2$ forminvariant.

**Corollary 4.2.2.** The interval between two infinitely close events of the frame $F_a^{(\text{II})}$

is written as $ds^2 = c^2dt^2 - dx^2$.

**Definition 4.2.3.** The pair $\langle F_a^{(\text{II})}, F_I \rangle$ there is a physical non-holonomic frame of reference (NHFR) called an (non-inertial) non-holonomic Minkowski frame of the type (II) (II-NHMF).

## IV.2.3. Non-holonomic Lorentz transformations of the type III.

**(III)** Let us consider the non-holonomic transformations of the type (**III**) of space and time between an inertial inertial Galilee frame $F_I = F_I(x',t')$ and an accelerated frame $F_a^{(\text{III})}(t,x) = F_{a(t',x)}(t,x)$:

$$dx = \gamma[\mathbf{V}(t',x)][dx' + \mathbf{V}(t',x)dt'],$$

$$dt = \gamma[\mathbf{V}(t',x)]\left[dt' + \frac{\mathbf{V}(t',x)}{c^2}dx'\right]. \quad (4.2.7)$$

**Theorem 4.2.3.** The non-holonomic transformations (4.2.7) keeping the Minkowski metric $ds^2 = c^2dt^2 - dx^2$ forminvariant.

**Proof.** From Eqs. (4.2.7) we obtain

$$ds^2 = c^2dt^2 - dx^2 =$$

$$= c^2\gamma^2[\mathbf{V}(t',x)]\left[dt' + \frac{\mathbf{V}(t',x)}{c^2}dx'\right]^2 - \gamma^2[\mathbf{V}(t',x)][dx' + \mathbf{V}(t',x)dt']^2 =$$

$$= \gamma^2[\mathbf{V}(t',x)]\left[c^2dt'^2 + \underbrace{2\mathbf{V}(t',x)dt'dx'}_{} + \frac{\mathbf{V}^2(t',x)}{c^2}dx'^2 - dx'^2\right.$$

$$\left.\underbrace{-2\mathbf{V}(t',x)dt'dx'}_{} - \mathbf{V}^2(t',x)dt'^2\right] =$$

$$= \gamma^2[\mathbf{V}(t',x)]\left[c^2\left(1 - \frac{\mathbf{V}^2(t',x)}{c^2}\right)dt'^2 - \left(1 - \frac{\mathbf{V}^2(t',x)}{c^2}\right)dx'^2\right] =$$

$$= c^2dt'^2 - dx'^2 = ds'^2.$$

(4.2.8)

Thus the non-holonomic transformetions (4.2.7) keeping the metric $ds^2 = c^2dt^2 - dx^2$ forminvariant.

**Corollary 4.2.3.** The interval between two infinitely close events of the frame $F_a^{(\text{III})}$
is written as $ds^2 = c^2dt^2 - dx^2$.

**Definition 4.2.4.** The pair $\langle F_a^{(\text{III})}, F_I \rangle$ there is a physical non-holonomic frame of reference (NHFR) called an (non-inertial) non-holonomic Minkowski frame of the type (III) (III-NHMF).

## IV.2.4. Non-holonomic Lorentz transformations of the type **IV**.

**(IV)** Let us consider the non-holonomic transformations of the type **(IV)** of space and time between an inertial inertial Galilee frame $F_I = F_I(x',t')$ and an accelerated frame $F_a^{(\text{IV})}(t,x) = F_{a(t,x')}(t,x)$ :

$$dx = \gamma[\mathbf{V}(t,x')][dx' + \mathbf{V}(t,x')dt'],$$

$$dt = \gamma[\mathbf{V}(t,x')]\left[dt' + \frac{\mathbf{V}(t,x')}{c^2}dx'\right].$$

(4.2.9)

**Theorem 4.2.4.** The non-holonomic transformations (4.2.7) keeping the Minkowski metric $ds^2 = c^2dt^2 - dx^2$ forminvariant.

**Proof.** From Eqs. (4.2.9) we obtain:

$$ds^2 = c^2dt^2 - dx^2 =$$

$$= c^2\gamma^2[\mathbf{V}(t,x')]\left[dt' + \frac{\mathbf{V}(t,x')}{c^2}dx'\right]^2 - \gamma^2[\mathbf{V}(t,x')][dx' + \mathbf{V}(t,x')dt']^2 =$$

$$= \gamma^2[\mathbf{V}(t,x')]\left[\underline{c^2dt'^2} + \underbrace{2\mathbf{V}(t,x')dt'dx'} + \frac{\mathbf{V}^2(t,x')}{c^2}dx'^2 - dx'^2\right.$$

$$\left. \underbrace{-2\mathbf{V}(t,x')dt'dx'} - \underline{\mathbf{V}^2(t,x')dt'^2}\right] = \quad (4.2.10)$$

$$= \gamma^2[\mathbf{V}(t,x')]\left[c^2\left(1 - \frac{\mathbf{V}^2(t,x')}{c^2}\right)dt'^2 - \left(1 - \frac{\mathbf{V}^2(t,x')}{c^2}\right)dx'^2\right] =$$

$$= c^2dt'^2 - dx'^2 = ds'^2.$$

Thus the non-holonomic transformetions (4.2.9) keeping the metric $ds^2 = c^2dt^2 - dx^2$ forminvariant.

**Corollary 4.2.4.** The interval between two infinitely close events of the frame $F_a^{(\mathbf{IV})}$
is written as $ds^2 = c^2dt^2 - dx^2$.

**Definition 4.2.5.** The pair $\langle F_a^{(\mathbf{IV})}, F_I \rangle$ there is a physical non-holonomic frame of reference (NHFR) called an (non-inertial) non-holonomic Minkowski frame of the type (IV) (IV-NHMF).

## IV.3. Local Lorentz length and time contractions.

**(I)** From the non-holonomic Lorentz transformations the type **(I)** (4.2.1) by usual way we obtain the local Lorentz length and time contractions:

$$dx = \gamma[\mathbf{V}(t',x')]dx' =$$
$$= \frac{dx'}{\sqrt{1 - \frac{\mathbf{V}^2(t',x')}{c^2}}}. \quad (4.3.1)$$

$$dt = \gamma[\mathbf{V}(t',x')]dt' =$$
$$= \frac{dt'}{\sqrt{1 - \frac{\mathbf{V}^2(t',x')}{c^2}}}. \quad (4.3.2)$$

By integration Eqs.(4.3.1),(4.3.2) we obtain:

$$L(t',x_1,x_2) = x_2(t') - x_1(t') = \int_{x_1'(t')}^{x_2'(t')} \gamma[\mathbf{V}(t',x')]dx' =$$

$$= \int_{x_1'(t')}^{x_2'(t')} \frac{dx'}{\sqrt{1 - \frac{\mathbf{V}^2(t',x')}{c^2}}}.$$

(4.3.3)

$$T(x',t_1,t_2) = t_2(x') - t_1(x') = \int_{t_1'(x')}^{t_2'(x')} \gamma[\mathbf{V}(t',x')]dt' =$$

$$= \int_{t_1'(x')}^{t_2'(x')} \frac{dt'}{\sqrt{1 - \frac{\mathbf{V}^2(t',x')}{c^2}}}.$$

(4.3.4)

**(II)** From the non-holonomic Lorentz transformations the type **(II)** (4.2.3) by usual way we obtain the local Lorentz length and time contractions:

$$dx = \gamma[\mathbf{V}(t,x)]dx', \text{i.e.}$$

$$dx\sqrt{1 - \frac{\mathbf{V}^2(t,x)}{c^2}} = dx'.$$

(4.3.5)

$$dt = \gamma[\mathbf{V}(t,x)]dt', \text{i.e.}$$

$$dt\sqrt{1 - \frac{\mathbf{V}^2(t,x)}{c^2}} = dt'.$$

(4.3.6)

By integration Eqs.(4.3.5),(4.3.6) we obtain:

$$\int_{x_1}^{x_2} \sqrt{1 - \frac{\mathbf{V}^2(t,x)}{c^2}} dx = \int_{x_1'}^{x_2'} dx' = x_2'(t,x_1) - x_1'(t,x_2) = L'(x_1',x_2'),$$

$$L(x_1,x_2,t) = \int_{x_1}^{x_2} \sqrt{1 - \frac{\mathbf{V}^2(t,x)}{c^2}} dx = x_2'(t,x_1) - x_1'(t,x_2) = L'(x_1',x_2').$$

(4.3.7)

$$\int_{t_1}^{t_2}\sqrt{1-\frac{\mathbf{V}^2(t,x)}{c^2}}\,dt = \int_{t'_1}^{t'_2} dt' = t'_2(t_2,x) - t'_1(t_1,x) = T'(t'_1,t'_2,x),$$

(4.3.8)

$$T(t_1,t_2,x) = \int_{t_1}^{t_2}\sqrt{1-\frac{\mathbf{V}^2(t,x)}{c^2}}\,dt == t'_2(t_2,x) - t'_1(t_1,x) = T'(t'_1,t'_2,x).$$

**(III)** From the non-holonomic Lorentz transformations the type **(III)** (4.2.5) by usual way we obtain the local Lorentz length and time contractions:

$$dx = \gamma[\mathbf{V}(t',x)]dx',\text{i.e.}$$
$$\sqrt{1-\frac{\mathbf{V}^2(t',x)}{c^2}}\,dx = dx'.$$

(4.3.9)

$$dt = \gamma[\mathbf{V}(t',x)]dt',\text{i.e.}$$
$$\sqrt{1-\frac{\mathbf{V}^2(t',x)}{c^2}}\,dt = dt'.$$

(4.3.10)

By integration Eqs.(4.3.9),(4.3.10) we obtain:

$$\int_{x_1}^{x_2}\sqrt{1-\frac{\mathbf{V}^2(t',x)}{c^2}}\,dx = \int_{x'_1}^{x'_2} dx'$$

(4.3.11)

$$\int_{t_1}^{t_2}\sqrt{1-\frac{\mathbf{V}^2(t',x)}{c^2}}\,dt = \int_{t'_1}^{t'_2} dt'$$

(4.3.12)

**(IV)** From the non-holonomic Lorentz transformations the type **(IV)** (4.2.7) by usual way we obtain the local Lorentz length and time contractions:

$$dx = \gamma[\mathbf{V}(t,x')]dx',\text{i.e.}$$

(4.3.13)

$$dt = \gamma[\mathbf{V}(t,x')]dt', \text{i.e.} \tag{4.3.14}$$

### IV.5. Hypothesis of Microlocality.

According to the standard theory of relativity, a noninertial observer is at each instant equivalent to an otherwise identical momentarily comoving inertial observer. This (*strong*) *hypothesis of locality* postulates a pointwise equivalence between noninertial and ideal inertial observers. The strong hypothesis of locality originates from Newtonian mechanics of point particles. The state of a classical particle is determined by its position and velocity at a given instant of time. If the force on the particle is turned off at some instant, the particle will follow the osculating straight line. Thus the assumption of strong (holonomic) locality is automatically satisfied in this case, since the noninertial and the ideal osculating inertial observer share the same state and are hence equivalent.

This is why the discussion of accelerated systems in classical mechanics does not require any new hypothesis. In classical electrodynamics, however, we need to deal with classical electromagnetic waves; their interactions can only be considered poinlike in the geometric optics limit. If all physical phenomena could be reduced to pointlike coincidences of classical particles and rays of radiation, then the hypothesis of locality would be exactly valid [38]. Imagine a background global inertial reference frame $K = K(t',x',y',z')$ with coordinates $t',x',y',z'$ and the class of fundamental observers in this frame. Each fundamental observer is by definition at rest in this frame and carries an orthonormal tetrad frame $\tilde{\lambda}^{\mu}_{(\alpha)} = \delta^{\mu}_{\alpha}, \alpha = 0,1,2,3$ such that $\tilde{\lambda}^{\mu}_{(0)}$ is tangent to its worldline and $\tilde{\lambda}^{\mu}_{(i)}, i = 1,2,3$ characterize its spatial frame. Consider now an accelerated observer $O_a$ following a worldline $x^{\alpha}_D(\tau)$ with four-velocity $u^{\alpha}_D(\tau) = dx^{\alpha}_D(\tau)/d\tau$ and translational acceleration $A^{\alpha}_D = du^{\alpha}_D(\tau)/d\tau$. Here $\tau$ is a temporal parameter along $x_D(\tau)$ defined by $d\tau/dt' = \gamma^{-1}(t')$, where $\gamma$ is the Lorentz factor, $\gamma^{-1}(t') = \sqrt{1 - \mathbf{v}^2(t')/c^2}$ and $\mathbf{v}(t')$ is the velocity of the accelerated device. Note that $\langle \mathbf{A}_D, \mathbf{u}_D \rangle = 0$ so that $\mathbf{A}_D$ is a spacelike vector such that $\langle \mathbf{A}_D, \mathbf{A}_D \rangle = w^2$ where $w$ is the magnitude of the translational acceleration.

**Axiom** *(**Strong Hypothesis of Locality**) According to the standard theory of relativity, a noninertial observer $O_w$ in (orthogonal i.e. $t \perp x \perp y \perp z$) accelerated frame $K_w = K_w(t,x,y,z)$ is at each instant equivalent to an otherwise identical momentarily comoving inertial observer $O_I$ in standard inertial frame $K_I = K_I(t_I,x_I,y_I,z_I)$ This hypothesis of locality postulates a pointwise physically equivalence between noninertial and ideal inertial observers.*

However, in general classical waves have intrinsic extensions in space and time characterized by their wavelengths and periods.For instance, to measure the frequency of an incident wave, a few oscillations of the wave must be observed before a reasonable determination of the frequency becomes possible.This situation must be compared with the intrinsic scales of length and time associated with an accelerated observer. That is, an accelerated observer has intrinsic length scales $L = \frac{c^2}{a}$ or $\frac{c}{\Omega}$ corresponding to its translational acceleration a or rotational frequency of its spatial frame and the relevant intrinsic time scales are then $\frac{c}{a}$ and $\frac{c}{\Omega}$. Let λ be the intrinsic length scale of the phenomenon under observation;then, the hypothesis of locality is valid if $\frac{\lambda}{L} \to 0$, i.e. the deviations from locality characterized by $\frac{\lambda}{L}$ are so small as to be below the sensitivity level of measurements by the accelerated observer [42],[43],[44]. It turns out that this is indeed the case for most situations of interest at present, since for Earth-based devices $c^2/g \simeq 1$ lyr and $c/\Omega_\oplus \simeq 28 AU$.

A noninertial observer passes through a continuous infinity of hypothetical momentarily comoving inertial observers along its worldline. This is mathematically analogous to the fact that a curved line is the envelope of the infinite class of straight lines tangent to it.Just as the replacement of a curve by its tangent is only a first approximation, one can show that the hypothesis of locality simply provides an estimate that is exact only in the eikonal limit. Once the limitations of the hypothesis of locality are recognized, it becomes possible to explore suitable *microlocal alternatives.*

## IV.6.Relativistic contraction of a rigid accelerated rod from non-holonomic Lorentz transformations.

Let us consider a rod whose velocity and acceleration are directed along its length, but are otherwise arbitrary functions of time. In particular if the rod is rigid and does not rotate, it is enough to know how one particular point $x_A(t'_A)$ of the rod labeled by $A$ changes its position with time $t'_A$ by the law $x_A(t'_A)$. Let $F_I = F_I(t_I, x_I) = F_I(t', x')$ be a stationary inertial frame and $F_{\mathbf{V}(t',x')}(t,x)$ the accelerated frame of an observer on the rod. We assume that we know the function $x(t'_A, \bar{L}_0(t'_A))$, where $x(t'_A, 0) = x_A(t'_A)$ and $0 \leq \bar{L}_0(t'_A) \leq \mathbf{L}_0(t'_A)$. So we also know the velocity:

$$\mathbf{V}(t'_A, \bar{L}_0) = \frac{dx_A(t'_A, \bar{L}_0)}{dt'_A}. \tag{4.5.1}$$

If, for instance, the rod, that is at rest for $t < 0$ in the unprimed inertial reference system and has infinite small length $d\bar{L}_0$, but at $t = t' = 0$ it starts moving with constant acceleration $\vec{a}$ along its length so that it turns on its engine which gives the constant acceleration $\vec{a}$ to the application point $A$ during a finite time interval $[0,T']$, then its length $dL'$ measured by observer in inertial frame $F_I(t',x')$ will decrease according to the law:

$$dL' = d\bar{L}_0 \sqrt{1 - \frac{\mathbf{V}^2(t'_A, \bar{L}_0)}{c^2}} \tag{4.5.2}$$

or, if Eq.(4.1.1) is taken into account, then one can express the velocity $\mathbf{V}(t'_A, L_0)$ via the acceleration $\vec{a}(\bar{L}_0)$

$$\mathbf{V}(t'_A, \bar{L}_0) = \frac{\vec{a}(\bar{L}_0) t'_A}{\sqrt{1 + \frac{|\vec{a}|^2(\bar{L}_0) t'^2_A}{c^2}}}, 0 \leq t'_A \leq T';$$

$$\mathbf{V}(T', \bar{L}_0) = \frac{\vec{a}(\bar{L}_0) T'}{\sqrt{1 + \frac{|\vec{a}|^2(\bar{L}_0) T'^2}{c^2}}}, t'_A \geq T'. \tag{4.5.3}$$

and obtain the expressions:

$$dL' = \frac{d\bar{L}_0}{\sqrt{1 + \frac{|\vec{a}|^2(\bar{L}_0) t'^2_A}{c^2}}}, 0 \leq t'_A \leq T';$$

$$dL' = \frac{d\bar{L}_0}{\sqrt{1 + \frac{|\vec{a}|^2(\bar{L}_0) T'^2}{c^2}}}, t'_A \geq T'. \tag{4.5.4}$$

By setting (see.[40] Eq.(27)), for the case when the force is applied to the back end of the rod (pushed rod)

$$\vec{a}(\bar{L}_0) = \frac{\vec{a}}{1 + \frac{|\vec{a}|\bar{L}_0}{c^2}}, 0 \leq \bar{L}_0 \leq \mathbf{L}_0, a > 0 \tag{4.5.5}$$

we obtain

$$dL = \frac{d\bar{L}_0}{\sqrt{1 + \frac{|\vec{a}|^2}{\left(1 + \frac{|\vec{a}|\bar{L}_0}{c^2}\right)^2} \frac{t'^2_A}{c^2}}},$$

$$0 \leq \bar{L}_0 \leq \mathbf{L}_0, 0 \leq t'_A \leq T';$$

$$dL = \frac{d\bar{L}_0}{\sqrt{1 + \frac{|\vec{a}|^2}{\left(1 + \frac{|\vec{a}|\bar{L}_0}{c^2}\right)^2} \frac{T'^2}{c^2}}},$$

$$0 \leq \bar{L}_0 \leq \mathbf{L}_0, t'_A \geq T'. \tag{4.5.6}$$

**Remark**  *Note that for a pulled rod Eq. (4.5.5) takes the form*

$$\vec{a}(\bar{L}_0) = \frac{-\vec{a}}{1 - \frac{|\vec{a}|\bar{L}_0}{c^2}}, \tag{4.5.7}$$

*And we see that the Eq.(4.5.7) for a pulled rod are meaningless if*

$$\frac{|\vec{a}|\mathbf{L}_0}{c^2} = 1, \tag{4.5.8}$$

*because in this case we obtain non-physical result:* $|\vec{a}|(\mathbf{L}_0) = \infty.$
By integration Eq.(4.5.6) we obtain:

$$\mathbf{L}(t'_A) = \int_0^{\mathbf{L}_0} \frac{d\bar{L}_0}{\sqrt{1 + \frac{|\vec{a}|^2}{\left(1 + \frac{|\vec{a}|\bar{L}_0}{c^2}\right)^2} \frac{t'^2_A}{c^2}}} =$$

$$= \int_0^{\mathbf{L}_0} \frac{d\bar{L}_0}{\sqrt{1 + \frac{\alpha}{(1 + \beta\bar{L}_0)^2}}}. \tag{4.5.9}$$

$$\alpha = \frac{|\vec{a}|^2 t'^2_A}{c^2}, \beta = \frac{|\vec{a}|}{c^2}.$$

By substitution

$$y = 1 + \beta\bar{L}_0,$$
$$dL_0 = \beta^{-1}dy, \tag{4.5.10}$$

we obtain:

$$\int \frac{d\bar{L}_0}{\sqrt{1 + \frac{\alpha}{(1 + \beta\bar{L}_0)^2}}} =$$

$$\beta^{-1} \int \frac{dy}{\sqrt{1 + \frac{\alpha}{y^2}}} = \beta^{-1} \int \frac{ydy}{\sqrt{y^2 + \alpha}}. \tag{4.5.11}$$

By substitution

$$y^2 = z,$$
$$dy^2 = dz, \tag{4.5.12}$$

we obtain:

$$\int \frac{ydy}{\sqrt{y^2 + \alpha}} = \frac{1}{2} \int \frac{dy^2}{\sqrt{y^2 + \alpha}} =$$

$$\frac{1}{2} \int \frac{dz}{\sqrt{z + \alpha}} = \sqrt{z + \alpha}. \tag{4.5.13}$$

Thus

$$\mathbf{L}(t'_A) = \int_0^{\mathbf{L}_0} \frac{d\bar{L}_0}{\sqrt{1 + \frac{\alpha}{(1+\beta\bar{L}_0)^2}}} = \left[\beta^{-1}\sqrt{(1+\beta\bar{L}_0)^2 + \alpha}\,\right]_0^{\mathbf{L}_0} =$$

$$\left[\frac{c^2}{a}\sqrt{\left(1 + \frac{a\bar{L}_0}{c^2}\right)^2 + \frac{a^2 t'^2_A}{c^2}}\,\right]_0^{\mathbf{L}_0} = \quad (4.5.14)$$

$$\frac{c^2}{a}\sqrt{\left(1 + \frac{a\mathbf{L}_0}{c^2}\right)^2 + \frac{a^2 t'^2_A}{c^2}} - \frac{c^2}{a}\sqrt{1 + \frac{a^2 t'^2_A}{c^2}}\,, 0 \le t'_A \le T'.$$

$$\mathbf{L}(t'_A) = \frac{c^2}{a}\sqrt{\left(1 + \frac{a\mathbf{L}_0}{c^2}\right)^2 + \frac{a^2 t'^2_A}{c^2}} - \frac{c^2}{a}\sqrt{1 + \frac{a^2 t'^2_A}{c^2}}\,, 0 \le t'_A \le T';$$

$$L(t'_A) = L_{\mathbf{f}}(t'_A) = \frac{\mathbf{L}_0}{\sqrt{1 + \frac{a^2 t'^2_A}{c^2}}}, t'_A \gg T'. \quad (4.5.15)$$

This result for the first time was obtained by Nikolić H. cf.[40] Eq.(21). Note that (4.5.15) differs from the result which one could expect from the naive generalization of the Lorentz-Fitzgerald formula:

$$\mathbf{L}(t') = \frac{\mathbf{L}_0}{\sqrt{1 + \frac{|\vec{a}|^2 t'^2}{c^2}}} \quad (4.5.16)$$

Formula (4.5.15) was obtained for the case when the force is applied to the back end of the rod. In other words, this is the result for a pushed rod. The analysis for a pulled rod is similar and the result is:

$$\mathbf{L}(t'_A) = \frac{c^2}{a}\sqrt{1 + \frac{a^2 t'^2_A}{c^2}} - \frac{c^2}{a}\sqrt{\left(1 - \frac{a\mathbf{L}_0}{c^2}\right)^2 + \frac{a^2 t'^2_A}{c^2}}\,, 0 \le t'_A \le T';$$

$$L(t'_A) = \mathbf{L}_{\mathbf{f}}(t'_A) = \frac{\mathbf{L}_0}{\sqrt{1 + \frac{a^2 t'^2_A}{c^2}}}, t'_A \gg T'. \quad (4.5.17)$$

We see that the results for a pulled rod are meaningless if

$$\frac{|a|\mathbf{L}_0}{c^2} > 1. \quad (4.5.18)$$

To understand why is that so, we calculate the velocity of the back end for a pulled rod. The result is

$$\mathbf{V}(t'_A, \bar{L}_0) = \frac{a(\bar{L}_0)t'_A}{\sqrt{1 + \frac{a^2(\bar{L}_0)t'^2_A}{c^2}}}, 0 \leq t'_A \leq T';$$

$$a(\bar{L}_0) = \frac{a}{1 - \frac{|a|\bar{L}_0}{c^2}},$$

$$\mathbf{V}(t'_A, \bar{L}_0) = \frac{at'_A}{\left(1 - \frac{a\bar{L}_0}{c^2}\right)\sqrt{1 + \frac{a^2 t'^2_A}{c^2\left(1 - \frac{|a|\bar{L}_0}{c^2}\right)^2}}} = \quad (4.5.19)$$

$$= \frac{at'_A}{\sqrt{\left(1 - \frac{|a|\bar{L}_0}{c^2}\right)^2 + \frac{a^2 t'^2_A}{c^2}}}, 0 \leq t'_A \leq T'.$$

We see that this velocity increases as $|a|L_0$ increases and reaches the velocity of light when

$$|a|\mathbf{L}_0 = c^2. \qquad (4.1.28)$$

We see that Eq.(4.1.28) coincides with Eq.(4.1.16).

**Remark** *This suggests that the rod cannot remain rigid under conditions expressed by means of the equation Eq.(4.1.15)*

An unaccelerated observer cannot actually know whether a rigid rod is pulled by an acceleration a, or is pushed by an acceleration â, given by

$$\hat{a} = \frac{a}{1 - \frac{|a|\bar{L}_0}{c^2}}. \qquad (4.1.29)$$

In particular, if the rod is pulled by acceleration $a = c^2/\mathbf{L}_0$ for an unaccelerated observer it looks the same as it is pushed by acceleration $\hat{a} = \infty$. Formula (4.1.29) can be generalized to an arbitrary point on the rod. If acceleration a is applied to the point $x'_A$, then this is the same as acceleration $a(x')$ is applied to the point $x'$, where

$$a(x') = \frac{a}{1 - \frac{(x' - x'_A)a}{c^2}}. \qquad (4.1.30)$$

This Eq. (4.1.30) is obtained in [40],[41], (see [40] Eq.(27)). The standard consequence of this is that an observer in a uniformly accelerated rocket does not feel a homogeneous inertial force, but rather an inertial force which decreases with $x'$, as given by (4.1.30).                      In real case no rigidity is assumed for the extended rods since it would contradict the relativity theory.The amount of the deformation of the rod along $x$ will depend on the nature of the rod and, specifically, on the stiffness of the material which it is made of.

# IV.7. Master equation for the case of an rigid accelerated rod.

**Theorem 4.7.1.(Master equation for the case of an rigid accelerated rod.)**
Suppose that equality (4.7.1) is satisfied:

$$\frac{\partial^2 L(t',\bar{L}_0)}{\partial \bar{L}_0 \partial t'} = \frac{\partial^2 L(t',\bar{L}_0)}{\partial t' \partial \bar{L}_0},$$
$$0 \leq \bar{L}_0 \leq \mathbf{L}_0,$$
$$0 \leq t' \leq T'.$$
(4.7.1)

Then for for the case of an rigid accelerated rod *master equation* (4.7.2) is satisfied.

$$\frac{\partial \vec{\mathbf{v}}(t',\bar{L}_0)}{\partial \bar{L}_0} = \frac{\partial^2 L(t',\bar{L}_0)}{\partial t' \partial \bar{L}_0}, 0 \leq \bar{L}_0 \leq \mathbf{L}_0, 0 \leq t' \leq T'. \qquad (4.7.2)$$

**Proof.** By definition function $L(t',\bar{L}_0)$ we obtain:

$$\vec{\mathbf{x}}(t',\bar{L}_0) = \vec{\mathbf{x}}(t',0) + L(t',\bar{L}_0), 0 \leq \bar{L}_0 \leq \mathbf{L}_0, 0 \leq t' \leq T'. \qquad (4.7.3)$$

By differentiation the equality (4.7.3) on the variable $t'$ we obtain:

$$\vec{\mathbf{v}}(t',\bar{L}_0) = \frac{\partial \vec{\mathbf{x}}(t',\bar{L}_0)}{\partial t'} = \frac{\partial \vec{\mathbf{x}}(t',0)}{\partial t'} + \frac{L(t',\bar{L}_0)}{\partial t'}. \qquad (4.7.4)$$

By differentiation the equality (4.7.4) on the variable $\bar{L}_0$ we obtain:

$$\frac{\partial \vec{\mathbf{v}}(t',\bar{L}_0)}{\partial \bar{L}_0} = \frac{\partial \vec{\mathbf{x}}(t',\bar{L}_0)}{\partial \bar{L}_0 \partial t'} = \frac{\partial \vec{\mathbf{x}}(t',0)}{\partial \bar{L}_0 \partial t'} + \frac{L(t',\bar{L}_0)}{\partial \bar{L}_0 \partial t'} = \frac{L(t',\bar{L}_0)}{\partial \bar{L}_0 \partial t'}. \qquad (4.7.5)$$

From the Eqs.(4.7.1),(4.7.5) we obtain:

$$\frac{\partial \vec{\mathbf{v}}(t',\bar{L}_0)}{\partial \bar{L}_0} = \frac{\partial^2 L(t',\bar{L}_0)}{\partial t' \partial \bar{L}_0}, 0 \leq \bar{L}_0 \leq \mathbf{L}_0, 0 \leq t' \leq T'.$$ It easy to check then for the case of constantly accelerated rod Eq.(4.7.2) is satisfied. From Eqs.(4.1.2),(4.1.7) we obtain:

$$\frac{\partial \vec{\mathbf{x}}(t',\bar{L}_0)}{\partial t'} = \vec{\mathbf{v}}(t',\bar{L}_0) = \frac{|\vec{a}|t'}{\sqrt{\left(1 + \frac{|\vec{a}|\bar{L}_0}{c^2}\right)^2 + \frac{|\vec{a}|^2 t'^2}{c^2}}}. \qquad (4.7.6)$$

From Eqs.(4.7.6) we obtain:

$$\frac{\partial \vec{\mathbf{v}}(t', \bar{L}_0)}{\partial \bar{L}_0} = -\frac{1}{2} \frac{2\left(1 + \frac{|\vec{a}|\bar{L}_0}{c^2}\right)\frac{|\vec{a}|^2}{c^2}t'}{\sqrt[3/2]{\left(1 + \frac{|\vec{a}|\bar{L}_0}{c^2}\right)^2 + \frac{|\vec{a}|^2 t'^2}{c^2}}} =$$

$$= -\frac{\left(1 + \frac{|\vec{a}|\bar{L}_0}{c^2}\right)\frac{|\vec{a}|^2}{c^2}t'}{\sqrt[3/2]{\left(1 + \frac{|\vec{a}|\bar{L}_0}{c^2}\right)^2 + \frac{|\vec{a}|^2 t'^2}{c^2}}}.$$

(4.7.7)

From Eq.(4.5.6) we obtain:

$$\frac{\partial L(t', \bar{L}_0)}{\partial \bar{L}_0} = \frac{1}{\sqrt{1 + \frac{|\vec{a}|^2}{\left(1 + \frac{|\vec{a}|\bar{L}_0}{c^2}\right)^2}\frac{t'^2}{c^2}}} = \frac{1 + \frac{|\vec{a}|\bar{L}_0}{c^2}}{\sqrt{\left(1 + \frac{|\vec{a}|\bar{L}_0}{c^2}\right)^2 + \frac{|\vec{a}|^2 t'^2}{c^2}}}. \quad (4.7.8)$$

By differentiation the equality (4.7.8) on the variable $t'$ we obtain:

$$\frac{\partial^2 L(t', \bar{L}_0)}{\partial t' \partial \bar{L}_0} = -\frac{1}{2} \frac{\left(1 + \frac{|\vec{a}|\bar{L}_0}{c^2}\right)\frac{2|\vec{a}|^2 t'}{c^2}}{\sqrt[3/2]{\left(1 + \frac{|\vec{a}|\bar{L}_0}{c^2}\right)^2 + \frac{|\vec{a}|^2 t'^2}{c^2}}} = -\frac{\left(1 + \frac{|\vec{a}|\bar{L}_0}{c^2}\right)\frac{|\vec{a}|^2}{c^2}t'}{\sqrt[3/2]{\left(1 + \frac{|\vec{a}|\bar{L}_0}{c^2}\right)^2 + \frac{|\vec{a}|^2 t'^2}{c^2}}}. \quad (4.7.9)$$

Thus master equation (4.7.2) is satisfied.

**Theorem 4.7.2.(Master equation for the case of an rigid accelerated rod.)**
Suppose that equality (4.7.1) is satisfied. Then for for the case of an rigid accelerated rod *master equation* (4.7.10) is satisfied:

$$\frac{\partial \vec{\mathbf{v}}(t', \bar{L}_0)}{\partial \bar{L}_0} = \frac{1}{c^2} \frac{\partial \vec{\mathbf{v}}(t', \bar{L}_0)}{\partial t'} \frac{\vec{\mathbf{v}}(t', \bar{L}_0)}{\sqrt{1 - \frac{|\vec{\mathbf{v}}|^2(t', \bar{L}_0)}{c^2}}},$$

$$\vec{\mathbf{v}}(0, \bar{L}_0) = 0, \vec{\mathbf{v}}(t', 0) = \vec{\mathbf{v}}(t'),$$

$$0 \le \bar{L}_0 \le \mathbf{L}_0, 0 \le t' \le T'.$$

(4.7.10)

**Proof.** By substitution Eq.(4.5.2) into Eq.(4.7.2) we obtain:

$$\frac{\partial \vec{\mathbf{v}}(t',\bar{L}_0)}{\partial \bar{L}_0} = \frac{\partial}{\partial t'}\sqrt{1 - \frac{|\vec{\mathbf{v}}|^2(t',\bar{L}_0)}{c^2}} =$$
$$= \frac{1}{c^2}\frac{\partial \vec{\mathbf{v}}(t',\bar{L}_0)}{\partial t'}\frac{\vec{\mathbf{v}}(t',\bar{L}_0)}{\sqrt{1 - \frac{|\vec{\mathbf{v}}|^2(t',\bar{L}_0)}{c^2}}}. \qquad (4.7.11)$$

# IV.8. Adding velocities in a non-holonomic case.

From the non-holonomic Lorentz transformations (4.1.4) we obtain

$$\frac{dx}{dt} = \frac{dx' + \mathbf{V}(t',x')dt'}{dt' + \frac{\mathbf{V}(t',x')}{c^2}dx'},$$

$$\frac{dy}{dt} = \frac{dy'\sqrt{1 - \frac{\mathbf{V}^2(t',x')}{c^2}}}{dt' + \frac{\mathbf{V}(t',x')}{c^2}dx'}, \qquad (4.8.1)$$

$$\frac{dz}{dt} = \frac{dz'\sqrt{1 - \frac{\mathbf{V}^2(t',x')}{c^2}}}{dt' + \frac{\mathbf{V}(t',x')}{c^2}dx'},$$

From (4.8.1) by usual way we obtain

$$\frac{dx}{dt} = \frac{\frac{dx'}{dt'} + \mathbf{V}(t',x')}{1 + \frac{\mathbf{V}(t',x')}{c^2}\frac{dx'}{dt'}},$$

$$\frac{dy}{dt} = \frac{\frac{dy'}{dt'}\sqrt{1 - \frac{\mathbf{V}^2(t',x')}{c^2}}}{1 + \frac{\mathbf{V}(t',x')}{c^2}\frac{dx'}{dt'}}, \qquad (4.8.2)$$

$$\frac{dz}{dt} = \frac{\frac{dz'}{dt'}\sqrt{1 - \frac{\mathbf{V}^2(t',x')}{c^2}}}{1 + \frac{\mathbf{V}(t',x')}{c^2}\frac{dx'}{dt'}}.$$

By setting

$$\mathbf{v}_x = \frac{dx}{dt}, \mathbf{v}'_x = \frac{dx'}{dt'},$$
$$\mathbf{v}_y = \frac{dy}{dt}, \mathbf{v}'_y = \frac{dy'}{dt'}, \quad (4.8.3)$$
$$\mathbf{v}_z = \frac{dz}{dt}, \mathbf{v}'_z = \frac{dz'}{dt'},$$

from (4.8.2) we obtain the formulae relating velocities:

$$\mathbf{v}_x = \frac{\mathbf{v}'_x + \mathbf{V}(t',x')}{1 + \frac{\mathbf{V}(t',x')\mathbf{v}'_x}{c^2}},$$

$$\mathbf{v}_y = \frac{\mathbf{v}'_y \sqrt{1 - \frac{\mathbf{V}^2(t',x')}{c^2}}}{1 + \frac{\mathbf{V}(t',x')\mathbf{v}'_y}{c^2}}, \quad (4.8.4)$$

$$\mathbf{v}_z = \frac{\mathbf{v}'_z \sqrt{1 - \frac{\mathbf{V}^2(t',x')}{c^2}}}{1 + \frac{\mathbf{V}(t',x')\mathbf{v}'_z}{c^2}}.$$

IV.9. General global transformations between an inertial Minkowski frame $F_I(t_I,x_I,y_I,z_I)$ and an uniformly accelerated frame $F_a(t,x,y,z)$.

Let us calculate general global transformations $\hat{L}(a,\mathbf{V})$ of space and time between an inertial Minkowski frame $F_I(t_I,x_I,y_I,z_I) = F_I(t',x',y',z')$ and a uniformly accelerated frame $F_a(t,x,y,z)$ with a metric (4.1.3) forming by the nonlinear transformations $\Lambda(a)$ (4.1.1). From [3] (see [3] Theorem in subsection I) we obtain:

$$\hat{L}(a,\mathbf{V}) = \Lambda(a) \circ \check{L}(\mathbf{V}).$$
$$\Lambda(a) : (t',x') \to (t,x)$$
$$\Lambda(a) : x = x' - \frac{c^2}{a}\left[\sqrt{1 + \frac{a^2 t'^2}{c^2}} - 1\right], t = t'. \quad (4.9.1.a)$$

$$\check{L}(\mathbf{V}) : \begin{cases} x' \to \gamma(\mathbf{V})(x' - \mathbf{V}t'), & (4.9.1.b) \\ t' \to \gamma(\mathbf{V})\left(t' - \frac{\mathbf{V}}{c^2}x'\right). & (4.9.1.c) \end{cases} \quad (4.9.1)$$

By substitution (4.9.1.b)-(4.9.1.c) into (4.9.1.a) we obtain the generalized Lorentz

transformations $\hat{L}(a, \mathbf{V})$ :

$$x = \gamma(\mathbf{V})(x' - \mathbf{V}t') - \frac{c^2}{a}\left[\sqrt{1 + \frac{a^2}{c^2}\gamma^2(\mathbf{V})\left(t' - \frac{\mathbf{V}}{c^2}x'\right)^2} - 1\right],$$

$$t = \gamma(\mathbf{V})\left(t' - \frac{\mathbf{V}}{c^2}x'\right). \qquad (4.9.2)$$

## V. General non-holonomic transformations for the case of the uniformly accelerated frames of reference.

### V.1. General non-holonomic transformations between an inertial Minkowski frame $F_I(t_I, x_I, y_I, z_I)$ and an uniformly accelerated frame $F_a(t, x, y, z)$.

Let us calculate general (local) non-holonomic transformations $d\hat{L}[a, \mathbf{V}(t', x')]$ of space and time between an inertial Minkowski frame $F_I(t_I, x_I, y_I, z_I) = F_I(t', x', y', z')$ and an uniformly accelerated frame $F_a(t, x, y, z)$ with a metric (4.1.3) forming by the nonlinear transformations $\Lambda(a)$ (4.1.1),(4.1.2). From [3] (see [3] Theorem in subsection I) we obtain:

$$d\hat{L}[a, \mathbf{V}(t',x')] = d\Lambda(a) \circ d\check{L}[\mathbf{V}(t',x')].$$

$$d\Lambda(a) : (dt', dx') \to (dt, dx),$$

$$d\Lambda(a) : dx = dx' - \frac{at'}{\sqrt{1 + \frac{a^2 t'^2}{c^2}}} dt', \quad dt' = dt.$$

$$dx = dx' - \mathbf{v}(t')dt', \qquad (5.1.1.a)$$

$$\mathbf{v}(t') = \frac{at'}{\sqrt{1 + \frac{a^2 t'^2}{c^2}}}. \qquad (5.1.1.a')$$

(5.1.1)

$$d\check{L}[\mathbf{V}(t',x')] : \begin{cases} dx' \to \gamma[\mathbf{V}(t',x')](dx' - \mathbf{V}(t',x')dt'), & (5.1.1.b) \\ dt' \to \gamma[\mathbf{V}(t',x')]\left(dt' - \frac{\mathbf{V}(t',x')}{c^2}dx'\right). & (5.1.1.c) \end{cases}$$

By substitution (5.1.1.b)-(5.1.1.c) into (5.1.1.a) we obtain the general non-holonomic transformations $d\hat{L}[a, \mathbf{V}(t',x')]$ :

$$dx = dx' - \mathbf{v}(t')dt' =$$

$$\gamma[\mathbf{V}(t',x')](dx' - \mathbf{V}(t',x')dt') - \mathbf{v}(t')\gamma[\mathbf{V}(t',x')]\left(dt' - \frac{\mathbf{V}(t',x')}{c^2}dx'\right) =$$

$$\gamma[\mathbf{V}(t',x')]\left[dx' - \mathbf{V}(t',x')dt' - \mathbf{v}(t')dt' + \frac{\mathbf{v}(t')\mathbf{V}(t',x')}{c^2}dx'\right] =$$

$$= \gamma[\mathbf{V}(t',x')]\left[\left(1 + \frac{\mathbf{v}(t')\mathbf{V}(t',x')}{c^2}\right)dx' - (\mathbf{v}(t') + \mathbf{V}(t',x'))dt'\right].$$

(5.1.2)

$$dt' = \gamma[\mathbf{V}(t',x')]\left(dt' - \frac{\mathbf{V}(t',x')}{c^2}dx'\right).$$

Thus

$$dx = \gamma[\mathbf{V}(t',x')]\left[\left(1 + \frac{\mathbf{v}(t')\mathbf{V}(t',x')}{c^2}\right)dx' - (\mathbf{v}(t') + \mathbf{V}(t',x'))dt'\right],$$

$$dt = \gamma[\mathbf{V}(t',x')]\left(dt' - \frac{\mathbf{V}(t',x')}{c^2}dx'\right).$$

(5.1.3)

By substitution (5.1.1.a′) into (5.1.3) we obtain the non-holonomic transformations:

$$dx = \gamma[\mathbf{V}(t',x')] \times$$

$$\times \left[\left(1 + \frac{at'}{\sqrt{1 + \frac{a^2 t'^2}{c^2}}} \frac{\mathbf{V}(t',x')}{c^2}\right)dx' - \left(\frac{at'}{\sqrt{1 + \frac{a^2 t'^2}{c^2}}} + \mathbf{V}(t',x')\right)dt'\right] =$$

$$= \gamma[\mathbf{V}(t',x')]\left[\left(1 + \frac{at'}{\sqrt{c^2 + a^2 t'^2}} \frac{\mathbf{V}(t',x')}{c}\right)dx' - \left(\frac{act'}{\sqrt{c^2 + a^2 t'^2}} + \mathbf{V}(t',x')\right)dt'\right],$$

(5.1.4)

$$dt' = \gamma[\mathbf{V}(t',x')]\left(dt' - \frac{\mathbf{V}(t',x')}{c^2}dx'\right).$$

For the case

$$\mathbf{V}(t',x') = \frac{at'}{\sqrt{1 + \frac{a^2 t'^2}{c^2}}}, 0 \leq t' < \infty.$$

(5.1.5)

from (5.1.4) we obtain the non-holonomic transformations:

$$dx = \sqrt{1+\left(\frac{at'}{c}\right)^2}\left[\left(1+\frac{1}{c^2}\frac{a^2t'^2}{1+\frac{a^2t'^2}{c^2}}\right)dx' - \frac{2at'}{\sqrt{1+\frac{a^2t'^2}{c^2}}}dt'\right] =$$

$$= \sqrt{1+\left(\frac{at'}{c}\right)^2}\left[\left(1+\frac{a^2t'^2}{c^2+a^2t'^2}\right)dx' - \frac{2act'}{\sqrt{c^2+a^2t'^2}}dt'\right]$$

$$dt' = \sqrt{1+\left(\frac{at'}{c}\right)^2}\left(dt' - \frac{1}{c^2}\frac{at'}{\sqrt{1+\frac{a^2t'^2}{c^2}}}dx'\right) =$$

$$= \sqrt{1+\left(\frac{at'}{c}\right)^2}\left(dt' - \frac{1}{c}\frac{at'}{\sqrt{c^2+a^2t'^2}}dx'\right).$$

(5.1.6)

For the case $\mathbf{V}(t',x') = \mathbf{V} = $ const from (5.1.4) we obtain the non-holonomic transformations:

$$dx = \gamma(\mathbf{V})\left[\left(1+\frac{at'}{\sqrt{1+\frac{a^2t'^2}{c^2}}}\frac{\mathbf{V}}{c^2}\right)dx' - \left(\frac{at'}{\sqrt{1+\frac{a^2t'^2}{c^2}}}+\mathbf{V}\right)dt'\right] =$$

$$= \gamma(\mathbf{V})\left[\left(1+\frac{at'}{\sqrt{c^2+a^2t'^2}}\frac{\mathbf{V}}{c}\right)dx' - \left(\frac{act'}{\sqrt{c^2+a^2t'^2}}+\mathbf{V}\right)dt'\right],$$

$$dt' = \gamma(\mathbf{V})\left(dt' - \frac{\mathbf{V}}{c^2}dx'\right).$$

(5.1.7)

For the case

$$\mathbf{v}(t') = \frac{aT'}{\sqrt{1+\frac{a^2T'^2}{c^2}}} = \mathbf{v}, t' \geq T',$$

$$\mathbf{V}(t',x') = \mathbf{V} = \text{const},$$

(5.1.8)

from (5.1.7) we obtain the transformations:

$$dx = \gamma(\mathbf{V})\left[\left(1 + \frac{\mathbf{v}\mathbf{V}}{c^2}\right)dx' - (\mathbf{v} + \mathbf{V})dt'\right],$$

$$dt = \gamma(\mathbf{V})\left(dt' - \frac{\mathbf{V}}{c^2}dx'\right). \tag{5.1.9}$$

Transformations (5.1.9) coincide with generalized Lorentz transformations (A.2.3).

## V.2. General non-holonomic transformations between uniformly accelerated frame frame $F_a(t,x,y,z)$ and an inertial Minkowski frame $F_I(t_I,x_I,y_I,z_I)$.

Let us calculate general (local) non-holonomic transformations $d\hat{L}^{-1}[a,\mathbf{V}(t',x')]$ of space and time between uniformly accelerated frame $F_a(t,x,y,z)$ with a metric (4.1.3) forming by the nonlinear transformations $\Lambda(a)$ (4.1.1),(4.1.2) and an inertial Minkowski frame $F_I(t_I,x_I,y_I,z_I) = F_I(t',x',y',z')$.

From [3] (see [3] Theorem I in subsection I) we obtain:

$$d\hat{L}^{-1}(\mathbf{v},\mathbf{V}) = d\check{L}(\mathbf{V}) \circ d\Lambda^{-1}(\mathbf{v}).$$
$$d\Lambda^{-1}(\mathbf{v}) : (dt,dx) \to (dt',dx')$$
$$dx' = dx + \mathbf{v}(t)dt, \quad dt = dt'. \quad (5.2.1.a)$$
$$\mathbf{v}(t) = \frac{at'}{\sqrt{1+\frac{a^2 t'^2}{c^2}}}. \quad (5.2.1.a')$$
$$d\Lambda^{-1}(a) : dx = dx' + \frac{at'}{\sqrt{1+\frac{a^2 t'^2}{c^2}}} dt', \quad dt' = dt. \tag{5.2.1}$$
$$d\check{L}(\mathbf{V}) : (dt',dx') \to L(\mathbf{V})(dt',dx').$$

$$d\check{L}[\mathbf{V}(t',x')] : \left\{ \begin{array}{l} dx' \to \gamma[\mathbf{V}(t',x')](dx' + \mathbf{V}(t',x')dt') \quad (5.2.1.b) \\ dt' \to \gamma[\mathbf{V}((t',x'))]\left(dt' + \dfrac{\mathbf{V}(t',x')}{c^2}dx'\right) \quad (5.2.1.c) \end{array} \right\}$$

By substitution (5.2.1.a) into (5.2.1.b)-(5.2.1.c) we obtain the general non-holonomic transformations $d\hat{L}^{-1}[a,\mathbf{V}(t',x')]$ :

$$dx' \to \gamma[\mathbf{V}(t',x')](dx' + \mathbf{V}(t',x')dt') =$$
$$= \gamma[\mathbf{V}(t,x')][(dx + \mathbf{v}(t)dt) + \mathbf{V}(t,x')dt'] =$$
$$= \gamma[\mathbf{V}(t,x')][dx + (\mathbf{v}(t) + \mathbf{V}(t,x'))dt]$$

$$dt' \to \gamma[\mathbf{V}(t',x')]\left(dt' + \frac{\mathbf{V}(t',x')}{c^2}dx'\right) = \quad (5.2.2)$$
$$= \gamma[\mathbf{V}(t,x')]\left(dt + \frac{\mathbf{V}(t,x')}{c^2}(dx + \mathbf{v}(t)dt)\right) =$$
$$\gamma[\mathbf{V}(t,x')]\left(\left(1 + \frac{\mathbf{v}(t)\mathbf{V}(t,x')}{c^2}\right)dt + \frac{\mathbf{V}(t',x')}{c^2}dx\right)$$

Thus

$$dx' = \gamma[\mathbf{V}(t,x')][dx + (\mathbf{v}(t) + \mathbf{V}(t,x'))dt],$$
$$dt' = \gamma[\mathbf{V}(t,x')]\left(\left(1 + \frac{\mathbf{v}(t)\mathbf{V}(t,x')}{c^2}\right)dt + \frac{\mathbf{V}(t,x')}{c^2}dx\right). \quad (5.2.3)$$

By substitution (5.2.1.a′) into (5.2.3) we obtain the non-holonomic transformations:

$$dx' = \gamma[\mathbf{V}(t,x')]\left[dx + \left(\frac{at'}{\sqrt{1 + \frac{a^2 t'^2}{c^2}}} + \mathbf{V}(t,x')\right)dt\right],$$
$$dt' = \gamma[\mathbf{V}(t,x')]\left(\left(1 + \frac{at'}{\sqrt{1 + \frac{a^2 t'^2}{c^2}}}\frac{\mathbf{V}(t',x')}{c^2}\right)dt + \frac{\mathbf{V}(t',x')}{c^2}dx\right). \quad (5.2.4)$$

For the case

$$\mathbf{V}(t,x') = \frac{at}{\sqrt{1 + \frac{a^2 t^2}{c^2}}}, 0 \leq t' < \infty. \quad (5.2.5)$$

from (5.2.4) we obtain the non-holonomic transformations:

$$dx' = \sqrt{1 + \left(\frac{at}{c}\right)^2}\left(dx + \frac{2atdt}{\sqrt{1 + \frac{a^2 t^2}{c^2}}}\right),$$
$$dt' = \sqrt{1 + \left(\frac{at}{c}\right)^2}\left[\left(1 + \frac{a^2 t^2}{c^2 + a^2 t^2}\right)dt + \frac{at}{c^2\sqrt{1 + \frac{a^2 t^2}{c^2}}}dx\right]. \quad (5.2.6)$$

For the case $\mathbf{V}(t',x') = \mathbf{V} =$ const from (5.2.4) we obtain the non-holonomic transformations:

$$dx' = \gamma(\mathbf{V})\left[dx + \left(\frac{at'}{\sqrt{1 + \frac{a^2 t'^2}{c^2}}} + \mathbf{V}\right)dt\right],$$

$$dt' = \gamma(\mathbf{V})\left(\left(1 + \frac{at'}{\sqrt{1 + \frac{a^2 t'^2}{c^2}}}\frac{\mathbf{V}}{c^2}\right)dt + \frac{\mathbf{V}}{c^2}dx\right).$$

(5.2.7)

For the case

$$\mathbf{v}(t) = \frac{aT}{\sqrt{1 + \frac{a^2 T^2}{c^2}}} = \mathbf{v}, t \geq T,$$

$$\mathbf{V}(t,x') = \mathbf{V} = \text{const},$$

(5.2.8)

from (5.7.7) we obtain the transformations:

$$dx' = \gamma(\mathbf{V})[dx + (\mathbf{v} + \mathbf{V})dt],$$

$$dt' = \gamma(\mathbf{V})\left[\left(1 + \frac{\mathbf{vV}}{c^2}\right)dt + \frac{\mathbf{V}}{c^2}dx\right].$$

(5.2.9)

Transformations (5.2.9) coincide with infinitesimal generalized Lorentz transformations (A.3.4).

# V.3. Adding velocities in a general non-holonomic case. Adding velocities in the case of the uniformly accelerated frame.

From the non-holonomic transformations (5.1.3) we obtain:

$$\frac{dx}{dt} = \frac{\left(1 + \frac{\mathbf{v}(t')\mathbf{V}(t',x')}{c^2}\right)dx' - (\mathbf{v}(t') + \mathbf{V}(t',x'))dt'}{dt' - \frac{\mathbf{V}(t',x')}{c^2}dx'},$$

$$\frac{dy}{dt} = \frac{dy'\sqrt{1 - \frac{\mathbf{V}^2(t',x')}{c^2}}}{dt' - \frac{\mathbf{V}(t',x')}{c^2}dx'}, \qquad (5.3.1)$$

$$\frac{dz}{dt} = \frac{dz'\sqrt{1 - \frac{\mathbf{V}^2(t',x')}{c^2}}}{dt' - \frac{\mathbf{V}(t',x')}{c^2}dx'},$$

From (5.3.1) by usual way we obtain

$$\frac{dx}{dt} = \frac{\left(1 + \frac{\mathbf{v}(t')\mathbf{V}(t',x')}{c^2}\right)\frac{dx'}{dt'} - (\mathbf{v}(t') + \mathbf{V}(t',x'))}{1 - \frac{\mathbf{V}(t',x')}{c^2}\frac{dx'}{dt'}},$$

$$\frac{dy}{dt} = \frac{\frac{dy'}{dt'}\sqrt{1 - \frac{\mathbf{V}^2(t',x')}{c^2}}}{1 - \frac{\mathbf{V}(t',x')}{c^2}\frac{dx'}{dt'}}, \qquad (5.3.2)$$

$$\frac{dz}{dt} = \frac{\frac{dz'}{dt'}\sqrt{1 - \frac{\mathbf{V}^2(t',x')}{c^2}}}{1 - \frac{\mathbf{V}(t',x')}{c^2}\frac{dx'}{dt'}}.$$

By setting

$$\mathbf{v}_x = \frac{dx}{dt}, \mathbf{v}'_x = \frac{dx'}{dt'},$$
$$\mathbf{v}_y = \frac{dy}{dt}, \mathbf{v}'_y = \frac{dy'}{dt'}, \qquad (5.3.3)$$
$$\mathbf{v}_z = \frac{dz}{dt}, \mathbf{v}'_z = \frac{dz'}{dt'},$$

from (5.3.2) we obtain the formulae relating velocities:

$$\mathbf{v}_x = \frac{\left(1 + \dfrac{\mathbf{v}(t')\mathbf{V}(t',x')}{c^2}\right)\mathbf{v}'_x - (\mathbf{v}(t') + \mathbf{V}(t',x'))}{1 - \dfrac{\mathbf{V}(t',x')\mathbf{v}'_x}{c^2}},$$

$$\mathbf{v}_y = \frac{\mathbf{v}'_y\sqrt{1 - \dfrac{\mathbf{V}^2(t',x')}{c^2}}}{1 - \dfrac{\mathbf{V}(t',x')\mathbf{v}'_y}{c^2}}, \qquad (5.3.4)$$

$$\mathbf{v}_z = \frac{\mathbf{v}'_z\sqrt{1 - \dfrac{\mathbf{V}^2(t',x')}{c^2}}}{1 - \dfrac{\mathbf{V}(t',x')\mathbf{v}'_z}{c^2}}.$$

For the case

$$\mathbf{V}(t',x') = \frac{at'}{\sqrt{1 + \dfrac{a^2 t'^2}{c^2}}}, \; 0 \leq t' < \infty. \qquad (5.3.5)$$

from (5.3.4)-(5.3.5) we obtain the formulae relating velocities:

$$\mathbf{v}_x = \frac{\left(1 + \dfrac{at'}{\sqrt{1 + \dfrac{a^2 t'^2}{c^2}}} \dfrac{\mathbf{v}(t')}{c^2}\right)\mathbf{v}'_x - \left(\mathbf{v}(t') + \dfrac{at'}{\sqrt{1 + \dfrac{a^2 t'^2}{c^2}}}\right)}{1 - \dfrac{at'}{\sqrt{1 + \dfrac{a^2 t'^2}{c^2}}} \dfrac{\mathbf{v}'_x}{c^2}},$$

$$\mathbf{v}_y = \frac{\mathbf{v}'_y}{\sqrt{1 + \dfrac{a^2 t'^2}{c^2}}\left(1 - \dfrac{at'}{\sqrt{1 + \dfrac{a^2 t'^2}{c^2}}} \dfrac{\mathbf{v}'_y}{c^2}\right)}, \qquad (5.3.6)$$

$$\mathbf{v}_z = \frac{\mathbf{v}'_z}{\sqrt{1 + \dfrac{a^2 t'^2}{c^2}}\left(1 - \dfrac{at'}{\sqrt{1 + \dfrac{a^2 t'^2}{c^2}}} \dfrac{\mathbf{v}'_z}{c^2}\right)}.$$

For the case

$$\mathbf{v}(t') = \frac{at'}{\sqrt{1 + \dfrac{a^2 t'^2}{c^2}}}. \qquad (5.3.6)$$

from (5.3.5)-(5.3.6) we obtain the formulae relating velocities:

$$\mathbf{v}_x = \frac{\left(1 + \frac{a^2 t'^2}{c^2 + a^2 t'^2}\right)\mathbf{v}'_x - \frac{2at'}{\sqrt{1 + \frac{a^2 t'^2}{c^2}}}}{1 - \frac{at'}{\sqrt{1 + \frac{a^2 t'^2}{c^2}}} \frac{\mathbf{v}'_x}{c^2}},$$

$$\mathbf{v}_y = \frac{\mathbf{v}'_y}{\sqrt{1 + \frac{a^2 t'^2}{c^2}} \left(1 - \frac{at'}{\sqrt{1 + \frac{a^2 t'^2}{c^2}}} \frac{\mathbf{v}'_y}{c^2}\right)}, \qquad (5.3.6)$$

$$\mathbf{v}_z = \frac{\mathbf{v}'_z}{\sqrt{1 + \frac{a^2 t'^2}{c^2}} \left(1 - \frac{at'}{\sqrt{1 + \frac{a^2 t'^2}{c^2}}} \frac{\mathbf{v}'_z}{c^2}\right)}.$$

From the non-holonomic transformations (5.1.3) we obtain:

$$\frac{dx'}{dt'} = \frac{dx + (\mathbf{v}(t') + \mathbf{V}(t,x'))dt}{\left(1 + \frac{\mathbf{v}(t)\mathbf{V}(t,x')}{c^2}\right)dt + \frac{\mathbf{V}(t,x')}{c^2}dx},$$

$$\frac{dy'}{dt'} = \frac{dy\sqrt{1 - \frac{\mathbf{V}^2(t,x')}{c^2}}}{\left(1 + \frac{\mathbf{v}(t)\mathbf{V}(t,x')}{c^2}\right)dt + \frac{\mathbf{V}(t,x')}{c^2}dx}, \qquad (5.3.7)$$

$$\frac{dz'}{dt'} = \frac{dz\sqrt{1 - \frac{\mathbf{V}^2(t,x')}{c^2}}}{\left(1 + \frac{\mathbf{v}(t)\mathbf{V}(t,x')}{c^2}\right)dt + \frac{\mathbf{V}(t,x')}{c^2}dx},$$

From (5.3.7) by usual way we obtain:

$$\frac{dx'}{dt'} = \frac{\frac{dx}{dt} + (\mathbf{v}(t) + \mathbf{V}(t,x'))}{\left(1 + \frac{\mathbf{v}(t)\mathbf{V}(t,x')}{c^2}\right) + \frac{\mathbf{V}(t,x')}{c^2}\frac{dx}{dt}},$$

$$\frac{dy'}{dt'} = \frac{\frac{dy}{dt}\sqrt{1 - \frac{\mathbf{V}^2(t,x')}{c^2}}}{\left(1 + \frac{\mathbf{v}(t)\mathbf{V}(t,x')}{c^2}\right) + \frac{\mathbf{V}(t,x')}{c^2}\frac{dx}{dt}}, \quad (5.3.8)$$

$$\frac{dz'}{dt'} = \frac{\frac{dz}{dt}\sqrt{1 - \frac{\mathbf{V}^2(t,x')}{c^2}}}{\left(1 + \frac{\mathbf{v}(t)\mathbf{V}(t,x')}{c^2}\right) + \frac{\mathbf{V}(t,x')}{c^2}\frac{dx}{dt}}.$$

By setting

$$\mathbf{v}_x = \frac{dx}{dt}, \mathbf{v}'_x = \frac{dx'}{dt'},$$
$$\mathbf{v}_y = \frac{dy}{dt}, \mathbf{v}'_y = \frac{dy'}{dt'}, \quad (5.3.9)$$
$$\mathbf{v}_z = \frac{dz}{dt}, \mathbf{v}'_z = \frac{dz'}{dt'},$$

from (5.3.8)-(5.3.9) we obtain the formulae relating velocities:

$$\mathbf{v}'_x = \frac{\mathbf{v}_x + (\mathbf{v}(t) + \mathbf{V}(t,x'))}{\left(1 + \frac{\mathbf{v}(t)\mathbf{V}(t,x')}{c^2}\right) + \frac{\mathbf{v}_x\mathbf{V}(t,x')}{c^2}},$$

$$\mathbf{v}'_y = \frac{\mathbf{v}'_y\sqrt{1 - \frac{\mathbf{V}^2(t,x')}{c^2}}}{\left(1 + \frac{\mathbf{v}(t)\mathbf{V}(t,x')}{c^2}\right) + \frac{\mathbf{v}_x\mathbf{V}(t,x')}{c^2}}, \quad (5.3.10)$$

$$\mathbf{v}'_z = \frac{\mathbf{v}'_z\sqrt{1 - \frac{\mathbf{V}^2(t,x')}{c^2}}}{\left(1 + \frac{\mathbf{v}(t)\mathbf{V}(t,x')}{c^2}\right) + \frac{\mathbf{v}_x\mathbf{V}(t,x')}{c^2}}.$$

For the case

$$\mathbf{V}(t,x') = \frac{at}{\sqrt{1 + \frac{a^2 t^2}{c^2}}}, 0 \leq t' < \infty. \quad (5.3.11)$$

from (5.3.10) - (5.3.11) we obtain the non-holonomic transformations:

$$\mathbf{v}'_x = \frac{\mathbf{v}_x + \left(\mathbf{v}(t) + \dfrac{at}{\sqrt{1+\dfrac{a^2 t^2}{c^2}}}\right)}{\left(1 + \dfrac{at}{\sqrt{1+\dfrac{a^2 t^2}{c^2}}}\dfrac{\mathbf{v}(t)}{c^2}\right) + \dfrac{at}{\sqrt{1+\dfrac{a^2 t^2}{c^2}}}\dfrac{\mathbf{v}_x}{c^2}},$$

$$\mathbf{v}'_y = \frac{\mathbf{v}'_y \sqrt{1 - \dfrac{\mathbf{V}^2(t,x')}{c^2}}}{\left(1 + \dfrac{at}{\sqrt{1+\dfrac{a^2 t^2}{c^2}}}\dfrac{\mathbf{v}(t)}{c^2}\right) + \dfrac{at}{\sqrt{1+\dfrac{a^2 t^2}{c^2}}}\dfrac{\mathbf{v}_x}{c^2}}, \qquad (5.3.12)$$

$$\mathbf{v}'_z = \frac{\mathbf{v}'_z \sqrt{1 - \dfrac{\mathbf{V}^2(t,x')}{c^2}}}{\left(1 + \dfrac{at}{\sqrt{1+\dfrac{a^2 t^2}{c^2}}}\dfrac{\mathbf{v}(t)}{c^2}\right) + \dfrac{at}{\sqrt{1+\dfrac{a^2 t^2}{c^2}}}\dfrac{\mathbf{v}_x}{c^2}}.$$

For the case

$$\mathbf{v}(t) = \frac{at}{\sqrt{1+\dfrac{a^2 t^2}{c^2}}}. \qquad (5.3.13)$$

from (5.3.12)-(5.3.13) we obtain the formulae relating velocities:

$$\mathbf{v}'_x = \frac{\mathbf{v}_x + \dfrac{2at}{\sqrt{1+\dfrac{a^2t^2}{c^2}}}}{\left(1+\dfrac{at}{\sqrt{1+\dfrac{a^2t^2}{c^2}}}\dfrac{\mathbf{v}(t)}{c^2}\right) + \dfrac{at}{\sqrt{1+\dfrac{a^2t^2}{c^2}}}\dfrac{\mathbf{v}_x}{c^2}},$$

$$\mathbf{v}'_y = \frac{\mathbf{v}'_y\sqrt{1-\dfrac{\mathbf{V}^2(t,x')}{c^2}}}{\left(1+\dfrac{at}{\sqrt{1+\dfrac{a^2t^2}{c^2}}}\dfrac{\mathbf{v}(t)}{c^2}\right) + \dfrac{at}{\sqrt{1+\dfrac{a^2t^2}{c^2}}}\dfrac{\mathbf{v}_x}{c^2}}, \quad (5.3.14)$$

$$\mathbf{v}'_z = \frac{\mathbf{v}'_z\sqrt{1-\dfrac{\mathbf{V}^2(t,x')}{c^2}}}{\left(1+\dfrac{at}{\sqrt{1+\dfrac{a^2t^2}{c^2}}}\dfrac{\mathbf{v}(t)}{c^2}\right) + \dfrac{at}{\sqrt{1+\dfrac{a^2t^2}{c^2}}}\dfrac{\mathbf{v}_x}{c^2}}.$$

# Chapter VI. Solution of the "Two-Spaceship Paradox"

## VI.1. Generalized Lorentz transformations. Physical interpretation.

Let us consider now generalized Lorentz transformations $\hat{L}(\mathbf{v})$ (Eqs.3.2.2) of space and time between an inertial (*ortogonal*) Minkowski's frame $\mathbf{S}_I(t_I,x_I,y_I,z_I) \triangleq \mathbf{S}_\perp(t_I,x_I,y_I,z_I)$ and a generalized (*non-orthogonal*) inertial frame $\mathbf{S_v}(t,x,y,z) \triangleq \mathbf{S}_{\angle}(t,x,y,z)$ with a metric given by Eq.(3.1.5) which was formed by the linear transformation $\Lambda(\mathbf{v})$ given by Eq. (3.1.4):

$$x = \gamma(\mathbf{v})\left[x_I\left(1 + \frac{\mathbf{v}^2}{c^2}\right) - 2\mathbf{v}t_I\right],$$

$$t = \gamma(\mathbf{v})\left(t_I - \frac{\mathbf{v}}{c^2}x_I\right).$$

(6.1.1)

By simple calculation from Eq.(6.1.1) one obtain:

$$x_I = \frac{x}{\gamma^*(\mathbf{v})} + \mathbf{v}^* t_I,$$

where

$$\gamma^*(\mathbf{v}) = \gamma(\mathbf{v})\left(1 + \frac{\mathbf{v}^2}{c^2}\right),$$

(6.1.2)

$$\mathbf{v}^* = \mathbf{v}_{ph.} = \frac{2\mathbf{v}}{\left(1 + \frac{\mathbf{v}^2}{c^2}\right)}.$$

Thus

$$\mathbf{v} = \frac{c^2}{\mathbf{v}^*} - \sqrt{\frac{c^4}{\mathbf{v}^{*2}} - c^2}.$$

(6.1.3)

Let us consider three inertial reference frames are involved $\mathbf{S}_\perp(XOY), \mathbf{S}'_\perp(X'O'Y')$ and $\mathbf{S}^0_\perp(X^0O^0Y^0)$ (**Pic.1**). $\mathbf{S}^0_\perp(X^0O^0Y^0)$ moves with velocity $\vec{u}_x$ relative to $\mathbf{S}_\perp(XOY)$, with velocity $\vec{u}'_x$ relative to $\mathbf{S}'_\perp(X'O'Y'), \mathbf{S}'_\perp(X'O'Y')$ moving at its turn with velocity $\vec{\mathbf{v}}$ relative to $\mathbf{S}_\perp(XOY)$. The axes of the reference frames are parallel to each other, the $OX, O'X'$ and $O^0X^0$ axes are overlapped and all the velocities show in theirs positive directions. Suppose that: (**a**) the two reference frames $\mathbf{S}_\perp(XOY)$ and $\mathbf{S}'_\perp(X'O'Y')$ have the same Minkowski's space-time origins and (**b**) the one reference frame $\mathbf{S}^0_\perp(X^0O^0Y^0)$ have the non-Minkowski's space-time origin with a non-ortogonal metric $\mathbf{g_v} = \mathbf{g_v}(t,x,y,z;\mathbf{v})$ given by Eq.(3.1.5):

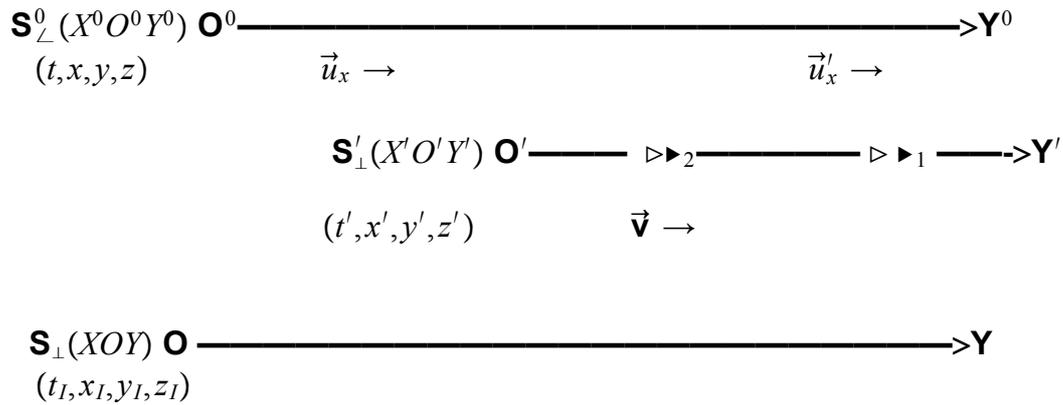

**Pic.1.** Three inertial reference frames are involved:
$\mathbf{S}_\perp(XOY), \mathbf{S}'_\perp(X'O'Y')$ and $\mathbf{S}^0_\perp(X^0O^0Y^0)$.

The velocities $u_x, u'_x$ and **v** mentioned above are related by relativistic velocity addition law:

$$u'_x = \frac{u_x - \mathbf{v}}{1 - \frac{u_x \mathbf{v}}{c^2}}. \qquad (6.1.4)$$

Suppose that $u_x = \mathbf{v}^* = \mathbf{v}_{ph}$. Thus

$$u'_x = \frac{\mathbf{v}^* - \mathbf{v}}{1 - \frac{\mathbf{v}^* \mathbf{v}}{c^2}} =$$

$$= \frac{\frac{2\mathbf{v}}{\left(1 + \frac{\mathbf{v}^2}{c^2}\right)} - \mathbf{v}}{1 - \frac{1}{c^2} \frac{2\mathbf{v}^2}{\left(1 + \frac{\mathbf{v}^2}{c^2}\right)}} = \mathbf{v}. \qquad (6.1.5)$$

$$u'_x = \mathbf{v}.$$

One can obtain (6.15) by calculation

$$\frac{\dfrac{2\mathbf{v}}{\left(1+\dfrac{\mathbf{v}^2}{c^2}\right)} - \mathbf{v}}{1 - \dfrac{1}{c^2}\dfrac{2\mathbf{v}^2}{\left(1+\dfrac{\mathbf{v}^2}{c^2}\right)}} = \mathbf{v} \Rightarrow \frac{\dfrac{2}{\left(1+\dfrac{\mathbf{v}^2}{c^2}\right)} - 1}{1 - \dfrac{1}{c^2}\dfrac{2\mathbf{v}^2}{\left(1+\dfrac{\mathbf{v}^2}{c^2}\right)}} = 1 \Rightarrow$$

$$\frac{2}{\left(1+\dfrac{\mathbf{v}^2}{c^2}\right)} - 1 = 1 - \dfrac{1}{c^2}\dfrac{2\mathbf{v}^2}{\left(1+\dfrac{\mathbf{v}^2}{c^2}\right)} \Rightarrow \qquad (6.1.6)$$

$$\frac{1}{\left(1+\dfrac{\mathbf{v}^2}{c^2}\right)} = 1 - \dfrac{\dfrac{\mathbf{v}^2}{c^2}}{\left(1+\dfrac{\mathbf{v}^2}{c^2}\right)} = \dfrac{1}{\left(1+\dfrac{\mathbf{v}^2}{c^2}\right)}.$$

Let us consider now three inertial reference frames are involved $\mathbf{S}_\perp(XOY), \mathbf{S}'_\perp(X'O'Y')$ and $\mathbf{S}^0_\angle(X^0O^0Y^0)$ (**Pic.2**). $\mathbf{S}^0_\angle(X^0O^0Y^0)$ moves with velocity $u_x = \mathbf{v}^*$ relative to $\mathbf{S}_\perp(XOY)$, with velocity $u'_x = \mathbf{v}$ relative to $\mathbf{S}'_\perp(X'O'Y')$, $\mathbf{S}'_\perp(X'O'Y')$ moving at its turn with velocity $\mathbf{v}$ relative to $\mathbf{S}_\perp(XOY)$. The axes of the reference frames are parallel to each other, the $OX, O'X'$ and $O^0X^0$ axes are overlapped and all the velocities show in theirs positive directions. Suppose that: (**a**) the two reference frames $\mathbf{S}_\perp(XOY)$ and $\mathbf{S}'_\perp(X'O'Y')$ have the same Minkowski's space-time origins and (**b**) the one reference frame $\mathbf{S}^0_\angle(X^0O^0Y^0)$ have the non-Minkowski's space-time origin with a non-ortogonal metric $\mathbf{g_v} = \mathbf{g_v}(t,x,y,z;\mathbf{v})$ given by Eq.(3.1.5):

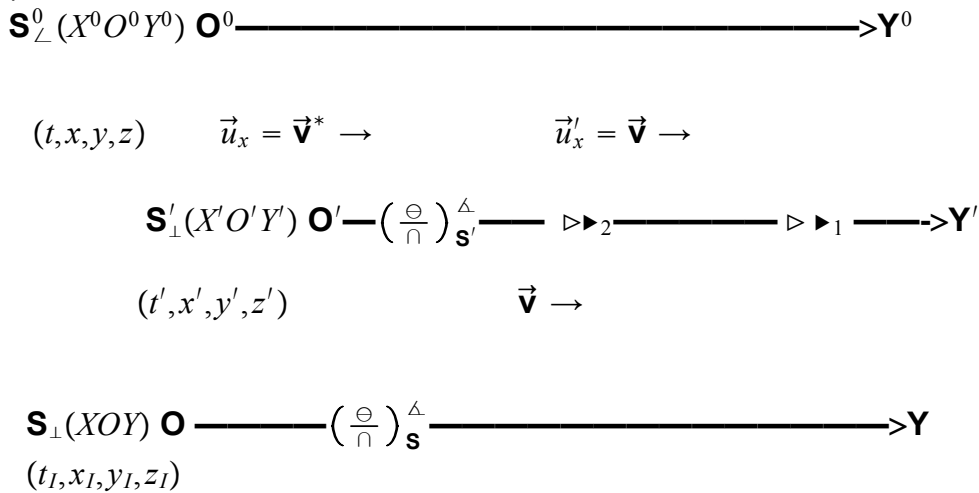

**Pic.2.** Three inertial reference frames are involved:
$\mathbf{S}_\perp(XOY), \mathbf{S}'_\perp(X'O'Y')$ and $\mathbf{S}^0_\angle(X^0O^0Y^0)$.

Let us consider two inertial reference frames are involved $\mathbf{S}_\perp(XOY), \mathbf{S}'_\perp(X'O'Y')$

and two rockets ▲$_1$, ▲$_2$ which at rest on the (comuvin) common frame $\mathbf{S}'_\perp(X'O'Y')$.

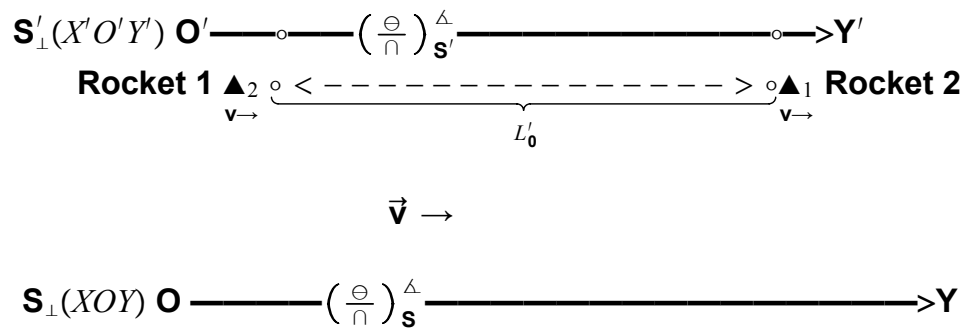

**Pic.3.** Two inertial reference frames are involved: $\mathbf{S}_\perp(XOY), \mathbf{S}'_\perp(X'O'Y')$ and two rockets ▲$_1$, ▲$_2$ which at rest on the common frame $\mathbf{S}'_\perp(X'O'Y')$. $L'_0$ is the initial distance beetwin rocket 1 and rocket 2 measured by inertial observer $\left(\frac{\ominus}{\cap}\right)^{\triangle}_{\mathbf{S}'}$ from inertial frame $\mathbf{S}'_\perp(X'O'Y')$.

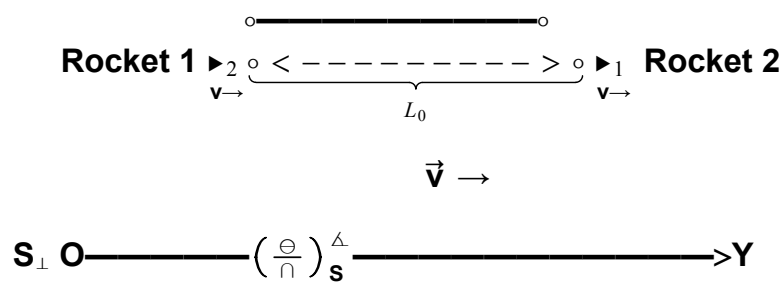

**Pic.4.** $L_0$ is the initial distance beetwin rocket 1 and rocket 2 is measured by inertial observer $\left(\frac{\ominus}{\cap}\right)^{\triangle}_{\mathbf{S}}$ from inertial frame $\mathbf{S}_\perp(XOY)$ in the any moment of the time $t$ such that $t < t_0$.

Thus using standard Lorentz contractions one obtain:

$$L_0(t < t_0) = L_0(t_0) = L'_0\sqrt{1 - \frac{\mathbf{v}^2}{c^2}}.  \qquad (6.1.7)$$

Suppose that at the some initial moment at time $t_I = t_0$ in the inertial frame $\mathbf{S}_\perp (X\,OY)$, both rockets start *simultaneously* (from the point of view of the observer $\left(\frac{\ominus}{\cap}\right)_\mathbf{S}^{\triangle}$ on the inertial frame $\mathbf{S}_\perp(X\,OY)$ ) with a constant velocity of the both rockets: $u_x = \mathbf{v}^*$ which is measured by inertial observer $\left(\frac{\ominus}{\cap}\right)_\mathbf{S}^{\triangle}$ from inertial frame $\mathbf{S}_\perp(X\,OY)$.

Following this reasoning above, the position coordinate $x_{I,1}(t_I), x_{I,2}(t_I)$ of each rocket as function of time $t_I$ is:

$$x_{I,1}(t_I \geq t_0) = L_0(t_0) + \mathbf{v}^* t_I,$$
$$\qquad (6.1.8)$$
$$x_{I,2}(t_I \geq t_0) = \mathbf{v}^* t_I.$$

**Remark**   6.1. Note that from Eq.(6.1.5) velocity of the both rockets which is measured by inertial observer $\left(\frac{\ominus}{\cap}\right)_{\mathbf{S}'}^{\triangle}$ are: $u'_x = \mathbf{v}$.

**Remark**   6.2. Note that from the point of view of the observer $\left(\frac{\ominus}{\cap}\right)_{\mathbf{S}'}^{\triangle}$ on the inertial frame $\mathbf{S}'_\perp(X'O'Y')$ rockets 1,2 does not start simultaneously.

Finally bouth rocets $\blacktriangleright_1, \blacktriangleright_2$ at rest on the (comuvin) common inertial frame $\mathbf{S}_\perp^0 (X^0 O^0 Y^0)$ which moves with velocity $u_x = \mathbf{v}^*$ relative to inertial frame $\mathbf{S}_\perp(XOY)$, see **(Pic.5)**.

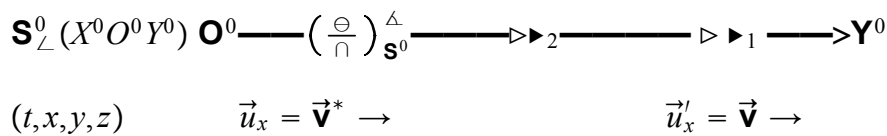

$\mathbf{S}_\perp^0(X^0O^0Y^0)$ $\mathbf{O}^0$——$\left(\frac{\ominus}{\cap}\right)_{\mathbf{S}^0}^{\triangle}$———▷▶$_2$————▷▶$_1$——>$Y^0$

$(t,x,y,z)$   $\vec{u}_x = \vec{\mathbf{v}}^* \rightarrow$   $\vec{u}'_x = \vec{\mathbf{v}} \rightarrow$

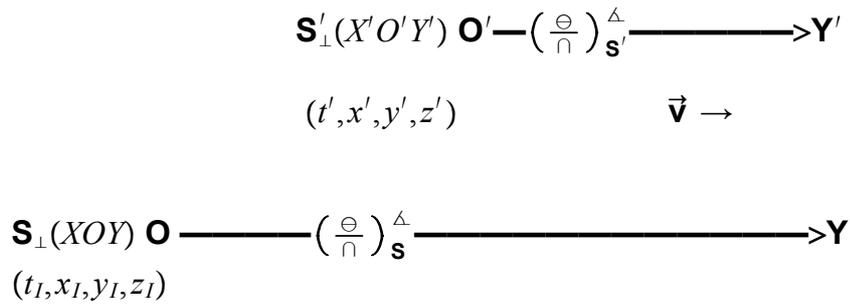

**Pic.5.** Finally bouth rocets at rest on the common inertial frame $S^0_\perp(X^0O^0Y^0)$.

(a) Suppose that $L^0$ is the distance beetwin rocket 1 and rocket 2 (which at rest on the common frame $S^0_\perp(X^0O^0Y^0)$) measured by inertial observer $\left(\frac{\ominus}{\cap}\right)^{\triangle}_{S^0}$ from inertial frame $S^0_\perp(X^0O^0Y^0)$, see (**Pic.6**).

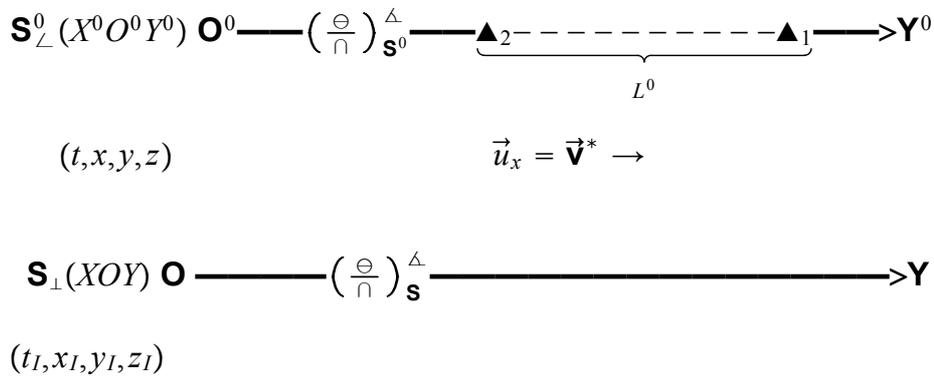

**Pic.6.** Two inertial reference frames are involved: $S_\perp(XOY), S^0_\perp(X^0O^0Y^0)$ and two rockets $\blacktriangle_1, \blacktriangle_2$ which at rest on the common frame $S^0_\perp(X^0O^0Y^0)$. $L^0$ is the distance beetwin rocket 1 and rocket 2 is measured by inertial observer $\left(\frac{\ominus}{\cap}\right)^{\triangle}_{S^0}$ from inertial frame $S^0_\perp(X^0O^0Y^0)$.

(b) Suppose that $L_I$ is the distance beetwin rocket 1 and rocket 2 (which at rest on the common frame $S^0_\perp(X^0O^0Y^0)$) measured by inertial observer $\left(\frac{\ominus}{\cap}\right)^{\triangle}_{S}$ from inertial frame $S_\perp(XOY)$, see (**Pic.7**).

$\mathbf{S}^0_\perp(X^0O^0Y^0)$ **O**$^0$ ────▷▶$_2$ ── ▷ ▶$_1$ ────>$Y^0$
$\underbrace{\qquad\qquad}_{L_I}$

$$\vec{u}_x = \vec{\mathbf{v}}^* \rightarrow$$

$\mathbf{S}_\perp(XOY)$ **O** ──── $\left(\begin{smallmatrix}\ominus\\\cap\end{smallmatrix}\right)^{\triangle}_{\mathbf{S}}$ ────────>Y

**Pic.7.** $L_I$ is the distance beetwin rocket 1 and rocket 2
is measured by inertial observer $\left(\begin{smallmatrix}\ominus\\\cap\end{smallmatrix}\right)^{\triangle}_{\mathbf{S}}$ from inertial frame
$\mathbf{S}_\perp(XOY)$ in the any moment of the time $t$ such that $t \geq t_0$.

Thus using generalized Lorentz contractions Eq.(3.2.6) one obtain:

$$L^0 = L_I(t \geq t_0)\frac{1 + \frac{\mathbf{v}^2}{c^2}}{\sqrt{1 - \frac{\mathbf{v}^2}{c^2}}}. \qquad (6.1.9)$$

Using the law of the position coordinate $x_{I,1}(t_I), x_{I,2}(t_I)$ of each
rocket as function of time $t_I$ Eq.(6.1.8) one obtain:

$$L_I(t \geq t_0) = L_0(t_0). \qquad (6.1.10)$$

Substitution Eq.(6.1.10) into Eq.(6.1.9) gives

$$L^0 = L_0(t_0)\frac{1 + \frac{\mathbf{v}^2}{c^2}}{\sqrt{1 - \frac{\mathbf{v}^2}{c^2}}}. \qquad (6.1.11)$$

Substitution Eq.(6.1.7) into Eq.(6.1.11) gives

$$L^0 = L_0(t_0) \frac{1 + \frac{\mathbf{v}^2}{c^2}}{\sqrt{1 - \frac{\mathbf{v}^2}{c^2}}} =$$

$$L^0 = L'_0 \sqrt{1 - \frac{\mathbf{v}^2}{c^2}} \; \frac{1 + \frac{\mathbf{v}^2}{c^2}}{\sqrt{1 - \frac{\mathbf{v}^2}{c^2}}} = \qquad (6.1.12)$$

$$= L'_0 \left(1 + \frac{\mathbf{v}^2}{c^2}\right).$$

$$L^0 = L'_0 \left(1 + \frac{\mathbf{v}^2}{c^2}\right).$$

**Conclusion** *Thus when switching the description to the co-moving frame, the distance between the rockets appears to increase only by the factor*

$$\gamma^\# = \left(1 + \frac{\mathbf{v}^2}{c^2}\right) \leq 2. \qquad (6.1.13)$$

# Appendix A.1.

Let the invariant $ds'^2$ in an inertial reference system in Galilean coordinates $(T, X)$ have the form

$$ds^2 = c^2 dT^2 - dX^2. \qquad (A.1.1)$$

Now, we pass to another inertial reference system

$$X = x + \mathbf{v}T. \qquad (A.1.2)$$

Then the invariant $ds^2$ assumes the form

$$ds^2 = c^2\left(1 - \frac{\mathbf{v}^2}{c^2}\right)dT^2 - 2\mathbf{v}dTdx - dx^2. \qquad (A.1.3)$$

Hence we have

$$ds^2 = c^2\left[\left(\sqrt{1-\frac{\mathbf{v}^2}{c^2}}\right)dT - \frac{\mathbf{v}dx}{\sqrt{1-\frac{\mathbf{v}^2}{c^2}}}\right]^2 - dx^2\left[1+\frac{\mathbf{v}^2}{c^2-\mathbf{v}^2}\right] \qquad (A.1.4)$$

Expression (A.1.4) can be written in the form

$$ds'^2 = c^2 dt'^2 - dx'^2, \qquad (A.1.5)$$

where

$$t' = T\sqrt{1-\frac{\mathbf{v}^2}{c^2}} - x\frac{\frac{\mathbf{v}}{c^2}}{\sqrt{1-\frac{\mathbf{v}^2}{c^2}}} = \frac{T - \frac{\mathbf{v}}{c^2}X}{\sqrt{1-\frac{\mathbf{v}^2}{c^2}}},$$

$$x' = \frac{x}{\sqrt{1-\frac{\mathbf{v}^2}{c^2}}} = \frac{x+\mathbf{v}T}{\sqrt{1-\frac{\mathbf{v}^2}{c^2}}}.$$

$$(A.1.6)$$

We see from expression (A.1.5) that the form-invariance of the invariant $ds^2$ is provided for by the Lorentz transformations (A.1.6).

## Appendix A.2.

## General transformations between an inertial Minkowski frame $F_I(t_I, x_I, y_I, z_I)$ and a generalized inertial frame $F_\mathbf{v}(t,x,y,z)$.

Let us calculate general transformations $\hat{L}(\mathbf{v},\mathbf{V})$ of space and time between an inertial Minkowski frame $F_I(t_I, x_I, y_I, z_I) = F_I(t', x', y', z')$ and a generalized inertial frame $F_\mathbf{v}(t,x,y,z)$ with a metric (3.1.5) forming by the linear (non-orthogonal) transformations $\Lambda(\mathbf{v})$ (3.1.4). From [3] (see [3] Theorem in subsection I) we obtain:

$$\hat{L}(\mathbf{v},\mathbf{V}) = \Lambda(\mathbf{v}) \circ L(\mathbf{V}).$$
$$\Lambda(\mathbf{V}) : (t',x') \to (t,x)$$
$$x = x' - \mathbf{v}t', t = t' \quad (A.2.1.a)$$
$$L(\mathbf{v}) : \begin{cases} x' \to \gamma(\mathbf{V})(x' - \mathbf{V}t') & (A.2.1.b) \\ t' \to \gamma(\mathbf{V})\left(t' - \frac{\mathbf{V}}{c^2}x'\right) & (A.2.1.c) \end{cases} \qquad (A.2.1)$$

By substitution (A.2.1.b)-(A.2.1.c) into (A.2.1.a) we obtain the generalized Lorentz transformations $\hat{L}(\mathbf{v},\mathbf{V})$:

$$x = x' - \mathbf{v}t' =$$
$$\gamma(\mathbf{V})(x' - \mathbf{V}t') - \mathbf{v}\gamma(\mathbf{V})\left(t' - \frac{\mathbf{V}}{c^2}x'\right) =$$
$$\gamma(\mathbf{V})\left[x' - \mathbf{V}t' - \mathbf{v}t' + \frac{\mathbf{vV}}{c^2}x'\right] =$$
$$\gamma(\mathbf{V})\left[\left(1 + \frac{\mathbf{vV}}{c^2}\right)x' - (\mathbf{v} + \mathbf{V})t'\right]$$

Thus

$$x = \gamma(\mathbf{V})\left[\left(1 + \frac{\mathbf{vV}}{c^2}\right)x' - (\mathbf{v} + \mathbf{V})t'\right],$$
$$t = \gamma(\mathbf{V})\left(t' - \frac{\mathbf{V}}{c^2}x'\right). \qquad (A.2.2)$$

Or in equivalent infinitesimal form:

$$dx = \gamma(\mathbf{V})\left[\left(1 + \frac{\mathbf{vV}}{c^2}\right)dx' - (\mathbf{v} + \mathbf{V})dt'\right],$$
$$dt = \gamma(\mathbf{V})\left(dt' - \frac{\mathbf{V}}{c^2}dx'\right). \qquad (A.2.3)$$

Substitution (A.2.3) into (3.1.5) gives:

$$ds^2 = c^2\left(1 - \frac{\mathbf{v}^2}{c^2}\right)dt^2 - 2\mathbf{v}dtdx - dx^2 =$$

$$c^2\left(1 - \frac{\mathbf{v}^2}{c^2}\right)\gamma^2(\mathbf{V})\left(dt' - \frac{\mathbf{V}}{c^2}dx'\right)^2 -$$

$$-2\mathbf{v}\gamma^2(\mathbf{V})\left(dt' - \frac{\mathbf{V}}{c^2}dx'\right)\left[\left(1 + \frac{\mathbf{v}\mathbf{V}}{c^2}\right)dx' - (\mathbf{v}+\mathbf{V})dt'\right] -$$

$$-\gamma^2(\mathbf{V})\left[\left(1 + \frac{\mathbf{v}\mathbf{V}}{c^2}\right)dx' - (\mathbf{v}+\mathbf{V})dt'\right]^2 =$$

$$\gamma^2(\mathbf{V})\Bigg\{c^2\left(1 - \frac{\mathbf{v}^2}{c^2}\right)\left(dt' - \frac{\mathbf{V}}{c^2}dx'\right)^2 -$$

$$-2\mathbf{v}\left(dt' - \frac{\mathbf{V}}{c^2}dx'\right)\left[\left(1 + \frac{\mathbf{v}\mathbf{V}}{c^2}\right)dx' - (\mathbf{v}+\mathbf{V})dt'\right] -$$

$$-\left[\left(1 + \frac{\mathbf{v}\mathbf{V}}{c^2}\right)dx' - (\mathbf{v}+\mathbf{V})dt'\right]^2\Bigg\} =$$

$$\gamma^2(\mathbf{V})\Bigg\{(c^2 - \mathbf{v}^2)\left(dt'^2 - 2\frac{\mathbf{V}}{c^2}dt'dx' + \frac{\mathbf{V}^2}{c^4}dx'^2\right) - 2\mathbf{v}\left(dt' - \frac{\mathbf{V}}{c^2}dx'\right)\left(1 + \frac{\mathbf{v}\mathbf{V}}{c^2}\right)dx' +$$

$$+2\mathbf{v}(\mathbf{v}+\mathbf{V})\left(dt' - \frac{\mathbf{V}}{c^2}dx'\right)dt' - \left(1 + \frac{\mathbf{v}\mathbf{V}}{c^2}\right)^2 dx'^2 + 2(\mathbf{v}+\mathbf{V})\left(1 + \frac{\mathbf{v}\mathbf{V}}{c^2}\right)dt'dx' - \quad (A.2.4)$$

$$-(\mathbf{v}+\mathbf{V})^2 dt'^2 =$$

$$\gamma^2(\mathbf{V})\Bigg\{c^2 dt'^2 - \boxed{2\mathbf{V}dt'dx'} + \underline{\frac{\mathbf{V}^2}{c^2}dx'^2} - \boxed{\mathbf{v}^2 dt'^2} + \boxed{2\frac{\mathbf{v}^2\mathbf{V}}{c^2}dt'dx'} - \boxed{\frac{\mathbf{v}^2\mathbf{V}^2}{c^4}dx'^2} -$$

$$-\boxed{2\mathbf{v}dt'dx'} + \boxed{2\frac{\mathbf{v}\mathbf{V}}{c^2}dx'^2} - \boxed{2\frac{\mathbf{v}^2\mathbf{V}}{c^2}dt'dx'} + \boxed{2\frac{\mathbf{v}^2\mathbf{V}^2}{c^4}dx'^2} +$$

$$+\boxed{2\mathbf{v}^2 dt'^2} + \boxed{2\mathbf{v}\mathbf{V}dt'^2} - \boxed{2\frac{\mathbf{v}^2\mathbf{V}}{c^2}dt'dx'} - \boxed{2\frac{\mathbf{v}\mathbf{V}^2}{c^2}dt'dx'} -$$

$$\underline{-dx'^2} - \boxed{2\frac{\mathbf{v}\mathbf{V}}{c^2}dx'^2} - \boxed{\frac{\mathbf{v}^2\mathbf{V}^2}{c^4}dx'^2} + \boxed{2\mathbf{v}dt'dx'} + \boxed{2\mathbf{V}dt'dx'} + \boxed{2\frac{\mathbf{v}^2\mathbf{V}}{c^2}dt'dx'} + \boxed{2\frac{\mathbf{v}\mathbf{V}^2}{c^2}dt'dx'} +$$

$$-\boxed{\mathbf{v}^2 dt'^2} - \boxed{2\mathbf{v}\mathbf{V}dt'^2} \underline{-\mathbf{V}^2 dt'^2}\Bigg\} =$$

$$\gamma^2(\mathbf{V})\left(c^2 dt'^2 - \mathbf{V}^2 dt'^2 - dx'^2 + \frac{\mathbf{V}^2}{c^2}dx'^2\right) = c^2 dt'^2 - dx'^2 = ds'^2.$$

Thus, the invariant interval $ds^2$ under the generalized Lorentz transformation (A.2.4) is

$$ds^2 = c^2\left(1 - \frac{\mathbf{v}^2}{c^2}\right)dt^2 - 2\mathbf{v}dtdx - dx^2 = c^2 dt'^2 - dx'^2 = ds'^2. \quad (A.2.5)$$

From the generalized Lorentz transformations (A.2.2) by usual way we obtain general length contractions:

$$x_2 - x_1 = \gamma(\mathbf{V})\left(1 + \frac{\mathbf{v}\mathbf{V}}{c^2}\right)(x'_2 - x'_1). \quad (A.2.6)$$

This result is similar to that in standard special relativity: i.e., roughly speaking, a "moving" meter stick (at rest in $F_\mathbf{v}$) appears to be shorter, as measured by observers in the frame $F_I$, as shown in (A.2.6).

Let us calculate now from generalized Lorentz transformations $\hat{L}(\mathbf{v},\mathbf{V})$ (A.2.3) the transformations $L^*(\mathbf{v},\mathbf{V})$, which keeping the metric (3.1.5) forminvariant. From [3] (see [3] Theorem in subsection I) we obtain: :

$$L^*(\mathbf{v},\mathbf{V}) = \check{L}(\mathbf{v},\mathbf{V}) \circ \Lambda^{-1}(\mathbf{v}),$$

$$\check{L}(\mathbf{v},\mathbf{V}) : (t',x') \to (t_1,x_1) = \hat{L}(\mathbf{v})(t',x').$$

$$\Lambda^{-1}(\mathbf{v}) : (t,x) \to (t',x')$$

$$x' = x + \mathbf{v}t, \quad t' = t. \qquad (A.2.7.a) \qquad (A.2.7)$$

$$\hat{L}(\mathbf{v},\mathbf{V}) : \begin{cases} x_1 = \gamma(\mathbf{V})\left[x'\left(1 + \frac{\mathbf{vV}}{c^2}\right) - (\mathbf{v}+\mathbf{V})t'\right], & (A.2.7.b) \\ t_1 = \gamma(\mathbf{V})\left(t' - \frac{\mathbf{V}}{c^2}x'\right). & (A.2.7.c) \end{cases}$$

By substitution (A.2.7.a) into (A.2.7.b)-(A.2.7.c) we obtain the transformations which keeping the metric (3.1.5) forminvariant, [were first obtained in [30], but in a much more complicated way — by solving a system of partial differential equations]:

$$x_1 = \gamma(\mathbf{V})\left[x'\left(1 + \frac{\mathbf{vV}}{c^2}\right) - (\mathbf{v}+\mathbf{V})t'\right] =$$

$$\gamma(\mathbf{V})\left[(x + \mathbf{v}t)\left(1 + \frac{\mathbf{vV}}{c^2}\right) - (\mathbf{v}+\mathbf{V})t\right] =$$

$$\gamma(\mathbf{V})\left(x + \frac{\mathbf{vV}}{c^2}x + \mathbf{v}t + \frac{\mathbf{v}^2\mathbf{V}}{c^2}t - \mathbf{v}t - \mathbf{V}t\right) =$$

$$\gamma(\mathbf{V})\left[\left(1 + \frac{\mathbf{vV}}{c^2}\right)x - \left(1 - \frac{\mathbf{v}^2}{c^2}\right)\mathbf{V}t\right]. \qquad (A.2.8)$$

$$t_1 = \gamma(\mathbf{V})\left(t' - \frac{\mathbf{V}}{c^2}x'\right) =$$

$$\gamma(\mathbf{V})\left[t - \frac{\mathbf{V}}{c^2}(x + \mathbf{v}t)\right] = \gamma(\mathbf{V})\left(t - \frac{\mathbf{V}}{c^2}x - \frac{\mathbf{vV}}{c^2}t\right) =$$

$$\gamma(\mathbf{V})\left[\left(1 - \frac{\mathbf{vV}}{c^2}\right)t - \frac{\mathbf{V}}{c^2}x\right].$$

Thus

$$\begin{aligned} x_1 &= \gamma(\mathbf{V})\left[\left(1 + \frac{\mathbf{vV}}{c^2}\right)x - \left(1 - \frac{\mathbf{v}^2}{c^2}\right)\mathbf{V}t\right], \\ t_1 &= \gamma(\mathbf{V})\left[\left(1 - \frac{\mathbf{vV}}{c^2}\right)t - \frac{\mathbf{V}}{c^2}x\right]. \end{aligned} \qquad (A.2.9)$$

Or in equivalent infinitesimal form:

$$\begin{aligned} dx_1 &= \gamma(\mathbf{V})\left[\left(1 + \frac{\mathbf{vV}}{c^2}\right)dx - \left(1 - \frac{\mathbf{v}^2}{c^2}\right)\mathbf{V}dt\right], \\ dt_1 &= \gamma(\mathbf{V})\left[\left(1 - \frac{\mathbf{vV}}{c^2}\right)dt - \frac{\mathbf{V}}{c^2}dx\right]. \end{aligned} \qquad (A.2.10)$$

## Appendix A.3.

## Generalized Lorentz transformations between a generalized inertial frame $F_\mathbf{v}(t,x,y,z)$ and an inertial Minkowski frame $F_I(t_I,x_I,y_I,z_I)$.

Let us calculate general transformations $\hat{L}^{-1}(\mathbf{v})$ of space and time between a generalized inertial frame $F_\mathbf{v}(t,x,y,z)$ with a metric (3.1.5) forming by the linear (non-orthogonal) transformations $\Lambda(\mathbf{v})$ (3.1.4) and an inertial Minkowski frame $F_I(t_I,x_I,y_I,z_I) = F_I(t',x',y',z')$. From [3] (see [3] Theorem in subsection I) we obtain:

$$\hat{L}^{-1}(\mathbf{v},\mathbf{V}) = \check{L}(\mathbf{V}) \circ \Lambda^{-1}(\mathbf{v}). \qquad (A.3.1.a)$$

$$\Lambda^{-1}(\mathbf{v}) : (t,x) \to (t',x')$$
$$x' = x + \mathbf{v}t, \, t = t'. \qquad (A.3.1.b)$$

$$\check{L}(\mathbf{V}) : (t',x') \to L(\mathbf{V})(t',x') \qquad (A.3.1)$$

$$L(\mathbf{V}) : \begin{cases} x' \to \gamma(\mathbf{V})(x' + \mathbf{V}t') \\ t' \to \gamma(\mathbf{V})\left(t' + \dfrac{\mathbf{V}}{c^2}x'\right) \end{cases} \quad (A.3.1.c)$$

By substitution (3.3.1.b) into (3.3.1.c) we obtain the generalized Lorentz transformations $\hat{L}^{-1}(\mathbf{v},\mathbf{V})$:

$$x' = \gamma(\mathbf{V})(x' + \mathbf{V}t') =$$
$$\gamma(\mathbf{V})(x + \mathbf{v}t + \mathbf{V}t) = \gamma(\mathbf{V})[x + (\mathbf{v} + \mathbf{V})t].$$
$$t' = \gamma(\mathbf{V})\left(t' + \frac{\mathbf{V}}{c^2}x'\right) = \gamma(\mathbf{V})\left[t + \frac{\mathbf{V}}{c^2}(x + \mathbf{v}t)\right] = \qquad (A.3.2)$$
$$\gamma(\mathbf{V})\left[\left(1 + \frac{\mathbf{v}\mathbf{V}}{c^2}\right)t + \frac{\mathbf{V}}{c^2}x\right].$$

Thus

$$x' = \gamma(\mathbf{V})[x + (\mathbf{v} + \mathbf{V})t],$$
$$t' = \gamma(\mathbf{V})\left[\left(1 + \frac{\mathbf{v}\mathbf{V}}{c^2}\right)t + \frac{\mathbf{V}}{c^2}x\right]. \qquad (A.3.3)$$

Or in the equivalent differential form:

$$dx' = \gamma(\mathbf{V})[dx + (\mathbf{v} + \mathbf{V})dt],$$
$$dt' = \gamma(\mathbf{V})\left[\left(1 + \frac{\mathbf{v}\mathbf{V}}{c^2}\right)dt + \frac{\mathbf{V}}{c^2}dx\right].$$
(A.3.4)

# Appendix A.4.

# General length contraction.

Again we consider two such inertial reference frame $F_\mathbf{v}(t,x,y,z)$ with a metric (3.1.5) forming by the linear (non-orthogonal) transformations $\Lambda(\mathbf{v})$ (3.1.4) and Minkowski frame $F_I(t_I, x_I, y_I, z_I) = F'_I(t', x', y', z')$, one of which $F_\mathbf{v}$ will be considered to be at rest, while another $F'_I$ one will move with respect to the first one.

Consider measurement, in a moving reference frame $F'_I$, of the length of a rod that is at rest in another reference frame $F_\mathbf{v}$. We first determine the method for measuring the length of a moving rod. Consider an observer in the moving reference frame $F'_I$, who records the ends of the rod, $x'_1$ and $x'_2$, at the same moment of time

$$t'_1 = t'_2,$$
(A.4.1)

this permits to reduce the Minkowski interval $s'^2_{1,2}$ in the moving reference frame $F'_I$ to the spatial part only

$$s'^2_{1,2} = -(x'_2 - x'_1)^2 = -l^2.$$
(A.4.2)

Thus, in our method of determining the length of a moving rod, by usual way to consider the quantity $l$ as its *length* in a moving reference frame $F'_I$.

The same interval in the reference frame at rest $F_\mathbf{v}$, where the rod is in the state of rest, is given as follows

$$s^2_{1,2} = c^2\left(1 - \frac{\mathbf{v}^2}{c^2}\right)(t_2 - t_1)^2 - 2\mathbf{v}(t_2 - t_1)(x_2 - x_1) - (x_2 - x_1)^2.$$
(A.4.3)

But, in accordance with the generalized Lorentz transformations (A.3.3) we have

$$t'_2 - t'_1 = \gamma(\mathbf{V})\left[\left(1 + \frac{\mathbf{v}\mathbf{V}}{c^2}\right)(t_2 - t_1) + \frac{\mathbf{V}}{c^2}(x_2 - x_1)\right].$$
(A.4.4)

whence for our case (A.4.1) we find

$$\left(1 + \frac{\mathbf{vV}}{c^2}\right)(t_2 - t_1) + \frac{\mathbf{V}}{c^2}(x_2 - x_1) = 0,$$

$$t_2 - t_1 = -\frac{\mathbf{V}}{c^2} \frac{x_2 - x_1}{1 + \frac{\mathbf{vV}}{c^2}} = -\frac{\mathbf{V}}{c^2} \frac{l_0}{1 + \frac{\mathbf{vV}}{c^2}},$$

$(A.4.5)$

where $l_0 = x_2 - x_1$ is length of the rod in the reference frame at rest. Substituting this expression into (A.4.3) we obtain:

$$s_{1,2}^2 = c^2\left(1 - \frac{\mathbf{v}^2}{c^2}\right)(t_2 - t_1)^2 - 2\mathbf{v}(t_2 - t_1)(x_2 - x_1) - (x_2 - x_1)^2 =$$

$$c^2\left(1 - \frac{\mathbf{v}^2}{c^2}\right)(t_2 - t_1)^2 - 2\mathbf{v}(t_2 - t_1)l_0 - l_0^2 =$$

$$\left(1 - \frac{\mathbf{v}^2}{c^2}\right)\frac{\mathbf{V}^2}{c^2}\frac{l_0^2}{\left(1 + \frac{\mathbf{vV}}{c^2}\right)^2} + 2\frac{\mathbf{vV}}{c^2}\frac{l_0^2}{1 + \frac{\mathbf{vV}}{c^2}} - l_0^2 =$$

$$l_0^2\left(1 + \frac{\mathbf{vV}}{c^2}\right)^{-2}\left[\left(1 - \frac{\mathbf{v}^2}{c^2}\right)\frac{\mathbf{V}^2}{c^2} + 2\frac{\mathbf{vV}}{c^2}\left(1 + \frac{\mathbf{vV}}{c^2}\right) - \left(1 + \frac{\mathbf{vV}}{c^2}\right)^2\right] =$$

$$l_0^2\left(1 + \frac{\mathbf{vV}}{c^2}\right)^{-2}\left[\boxed{\frac{\mathbf{V}^2}{c^2}} - \boxed{\frac{\mathbf{v}^2\mathbf{V}^2}{c^4}} + \boxed{2\frac{\mathbf{vV}}{c^2}} + \boxed{2\frac{\mathbf{v}^2\mathbf{V}^2}{c^4}} - 1 - \boxed{2\frac{\mathbf{vV}}{c^2}} - \boxed{\frac{\mathbf{v}^2\mathbf{V}^2}{c^4}}\right] =$$

$(A.4.6)$

$$l_0^2\left(1 + \frac{\mathbf{vV}}{c^2}\right)^{-2}\left(\frac{\mathbf{V}^2}{c^2} - 1\right) = -l_0^2 \frac{1 - \frac{\mathbf{V}^2}{c^2}}{\left(1 + \frac{\mathbf{vV}}{c^2}\right)^2}.$$

$$s_{1,2}^2 = -l_0^2 \frac{1 - \frac{\mathbf{V}^2}{c^2}}{\left(1 + \frac{\mathbf{vV}}{c^2}\right)^2}.$$

Comparing (A.4.2) and (A.4.6) we find

$$l = l_0 \frac{\sqrt{1 - \frac{\mathbf{V}^2}{c^2}}}{1 + \frac{\mathbf{vV}}{c^2}}.$$

$(A.4.7)$

# Appendix A.5.

# Relativistic motion with constant acceleration.

Relativistic motion with constant acceleration is a motion under the influence of a force $\vec{f}$, that is constant in value and direction. According to equations of relativistic mechanics we have

$$\frac{d}{dt}\left(\frac{\vec{v}}{\sqrt{1-\frac{v^2}{c^2}}}\right) = \frac{\vec{f}}{m} = \vec{a} \qquad (A.5.1)$$

Integrating equation (A.5.1) over time, we obtain

$$\frac{\vec{v}}{\sqrt{1-\frac{v^2}{c^2}}} = \vec{a}\cdot t + \vec{v}_0,$$

$$\vec{v}_0 = \frac{\vec{v}(0)}{\sqrt{1-\frac{v^2(0)}{c^2}}}. \qquad (A.5.2)$$

From Eq.(A.5.2) we find after squaring

$$\frac{1}{1-\frac{v^2}{c^2}} = 1 + \frac{\vec{a}\cdot t + \vec{v}_0}{c^2}. \qquad (A.5.3)$$

Taking into account Eqs.(A.5.2),Eq.(A.5.3),we obtain

$$\vec{v}(t) = \frac{d\vec{r}(t)}{dt} = \frac{\vec{a}\cdot t + \vec{v}_0}{\sqrt{1+\frac{(\vec{a}\cdot t + \vec{v}_0)^2}{c^2}}}. \qquad (A.5.4)$$

Integrating Eq.(A.5.4),we find

$$\vec{r}(t) = \vec{r}_0 + \frac{\vec{a}c^2}{|\vec{a}|^2}\left[\sqrt{1+\frac{(\vec{a}\cdot t + \vec{v}_0)^2}{c^2}} - 1\right] +$$

$$+\frac{c}{a}\left(\vec{v}_0 - \vec{a}\frac{(\vec{a}\cdot\vec{v}_0)}{|\vec{a}|^2}\right)\ln\left[\frac{|\vec{a}|\cdot t}{c} + \frac{\vec{a}\cdot\vec{v}_0}{c|\vec{a}|} + \sqrt{1+\frac{(\vec{a}\cdot t + \vec{v}_0)^2}{c^2}}\right]. \qquad (A.5.5)$$

# Appendix A.6.

# Relativistic contraction of an accelerated rod.

Let us consider a rod whose velocity and acceleration are directed along its length, but are otherwise arbitrary functions of time.
We assume in general:

(1) That the accelerated rod is rigid, which means that an observer located on the rod does not observe any change of the rod's length.

(2) (**Hypothesis of Locality**) According to the standard theory of relativity, a noninertial observer $O_w$ in (orthogonal i.e. $t \perp x \perp y \perp z$) accelerated frame $K_w = K_w(t,x,y,z)$ is at each instant equivalent to an otherwise identical momentarily comoving inertial observer $O_I$ in standard inertial frame $K_I = K_I(t_I,x_I,y_I,z_I)$ This hypothesis of locality postulates a pointwise physically equivalence between noninertial and ideal inertial observers.(see Appendix 7)

Let us consider a rod whose velocity and acceleration are directed along its length, but are otherwise arbitrary functions of time.

Since the rod is rigid and does not rotate, it is enough to know how one particular point of the rod labeled by $A$ changes its position with time.

Let $F_I = F_I(t_I,x_I) = F_I(t',x')$ be a stationary inertial frame and $F_{\mathbf{V}(t',x')}(t,x)$ the accelerated frame of an observer on the rod. We assume that we know the function $x_A(t'_A)$, so we also know the velocity:

$$\mathbf{V}(t'_A) = \frac{dx_A(t'_A)}{dt'_A}. \qquad (A.6.1)$$

The function (A.6.1) defines the infinite sequence of comoving inertial frames $\mathbf{S}^c_I(t_A)$. The rod is instantaneously at rest for an observer in this frames $\mathbf{S}^c_I(t_A)$. This means that he observes no contraction, i.e., $L_0 = x_B - x_A$, where $L_0$ is the proper length of the rod, while $A$ and $B$ label the back and front ends of the rod, respectively. He observes both ends at the same instant, so $t_B - t_A = 0$. From the Lorentz transformations

$$\begin{aligned}
x'_A &= \gamma(t'_A)[x_A + \mathbf{V}(t'_A)t_A], \\
x'_B &= \gamma(t'_A)[x_B + \mathbf{V}(t'_B)t_B], \\
t'_A &= \gamma(t'_A)\left[t_A + \frac{\mathbf{V}(t'_A)}{c^2}x_A\right], \\
t'_B &= \gamma(t'_A)\left[t_B + \frac{\mathbf{V}(t'_A)}{c^2}x_B\right], \\
\gamma(t'_A) &= \left(1 - \frac{\mathbf{V}^2(t'_A)}{c^2}\right)^{-1/2},
\end{aligned} \qquad (A.6.2)$$

we obtain

$$x'_B - x'_A = \gamma(t'_A)(x_B - x_A) = L_0\gamma(t'_A), \quad (A.6.3)$$

$$t'_A - t'_B = \gamma(t'_A)\frac{\mathbf{V}(t'_A)}{c^2}(x_B - x_A) = L_0\gamma(t'_A)\frac{\mathbf{V}(t'_A)}{c^2}. \quad (A.6.4)$$

From (A.6.3) and the known functions $x'_A(t'_A)$ and (A.6.1) we can find the function $x'_B(t'_A) = x'_A(t'_A) + L_0\gamma(t'_A)$. From (A.6.4) and the known function (A.6.1) we can find the function $t'_A(t'_B)$. Thus we find the function

$$\begin{aligned} x'_B(t'_A(t'_B)) &= x'_A(t'_A(t'_B)) + L_0\gamma(t'_A(t'_B)), \\ x'_B(t'_A(t'_B)) &\triangleq \tilde{x}'_B(t'_B). \end{aligned} \quad (A.6.5)$$

To determine how the rod's length changes with time for an observer in $F_I(t',x')$, both ends of the rod must be observed at the same instant, so $t'_B = t'_A \equiv t'$. Thus the length as a function of time is given by

$$L(t) = \tilde{x}'_B(t') - x_A(t'). \quad (A.6.6)$$

Let us now see how velocity (A.6.1) can be found if the force $F(t_A)$ applied to the point $A$ is known. $F(t_A)$ is the force as seen by an observer in $\mathbf{S}^c_I(t_A)$. We introduce the quantity

$$a(t_A) = \frac{F(t_A)}{m(t_A)}, \quad (A.6.7)$$

which we call acceleration, having in mind that this would be the second time derivative of a position only in the nonrelativistic limit. Here $m(t_A)$ is the proper mass of the rod, which, in general, can also change with time, for example by loosing the fuel of a rocket engine. As shown in [39], if there is only one force, applied to a specific point on an elastic body, and if $F$ and $m$ do not vary with time, then this point moves in the same way as it would move if all mass of the body were concentrated in this point. If acceleration changes with time slowly enough, then this is approximately true for a time-dependent acceleration as well. Here we assume that these conditions are fulfilled. The application point is labeled by $A$. Thus, by a straightforward application of the velocity addition formula (in non-holonomic case), we find that the infinitesimal change of velocity is given by

$$u(t_A + dt_A) = \frac{u(t_A) + a(t_A)dt_A}{1 + \frac{u(t_A)a(t_A)dt_A}{c^2}} = u(t_A) + \left(1 - \frac{u(t_A)}{c^2}\right)a(t_A)dt_A, \quad (A.6.8)$$

where $u(t_A)$ is velocity defined in such a way that $u(t_A(t'_A)) = \mathbf{v}(t'_A)$. Since $u(t_A + dt_A) = u(t_A) + du$, this leads to the differential equation

$$\frac{du(t_A)}{dt_A} = \left(1 - \frac{u^2(t_A)}{c^2}\right)a(t_A). \quad (A.6.9)$$

which can be easily integrated, since $a(t_A)$ is the known function by assumption. Thus we find the function $u(t_A)$. To find the function $\mathbf{v}(t_A)$, we must find the function

$t_A(t'_A)$. We find this from the infinitesimal (non-holonomic) Lorentz transformation

$$dt'_A = \frac{dt_A + \frac{u(t_A)}{c^2} dx_A}{\sqrt{1 - \frac{u^2(t_A)}{c^2}}} \quad (A.6.10)$$

The point on the rod labeled by A does not change, i.e., $dx_A = 0$, so (A.6.10) can be integrated as

$$t'_A = \int_0^{t_A} \frac{d\bar{t}_A}{\sqrt{1 - \frac{u^2(\bar{t}_A)}{c^2}}}. \quad (A.6.11)$$

which gives a function

$$t'_A = f(t_A) \quad (A.6.12)$$

and thus

$$t_A = f^{-1}(t'_A). \quad (A.6.13)$$

Now we consider the case of a rod which is at rest for $t' < 0$, but at $t' = 0$ it turns on its engine which gives the constant acceleration a to the application point during a finite time interval $T$, after which the engine turns off. From (A.6.9) and (A.6.11) for $t'_A < T'$ we find

$$u(t_A) = c\tanh\left(\frac{at_A}{c}\right) \quad (A.6.14)$$

$$t'_A(t_A) = \frac{c}{a}\sinh\left(\frac{at_A}{c}\right) \quad (A.6.15)$$

and thus

$$\mathbf{v}(t'_A) = \begin{cases} \dfrac{at'_A}{\sqrt{1 + \left(\frac{at'_A}{c}\right)^2}}, & 0 \le t'_A \le T'; \\[2ex] \dfrac{aT'_A}{\sqrt{1 + \left(\frac{aT'}{c}\right)^2}}, & t'_A \ge T'. \end{cases} \quad (A.6.16) \text{ where}$$

$$T' = \frac{c}{a}\sinh\left(\frac{aT}{c}\right). \quad (A.6.17) \text{ From the initial condition}$$

$x'_A(t'_A = 0) = 0$ we obtain

$$x'_A(t'_A) = \begin{cases} \sqrt{\left(\frac{c^2}{a}\right)^2 + (ct'_A)^2} - \frac{c^2}{a}, 0 \leq t'_A \leq T'; \\ \\ \dfrac{aT't'_A}{\sqrt{1+\left(\frac{aT'}{c}\right)^2}} + \dfrac{c^2}{a}\left(\dfrac{1}{\sqrt{1+\left(\frac{aT'}{c}\right)^2}} - 1\right), t'_A \geq T'. \end{cases} \quad (A.6.18)$$

Thus we find

$$t'_A(t'_B) = \begin{cases} \dfrac{t'_B}{1+\frac{aL_0}{c^2}}, 0 \leq t'_B \leq T'_+; \\ \\ t'_B - \dfrac{aL_0 T'}{c^2}, t'_B \geq T'_+. \end{cases} \quad (A.6.19)$$

$$\tilde{x}'_B(t'_B) = \begin{cases} \left(\dfrac{c^2}{a}+L_0\right)\sqrt{1+\dfrac{\left(\frac{at'_B}{c}\right)^2}{\left(1+\frac{aL_0}{c^2}\right)^2}} - \dfrac{c^2}{a}, 0 \leq t'_B \leq T'_+; \\ \\ \dfrac{\left(\frac{c^2}{a}+L_0+aT't'_B\right)}{\sqrt{1+\left(\frac{aT'}{c}\right)^2}}, t'_B \geq T'_+. \end{cases} \quad (A.6.20)$$

$$L(t') = \begin{cases} \left(\dfrac{c^2}{a}+L_0\right)\sqrt{1+\dfrac{\left(\frac{at'}{c}\right)^2}{\left(1+\frac{aL_0}{c^2}\right)^2}} - \dfrac{c^2}{a}\sqrt{1+\left(\frac{at'}{c}\right)^2}, 0 \leq t' \leq T'; \\ \\ \left(\dfrac{c^2}{a}+L_0\right)\sqrt{1+\dfrac{\left(\frac{at'}{c}\right)^2}{\left(1+\frac{aL_0}{c^2}\right)^2}} - \\ -\dfrac{1}{\sqrt{1+\left(\frac{aT'}{c}\right)^2}}\left(\dfrac{c^2}{a}+\dfrac{aT't'}{c^2}\right), T' \leq t' \leq T'_+; \\ \\ \dfrac{L_0}{\sqrt{1+\left(\frac{aT'}{c}\right)^2}}, t' \geq T'_+. \end{cases} \quad (A.6.21)$$

Or in equivalent form

$$\tilde{x}'_B(t'_B) = \begin{cases} \frac{a}{c^2}\sqrt{\left(1+\frac{aL_0}{c^2}\right)^2 + \left(\frac{at'_B}{c}\right)^2} - \frac{c^2}{a}, 0 \leq t'_B \leq T'_+; \\ \dfrac{\left(\frac{c^2}{a} + L_0 + aT't'_B\right)}{\sqrt{1+\left(\frac{aT'}{c}\right)^2}}, t'_B \geq T'_+. \end{cases} \qquad (A.6.22)$$

$$L(t') = \begin{cases} \frac{a}{c^2}\sqrt{\left(1+\frac{aL_0}{c^2}\right)^2 + \left(\frac{at'}{c}\right)^2} - \frac{c^2}{a}\sqrt{1+\left(\frac{at'}{c}\right)^2}, 0 \leq t' \leq T'; \\ \frac{a}{c^2}\sqrt{\left(1+\frac{aL_0}{c^2}\right)^2 + \left(\frac{at'}{c}\right)^2} - \\ -\dfrac{1}{\sqrt{1+\left(\frac{aT'}{c}\right)^2}}\left(\frac{c^2}{a}+\frac{aT't'}{c^2}\right), T' \leq t' \leq T'_+; \\ \dfrac{L_0}{\sqrt{1+\left(\frac{aT'}{c}\right)^2}}, t' \geq T'_+. \end{cases} \qquad (A.6.23)$$

# Appendix A.7.

# Hypothesis of locality.

According to the standard theory of relativity, a noninertial observer is at each instant equivalent to an otherwise identical momentarily comoving inertial observer. This *hypothesis of locality* postulates a pointwise equivalence between noninertial and ideal inertial observers. The hypothesis of locality originates from Newtonian mechanics of point particles. The state of a classical particle is determined by its position and velocity at a given instant of time. If the force on the particle is turned off at some instant, the particle will follow the osculating straight line. Thus the assumption of locality is automatically satisfied in this case, since the noninertial and the ideal osculating inertial observer share the same state and are hence equivalent.

This is why the discussion of accelerated systems in classical mechanics does not require any new hypothesis. In classical electrodynamics, however, we need to deal with classical electromagnetic waves; their interactions can only be considered poinlike in the geometric optics limit. If all physical phenomena could be reduced to pointlike coincidences of classical particles and rays of radiation, then the hypothesis of locality would be exactly valid [38]. However, in general classical waves have intrinsic extensions in space and time characterized by their wavelengths and periods.

Imagine a background global inertial reference frame $K = K(t',x',y',z')$ with coordinates $t',x',y',z'$ and the class of fundamental observers in this frame. Each fundamental observer is by definition at rest in this frame and carries an orthonormal tetrad frame $\tilde{\lambda}^\mu_{(\alpha)} = \delta^\mu_\alpha, \alpha = 0,1,2,3$ such that $\tilde{\lambda}^\mu_{(0)}$ is tangent to its worldline and $\tilde{\lambda}^\mu_{(i)}, i = 1,2,3$ characterize its spatial frame. Consider now an accelerated observer $O_a$ following a worldline $x^\alpha_D(\tau)$ with four-velocity $u^\alpha_D(\tau) = dx^\alpha_D(\tau)/d\tau$ and translational acceleration $A^\alpha_D = du^\alpha_D(\tau)/d\tau$. Here $\tau$ is a temporal parameter along $x_D(\tau)$ defined by $d\tau/dt' = \gamma^{-1}(t')$, where $\gamma$ is the Lorentz factor, $\gamma^{-1}(t') = \sqrt{1 - \mathbf{v}^2(t')/c^2}$ and $\mathbf{v}(t')$ is the velocity of the accelerated device. Note that $\langle \mathbf{A}_D, \mathbf{u}_D \rangle = 0$ so that $\mathbf{A}_D$ is a spacelike vector such that $\langle \mathbf{A}_D, \mathbf{A}_D \rangle = w^2$ where $w$ is the magnitude of the translational acceleration.

**Axiom** *(**Hypothesis of Locality**) According to the standard theory of relativity, a noninertial observer $O_w$ in (orthogonal i.e. $t \perp x \perp y \perp z$) accelerated frame $K_w = K_w(t,x,y,z)$ is at each instant equivalent to an otherwise identical momentarily comoving inertial observer $O_I$ in standard inertial frame $K_I = K_I(t_I, x_I, y_I, z_I)$ This hypothesis of locality postulates a pointwise physically equivalence between noninertial and ideal inertial observers.*

**Remark** *1. The hypothesis of locality originates from Newtonian mechanics of point particles. The state of a classical particle is determined by its position and velocity at a given instant of time. If the force on the particle is turned off at some instant, the particle will follow the osculating straight line. Thus the assumption of locality is automatically satisfied in this case, since the noninertial and the ideal osculating inertial observer share the same state and are hence physically equivalent. It follows from the hypothesis of locality that the accelerated device is at each instant endowed with an orthonormal tetrad frame $\lambda^\mu_{(\alpha)}$ as well such that $\lambda^\mu_{(0)} = u^\mu_D$ and $\eta_{\mu\nu}\lambda^\mu_{(\alpha)}\lambda^\nu_{(\beta)} = \eta_{\alpha\beta}$, where $\eta_{\mu\nu}$ is the Minkowski metric tensor with signature $+2$. Moreover, the application of the hypothesis of locality to the measurement of time by the accelerated observer implies that $\tau$ is the proper time along $x_D$. The variation of the orthonormal tetrad along the path of the reference observer is given by*

$$\frac{d\lambda^\mu_{(\alpha)}(\tau)}{d\tau} = \Phi^\beta_\alpha(\tau)\lambda^\mu_{(\beta)} \qquad (A.7.1)$$

*where $\Phi_{\alpha\beta}$ is an antisymmetric acceleration tensor with "electric" part $\Phi_{0i} = a_i$ and "magnetic" part $\Phi_{ij} = \epsilon_{ijk}\Omega^k$ Here $\mathbf{a}$ and $\mathbf{\Omega}$ are spacetime scalars that represent respectively the translational acceleration, $a_i = \mathbf{A}_D \cdot \lambda_{(i)}$, and the rotational frequency of the local spatial frame with respect to the local nonrotating (i.e. Fermi-Walker transported) frame. It is useful to consider the invariants*

$$I = \frac{1}{2}\Phi_{\alpha\beta}\Phi^{\alpha\beta},$$
$$I^* = \frac{1}{2}\Phi^*_{\alpha\beta}\Phi^{\alpha\beta},$$
(A.7.2)

where $\Phi^*_{\alpha\beta}$ is the dual acceleration tensor. The significance of $I = \mathbf{a}^2 + \mathbf{\Omega}^2, I^* = -\mathbf{a}\cdot\mathbf{\Omega}$ lies in the fact that these depend merely on the acceleration, while depends on the velocity as well as the acceleration of the reference observer. At any given instant of proper time $\tau$, $I$ and $I^*$ are independent of any local Lorentz tranformations of the tetrad frame in eq. (A.7.1); therefore, they represent the velocityindependent content of the acceleration tensor . The proper acceleration scales can then be defined in terms of $I$ and $I^*$, i.e. $|I|^{-1/2}$ and $|I^*|^{-1/2}$. Consider now a geodesic coordinate system established along the worldline $x_D(\tau)$ of the fiducial observer. At any given instant $\tau$ along $x_D(\tau)$, the straight spacelike geodesiclines orthogonal to $x_D(\tau)$ span a hyperplane that is Euclidean space. Let $x^\mu$ be the coordinates of a point on such a hypersurface and let $X^\mu$ be the corresponding geodesic coordinates. Then

$$\tau = X^0, x^\mu = x_D^\mu(\tau) + X^i \lambda^\mu_{(i)}(\tau) \qquad (A.7.3)$$

completely characterize the transformation to the new geodesic coordinates. Writing the metric of the background system as $ds^2 = \eta_{\mu\nu}dx^\mu dx^\nu$ and differentiating system (A.7.2), one finds with the help of equation (A.7.1) that $ds^2 = g_{\mu\nu}dX^\mu dX^\nu$, where

$$g_{00} = -S,$$
$$g_{0i} = U_i, \qquad (A.7.4)$$
$$g_{ij} = \delta_{ij}.$$

Here $S$ and $\mathbf{U}$ are given by

$$S = (1 + \mathbf{a}\cdot\mathbf{X})^2 - \mathbf{U}^2,$$
$$\mathbf{U} = \mathbf{\Omega}\times\mathbf{X}. \qquad (A.7.5)$$

where $\mathbf{a}$ and $\mathbf{\Omega}$ are in general functions of $X^0$. One can show that $\det(g_{\mu\nu}) = g$ is given by

| Notation | Axiom | Remark | Corollary |

$$g = (1 + \mathbf{a}\cdot\mathbf{X})^2 \qquad (A.7.6)$$

so that the inverse metric tensor can be expressed as

$$g^{00} = g^{-1},$$
$$g^{0i} = -U^i g^{-1}, \qquad (A.7.7)$$
$$g^{ij} = \delta^{ij} + U^i U^j g^{-1}.$$

A detailed examination of the geodesic coordinate system shows that these coordinates are admissible as long as $g_{00} < 0$. The boundary of the admissible region is characterized by $S = 0$. At each instant of time $X^0, S = 0$ is a quadratic equation in the spatial coordinates and represents a surface. Such surfaces have been classified under the Euclidean group into seventeen standard forms called quadric surfaces. Specifically,

$$S(X^0, X) = 1 + 2a_i(X^0)X^i + M_{ij}(X^0)X^i X^j = 0, \qquad (A.7.8)$$

where $\mathbf{M} = (M_{ij})$ is a symmetric matrix with components

$$M_{ij} = a_i a_j + \Omega_i \Omega_j - 2\delta_{ij}. \qquad (A.7.9)$$

It is possible to show that $\mathbf{M}$ has eigenvalues

$$\mu_0 = -\Omega^2, \mu_\pm = -1 \pm \sqrt{I^2 + I^{*2}}, \qquad (A.7.10)$$

so that $\mu_+ \geq 0, \mu_0 \leq 0, \mu_- \leq 0$ and

$$det(M_{ij}) = \mu_+ \mu_0 \mu_- = \Omega^2 (\mathbf{a} \cdot \mathbf{\Omega})^2. \qquad (A.7.11)$$

Consider first the general case in which $det(M_{ij}) \neq 0$. It follows that $\mathbf{M}$ has an inverse and it is possible to show that

$$\left(M^{-1}\right)_{ij} a^i a^j = 1 \qquad (A.7.12)$$

matrix $M$ can be diagonalized at any instant $X_0$ by a rotation of spatial coordinates. The standard form of the quadric surface represented by $S = 0$ is then achieved by completing the squares in eq. (A.7.8) followed by a translation to new coordinates. More explicitly, let $R$ be the orthogonal matrix such that $R^{-1}MR$ is diagonal with diagonal elements $(\mu_+, \mu_0, \mu_-)$. Using the rotated spatial coordinates $\hat{X} = R^{-1}X$ and corresponding parameters $\hat{a} = R^{-1}a$, the translations

$$\xi = \hat{X}_1 + \frac{\hat{a}_1}{\mu_+}, \eta = \hat{X}_2 + \frac{\hat{a}_2}{\mu_0}, \zeta = \hat{X}_3 + \frac{\hat{a}_3}{\mu_-}, \qquad (A.7.13)$$

define a new spatial coordinate system $(\xi, \eta, \zeta)$. In terms of these new coordinates, Eq.(A.7.8) then takes the form

$$|\mu_+|\xi^2 - |\mu_0|\eta^2 - |\mu_-|\zeta^2 = 0, \qquad (A.7.14)$$

which represents a real quadric cone (i.e. an elliptic cone) in general. In deriving this result, the relation

$$\frac{\hat{a}_1}{\mu_+} + \frac{\hat{a}_2}{\mu_0} + \frac{\hat{a}_3}{\mu_-} = 1, \qquad (A.7.15)$$

which follows from Eq. (A.7.12), has been employed. An important feature of Eq. (A.7.14) should be noted: the extent of validity of the admissible coordinates is determined by the acceleration lengths that are implicit in the eigenvalues of $M$.

If $M$ is a singular matrix, then either $\Omega = 0$, in which case the quadric surface degenerates to coincident planes, or $\Omega \neq 0$ but $\mathbf{a} \cdot \mathbf{\Omega} = 0$, in which case the quadric surface is a cylinder. This cylinder is hyperbolic for $\Omega^2 < a^2$ and parabolic for $\Omega^2 = a^2$. It is a real elliptic cylinder for $\Omega^2 > a^2$. These assertions can be simply demonstrated by working in a system of coordinates that is obtained from the $(X_1, X_2, X_3)$ system by a rotation such that in the new system one coordinate axis is parallel to a and another is parallel to $\mathbf{\Omega}$. For $a = 0$, Eq. (A.7.8) reduces to a circular cylinder of radius $\Omega^{-1}$.